\documentclass[twocolumn]{aastex631}
\usepackage{xcolor}
\usepackage{amsmath}

\def\at{AT\,2024wpp}

\def\xmm{\emph{XMM-Newton}}

\def\chandra{\emph{CXO}}
\def\nustar{\emph{NuSTAR}}

\begin{document}

\title{The Most Luminous Known Fast Blue Optical Transient AT\,2024wpp: Unprecedented Evolution and Properties in the X-rays and Radio}

\correspondingauthor{Nayana A.J.}
\email{nayana@berkeley.edu}

\author[0000-0002-8070-5400]{Nayana A. J.}
\affiliation{Department of Astronomy, University of California, Berkeley, CA 94720-3411, USA}
\affiliation{Berkeley Center for Multi-messenger Research on Astrophysical Transients and Outreach (Multi-RAPTOR), University of California, Berkeley, CA 94720-3411, USA}

\author[0000-0003-4768-7586]{Raffaella Margutti}
\affiliation{Department of Astronomy, University of California, Berkeley, CA 94720-3411, USA}
\affiliation{Department of Physics, University of California, 366 Physics North MC 7300,
Berkeley, CA 94720, USA}
\affiliation{Berkeley Center for Multi-messenger Research on Astrophysical Transients and Outreach (Multi-RAPTOR), University of California, Berkeley, CA 94720-3411, USA}

\author[0000-0000-0000-0000]{Eli Wiston}
\affiliation{Department of Astronomy, University of California, Berkeley, CA 94720-3411, USA}
\affiliation{Berkeley Center for Multi-messenger Research on Astrophysical Transients and Outreach (Multi-RAPTOR), University of California, Berkeley, CA 94720-3411, USA}

\author[0000-0003-1792-2338]{Tanmoy Laskar}
\affiliation{Department of Physics \& Astronomy, University of Utah, Salt Lake City, UT 84112, USA}

\author[0000-0002-0786-7307]{Giulia Migliori}
\affiliation{INAF Istituto di Radioastronomia, via Gobetti 101, 40129 Bologna, Italy}

\author[0000-0002-7706-5668]{Ryan Chornock} 
\affiliation{Department of Astronomy, University of California, Berkeley, CA 94720-3411, USA}
\affiliation{Berkeley Center for Multi-messenger Research on Astrophysical Transients and Outreach (Multi-RAPTOR), University of California, Berkeley, CA 94720-3411, USA}

\author[0000-0002-2801-766X]{Timothy J. Galvin}
\affiliation{ATNF, CSIRO Space \& Astronomy, PO Box 1130, Bentley, WA 6102, Australia}

\author[0000-0002-2249-0595]{Natalie LeBaron}
\affiliation{Department of Astronomy, University of California, Berkeley, CA 94720-3411, USA}
\affiliation{Berkeley Center for Multi-messenger Research on Astrophysical Transients and Outreach (Multi-RAPTOR), University of California, Berkeley, CA 94720-3411, USA}

\author[0000-0003-2349-101X]{Aprajita Hajela}
\affiliation{DARK, Niels Bohr Institute, University of Copenhagen, Jagtvej 155, 2200 Copenhagen, Denmark}

\author[0000-0003-0528-202X]{Collin T. Christy}
\affiliation{Department of Astronomy/Steward Observatory, 933 North Cherry Avenue, Room N204, Tucson, AZ 85721-0065, USA}

\author[0000-0003-0466-3779]{Itai Sfaradi}
\affiliation{Department of Astronomy, University of California, Berkeley, CA 94720-3411, USA}
\affiliation{Berkeley Center for Multi-messenger Research on Astrophysical Transients and Outreach (Multi-RAPTOR), University of California, Berkeley, CA 94720-3411, USA}

\author[0000-0002-6347-3089]{Daichi Tsuna}
\affiliation{TAPIR, Mailcode 350-17, California Institute of Technology, Pasadena, CA 91125, USA}
\affiliation{Research Center for the Early Universe (RESCEU), School of Science, The University of Tokyo, 7-3-1 Hongo, Bunkyo-ku, Tokyo 113-0033, Japan}

\author[0000-0001-5674-8403]{Olivia Aspegren}
\affiliation{Department of Astronomy, University of California, Berkeley, CA 94720-3411, USA}

\author[0000-0002-3137-4633]{Fabio De Colle}
\affiliation{Instituto de Ciencias Nucleares, Universidad Nacional Aut\'{o}noma de M\'{e}xico, Apartado Postal 70-264, 04510 M\'{e}xico, CDMX, Mexico} 

\author[0000-0002-4670-7509]{Brian D. Metzger}
\affiliation{Department of Physics and Columbia Astrophysics Laboratory, Columbia University, New York, NY 10027, USA} 
\affil{Center for Computational Astrophysics, Flatiron Institute, 162 5th Ave, New York, NY 10010, USA} 

\author[0000-0002-1568-7461]{Wenbin Lu}
\affiliation{Department of Astronomy, University of California, Berkeley, CA 94720-3411, USA}
\affiliation{Theoretical Astrophysics Center, University of California, Berkeley, CA 94720-3411, USA} 
\affiliation{Berkeley Center for Multi-messenger Research on Astrophysical Transients and Outreach (Multi-RAPTOR), University of California, Berkeley, CA 94720-3411, USA}

\author[0000-0001-7833-1043]{Paz Beniamini}
\affiliation{Department of Natural Sciences, The Open University of Israel, P.O Box 808, Ra’anana 4353701, Israel } 

\author[0000-0002-5981-1022]{Daniel Kasen}
\affiliation{Department of Physics and Astronomy, University of California, Berkeley, CA 94720, USA} 
\affiliation{Nuclear Science Division, Lawrence Berkeley National Laboratory, 1 Cyclotron Road, Berkeley, CA 94720, USA}
\affiliation{Berkeley Center for Multi-messenger Research on Astrophysical Transients and Outreach (Multi-RAPTOR), University of California, Berkeley, CA 94720-3411, USA}

\author[0000-0002-9392-9681]{Edo Berger}
\affiliation{Center for Astrophysics \textbar{} Harvard \& Smithsonian, 60 Garden Street, Cambridge, MA 02138-1516, USA}

\author[0000-0002-1984-2932]{Brian W. Grefenstette}
\affiliation{Space Radiation Laboratory
California Institute of Technology 
1200 E California Blvd 
Pasadena, CA 91125, USA}

\author[0000-0002-8297-2473]{Kate~D.~Alexander}
\affiliation{Department of Astronomy/Steward Observatory, 933 North Cherry Avenue, Rm. N204, Tucson, AZ 85721-0065, USA}

\author[0000-0003-3533-7183]{G. C. Anupama}
\affiliation{Indian Institute of Astrophysics, II Block, Koramangala, Bangalore 560034, India}

\author[0000-0001-5126-6237]{Deanne L. Coppejans}
\affiliation{Department of Physics, University of Warwick, Coventry CV4 7AL, UK}

\author[0000-0001-5576-2254]{Luigi F. Cruz}
\affiliation{SETI Institute, 339 Bernardo Ave, Suite 200 Mountain View, CA 94043, USA}

\author[0000-0003-3197-2294]{David R DeBoer}
\affiliation{Radio Astronomy Laboratory, University of California, Berkeley, CA, 94720 USA}
\affiliation{Sub-department of Astrophysics, University of Oxford, Oxford, OX1-3RH, UK}

\author[0000-0001-7081-0082]{Maria R. Drout}
\affiliation{David A. Dunlap Department of Astronomy \& Astrophysics, University of Toronto, 50 St. George Street, Toronto, ON M5S 3H4, Canada }

\author[0000-0002-0161-7243]{Wael Farah}
\affiliation{Department of Astronomy, University of California, Berkeley, CA 94720-3411, USA}
\affiliation{SETI Institute, 339 Bernardo Ave, Suite 200 Mountain View, CA 94043, USA}
\affiliation{Berkeley SETI Research Centre, University of California, Berkeley, CA 94720, USA}

\author[0000-0003-2868-489X]{Xiaoshan Huang}
\affiliation{California Institute of Technology, TAPIR, Mail Code 350-17, Pasadena, CA 91125, USA }

\author[0000-0002-3934-2644]{W.~V.~Jacobson-Gal\'{a}n}
\altaffiliation{NASA Hubble Fellow}
\affil{Department of Astronomy and Astrophysics, California Institute of Technology, Pasadena, CA 91125, USA}

\author[0000-0002-0763-3885]{Dan Milisavljevic}
\affiliation{Purdue University, Department of Physics and Astronomy, 525 Northwestern Ave, West Lafayette, IN 47907 }
\affiliation{Integrative Data Science Initiative, Purdue University, West Lafayette, IN 47907, USA}

\author[0000-0002-3430-7671]{Alexander~W.~Pollak}
\affiliation{SETI Institute, 339 Bernardo Ave, Suite 200 Mountain View, CA 94043, USA}

\author[0000-0002-6006-9574]{Nathan~J.~Roth}
\affiliation{Lawrence Livermore National Laboratory, P.O. Box 808, Livermore, CA 94550, USA }
\affiliation{Department of Physics, American University, 4400 Massachusetts Avenue NW, Washington, DC 20016, USA}

\author[0000-0001-8023-4912]{Huei Sears}
\affiliation{Department of Physics and Astronomy, Rutgers, the State University of New Jersey, 136 Frelinghuysen Road, Piscataway, NJ 08854-8019, USA}

\author[0000-0003-2828-7720]{Andrew Siemion}
\affiliation{Astrophysics, Department of Physics, University of Oxford, Keble Road, Oxford, OX1 3RH, UK}
\affiliation{Breakthrough Listen, Astrophysics, Department of Physics, The University of Oxford, Keble Road, Oxford OX1 3RH, UK}
\affiliation{SETI Institute, 339 Bernardo Ave, Suite 200 Mountain View, CA 94043, USA}
\affiliation{Berkeley SETI Research Centre, University of California, Berkeley, CA 94720, USA}
\affiliation{Department of Physics and Astronomy, University of Manchester, UK}
\affiliation{University of Malta, Institute of Space Sciences and Astronomy, Msida, MSD2080, Malta}

\author[0000-0001-7057-4999]{Sofia Z. Sheikh}
\affiliation{SETI Institute, 339 Bernardo Ave, Suite 200 Mountain View, CA 94043, USA}
\affiliation{Berkeley SETI Research Centre, University of California, Berkeley, CA 94720, USA}

\author[0000-0002-5872-6061]{James F. Steiner}
\affiliation{Center for Astrophysics \textbar{} Harvard \& Smithsonian, 60 Garden Street, Cambridge, MA 02138-1516, USA}

\author[0000-0003-1336-4746]{Indrek Vurm}
\affiliation{Tartu Observatory, University of Tartu, Tõravere, 61602 Tartumaa, Estonia }
 
\begin{abstract}
We present X-ray (0.3--79 keV) and radio (0.25--203\,GHz) observations of the most luminous Fast Blue Optical Transient (LFBOT) AT\,2024wpp at $z=0.0868$, spanning 2--280\,days after first light. AT\,2024wpp shows luminous ($L_{\rm X} \approx 1.5 \times 10^{43}\, \rm erg\,s^{-1}$), variable X-ray emission with a Compton hump peaking at $\delta t \approx 50$\,days. The X-ray spectrum evolves from a soft ($F_{\nu} \propto \nu^{-0.6}$) to an extremely hard state ($F_{\nu} \propto \nu^{1.26}$) accompanied by a re-brightening at $\delta t \approx 50$\,days. The X-ray emission properties favor an embedded high-energy source shining through asymmetric expanding ejecta. We detect radio emission peaking at $L_{\rm 9\,GHz} \approx 1.7 \times 10^{29}\,\rm erg\,s^{-1}\,Hz^{-1}$ at $\delta t \approx 73$\,days. The spectral evolution is unprecedented: the early millimeter fluxes rise nearly an order of magnitude during $\delta t \approx 17-32$\,days followed by a decline in spectral peak fluxes. We model the radio emission as synchrotron radiation from an expanding blast wave interacting with a dense environment ($\dot{M} \sim 10^{-3}\, \rm M_{\odot}\,yr^{-1}$ for $v_{\rm w} = 1000\,\rm km\,s^{-1}$). The inferred outflow velocities increase from $\Gamma \beta c \approx 0.07\, \rm to\,0.42c$ during $\delta t \approx 32-73$\,days, indicating an accelerating blast-wave. We interpret these observations as a  shock propagating through a dense shell of radius $\approx 10^{16}$\,cm, then accelerating into a steep density profile $\rho_{\rm CSM}(r) \propto r^{-3.1}$. All radio-bright LFBOTs exhibit similar circumstellar medium (CSM) density profiles ($\rho_{\rm CSM} \propto r^{-3}$), suggesting similar progenitor processes. The X-ray and radio properties favor a progenitor involving super-Eddington accretion onto a compact object launching mildly-relativistic disk-wind outflows. 
\end{abstract}


\section{Introduction} \label{sec:intro}
High cadence, wide-field optical transient surveys have been populating the phase-space of transients with a variety of new fast-timescale events. A note-worthy class is Fast Blue Optical Transients (FBOTs) characterized by rapid rise to peak brightness ($\lesssim 10$\,d), persistent blue colors, and peak optical luminosities reaching $L_{\rm pk} \gtrsim 10^{45}\, \rm erg\,s^{-1}$. These properties are difficult to reconcile in traditional supernova (SN) models \citep{Drout2014,Arcavi2016,Tanaka2016,Pursiainen2018,Rest2018}. FBOTs are intrinsically rare with a rate between 1--10\% of the core-collapse supernovae (CCSNe) rate in the local Universe \citep{Drout2014,Pursiainen2018,Tampo2020,Li2011}. Luminous FBOTs (LFBOTs), a sub-class of FBOTs with $L_{\rm pk} > 10^{43}\, \rm erg\,s^{-1}$, show bright X-ray and radio emission. LFBOTs are even rarer with an intrinsic rate of only $\lesssim 0.1\%$ of that of CCSNe \citep{coppejans2020,Ho2020-AT2018lug,Ho2023-FBOT-rate}. 

While tens of LFBOTs have been detected in optical surveys, only a small fraction have been followed up in X-ray and radio bands. To date, there are only seven LFBOTs with detailed X-ray and radio observations: the prototypical LFBOT AT\,2018cow \citep{Ho2019-cow,Margutti2019-cow}, AT\,2018lug \citep{Ho2020-AT2018lug}, the first LFBOT with X-ray and radio detection CSS161010 \citep{coppejans2020}, AT\,2020xnd \citep{Bright2021-xnd,Ho2022-xnd}, AT\,2020mrf \citep{Yao2021-mrf}, AT\,2022tsd \citep{Ho2023-at2022tsd,Matthews23}, and AT\,2023fhn \citep{Chrimes2024a-fhn,Chrimes2024b-fhn}. These studies resulted in various critical insights into the nature of LFBOTs. Luminous and variable X-ray emission from AT\,2018cow was interpreted as evidence for a central engine powering the transient \citep{Margutti2019-cow}. AT\,2020mrf \citep{Yao2021-mrf} and AT\,2022tsd \citep{Ho2023-at2022tsd} showed extremely luminous X-ray emission, exceeding that of AT\,2018cow by an order of magnitude  and comparable to that of cosmological GRBs. Meanwhile, radio observations have revealed a range of outflow velocities; mildly-relativistic in the case of CSS161010 \citep{coppejans2020}, AT\,2020lug \citep{Ho2020-AT2018lug}, and AT\,2023fhn \citep{Chrimes2024b-fhn}, and non-relativistic in the case of AT\,2018cow \citep{Ho2019-cow,Margutti2019-cow}, AT\,2020mrf \citep{Yao2021-mrf}, AT\,2020xnd \citep{Bright2021-xnd,Ho2022-xnd}, and AT\,2022tsd \citep{Ho2023-at2022tsd}.  Although these studies on individual events have significantly advanced our understanding of FBOTs, the overall sample remains small. Hence, it is still unclear whether the presence of a central engine, the origin of high-energy emission, and the observed range in outflow velocities are generic features of FBOTs or reflect heterogeneity in the population. This highlights the need for detailed X-ray and radio observations of additional FBOTs.

AT\,2024wpp is the newest addition to the class of LFBOTs with detailed multiwavelength observations. It was discovered on 2024 September 25.44 UT (MJD 60578.4) in the ZTF survey data \citep{Ho2024-AT2024wpp} at a redshift of $z=0.0868$ \citep{Perley2024}. The transient brightened by $\approx$3 mag in a day with blue colors $g-r=-0.4$ mag. X-ray emission was detected in the subsequent \textit{Swift} \citep{Srinivasaragavan2024} and \textit{NuSTAR} \citep{Margutti2024-wpp-Nustar} follow-up observations. Radio emission was detected in the Very Large Array (VLA) X (10 GHz) and Ku (15 GHz) band observations with a spectral luminosity of $L_{\rm 10\,GHz} \approx 5 \times 10^{28}\, \rm erg\,s^{-1}\,Hz^{-1}$ at $\approx 29$\,d post discovery \citep{Schroeder2024}. \cite{Ofek2025} reported non-detection of any minute time-scale optical flares in \at{} with a $2\sigma$ upper limit of $< 0.02$ on the flare's duty cycle. Our detailed UV-optical-NIR campaign of \at{} in the first 100\,days is presented in our companion paper (LeBaron et al. 2025, hereafter Paper I) and we summarize the main observational findings below. \at{} is the most luminous FBOT to date with peak luminosity $L_{\rm pk} \approx (2-4) \times 10^{45}$\,ergs (in agreement with \citealt{Pursiainen25}), with a detailed UV lightcurve that samples the pre-peak phase of an FBOT for the first time. The UV-optical spectrum remains featureless and dominated by blue thermal continuum emission for weeks. The black-body temperature at optical peak is $T > 30000$\,K and remains at $T \gtrsim 20000$\,K for weeks. At $\delta t \approx 35$\,days, faint H and He spectral features are detected with some similarities to the phenomenology of AT\,2018cow \citep{Margutti2019-cow,Perley2019}. Finally, we find evidence for a NIR excess of emission, which might be related to pre-existing dust or free-free emission in a high-density medium. Despite being at a distance of 411 Mpc, the exceptional luminosity of \at{} allowed us to carry out unprecedented multiwavelength follow-up observations in the optical, UV, NIR, X-ray, and radio bands. This extensive data set of \at{} is superior to even the prototypical FBOT AT\,2018cow and provides us the unique opportunity to study this transient in exquisite detail. 
 
In this paper, we present extensive X-ray (soft to hard) and radio (sub-GHz to millimeter band) follow-up observations of AT\,2024wpp spanning $\delta t \approx 2-280$\,days after first light. We refer to Paper I where applicable to place our results in a broader context and build a comprehensive physical picture.  The paper is structured as follows. We present the X-ray and radio observations in \S \ref{Sec:Xray} and \S \ref{Sec:Radio}, respectively. \S \ref{Sec:Xraymodeling} presents inferences based on X-ray observations. The properties of the shock wave and the environment are presented in \S \ref{Sec:radio-modeling} and \S \ref{sec:origin of mm emission} based on radio observations. We discuss the properties of \at{} in the context of other LFBOTs in \S \ref{Sec:comparisonRadio-Xray}. Plausible physical scenarios are discussed in \S \ref{Sec:physical models} and conclusions are drawn in \S \ref{Sec:conclusions}.
We adopt the cosmological parameters of Lambda cold dark matter $\Lambda$CDM $H_0=67.4\,\rm{km\,s^{-1}Mpc^{-1}}$, $\Omega_m=0.315$, $\Omega_{\Lambda}=0.685$ \citep{Cosmology20}. For these parameters, the redshift $z=0.0868$ of \at\, 
\citep{Perley24wppredshift} corresponds to a luminosity distance of  411 Mpc. Following Paper I, we adopt MJD 60578.3 as our reference time t=0 days. Times are in the observer frame unless noted otherwise. 
\section{Broad-band X-ray Observations} \label{Sec:Xray}
\subsection{Swift-XRT (0.3--10 keV)}  \label{SubSec:XRT}
Prompt observations of \at\, with the X-Ray Telescope (XRT; \citealt{Burrows05}) onboard the Neil Gehrels \emph{Swift} observatory \citep{Gehrels04} were obtained starting on 2024-09-27, 09:45:24\,UT ($\delta t=$2.1\,days, PI: Coughlin, exposure time of $\approx$2.2\,ks, target ID 16843). We triggered an intense campaign under our \emph{Swift} Guest Observer program (PI: Margutti, total of 116\,ks, target IDs: 16848, 18973) covering the time interval $\delta t = 4.7-118.7$\,days.  We processed the  XRT data with HEASoft v6.34 and corresponding calibration files. We extracted a 0.3--10 keV count-rate light-curve and rebinned to have a minimum number of 5 counts per bin following standard procedures (\citealt{Evans09,Margutti13}).

We extracted several XRT spectra at salient phases of the FBOT X-ray light-curve (e.g., initial ``plateau'' at $\delta t=2.1\rm{days\,}-6.6$\,days; decay phase at $\delta t=6.7\rm{days\,}-34.5$\,days;  late-time ``flare'' at $\delta t=$39-59\,days; and post-flare phase at $\delta t>60$\,days, see Fig. \ref{Fig:SoftXrays}), and around the time of acquisition of \nustar\, observations. 
 We employed W-statistics to fit the spectra; we set the metal abundances to Solar values with \texttt{aspl} and we adopted the \texttt{tbabs} cross-sections within \texttt{Xspec}. In all cases, the spectra are well modeled with an absorbed power-law model. We find no statistical evidence for intrinsic absorption in any individual spectrum or in joint spectral analysis once the fit solutions are verified with the \texttt{steppar} command: a neutral absorption column $\rm{NH_{int}}=0\,\rm{cm^{-2}}$ is a statistically acceptable solution in all cases. We thus proceed with $\rm{NH_{int}}=0\,\rm{cm^{-2}}$ in our following spectral fits and correct for the neutral-hydrogen absorption component from the Galaxy only,  which is $\rm{NH_{MW}}=2.6\times 10^{20}\,\rm{cm^{-2}}$ \citep{HI4PI}. 

 We find clear evidence for spectral hardening of the source with time that becomes extreme at the time of the flare, followed by softening of the emission. These findings are confirmed and strengthened by our deep \textit{Chandra}, \textit{XMM}, and coordinated \nustar\, observations. We end by noting (i) the presence of the well-known spurious correlation between the inferred $\rm{NH_{int}}$ and photon index that is due to the degeneracy between these two parameters in soft-X-ray only fits. This degeneracy is lifted by modeling broad-band observations that span the soft and hard X-ray spectral range of \S\ref{Sec:Xraymodeling}. (ii) We also note that with the exception of the flare peak time interval, the spectral fits that model the XRT data only tend to return softer photon indices than those from XRT+\nustar\, data (Table \ref{Tab:Xrayspecparameters}). 

\subsection{Chandra X-ray Observatory (CXO) (0.3--10 keV)} \label{SubSec:CXO}
We obtained four epochs of \chandra\, ACIS-S observations of \at\, under two Director’s Discretionary Time (DDTs) programs (25509013 and 25509020; PI: Margutti). Acquired in the time period $\delta t=17.8-75.4$\,days, the \textit{CXO} observations provide key information during the late decay phase, flare and post-flare phases (Table \ref{Tab:Xraylog}). The \chandra\,ACIS-S data have been reduced following standard practice with \texttt{CIAO v4.16} and corresponding calibration files. A bright X-ray source is detected with high confidence at the location of \at\, with \texttt{wavdetect}. For each epoch, we extracted a spectrum with \texttt{specextract} using a source region with radius of $2\arcsec$ and source-free background region with radius  $>35\arcsec$. \chandra\, observations are not sensitive to the $\rm{NH_{int}}$ parameter because of the limited effective area below $\approx$1\,keV. We modeled the spectra with an absorbed power-law model as in \S\ref{SubSec:XRT}. Remarkably, \chandra\, observations acquired at the time of the flare peak at $\delta t \approx 50$\,days indicate a rising $F_{\nu}\propto \nu^{1.26}$ spectrum. Spectral modeling is described in \S\ref{Sec:Xraymodeling}.
\subsection{XMM-Newton (0.2--12 keV) (XMM)} \label{SubSec:XMM}
We acquired a sequence of three late-time, deep \xmm\, observations of \at{}\, at $\delta
t= 98.9-99.5$\,days, $\delta t=139.7-140.2$\,days and $\delta t= 279.1-279.7$\,days  under a Guest Investigator program (\#090332, PI Margutti, Table
\ref{Tab:Xraylog}). The data from the three European Photon Imaging Camera
(EPIC)-pn, MOS1, and MOS2 have been reduced with the Scientific Analysis
System (SAS) v.20.0.0 and corresponding calibration files (CALDB 3.13).
We filtered out time intervals with enhanced background due to proton flaring,
which led to a significant reduction of the exposure time in the second XMM
observation, especially for the pn camera. To assess the significance of the detection, for each observation we ran the EPIC source detection tasks 
\texttt{emosaic\_prep} and \texttt{emosaicproc} for all the three cameras
simultaneously over the full 0.3--12 keV (0.2--12 keV) energy band, and in three
standard sub-energy bands (i.e., 0.3--1.0 keV for the pn and 0.2--1 for the MOS,
1.0--7.5 keV, 7.5--12.0 keV).

A source is significantly detected at the location of
\at{} in the full band in the first two  observations with a resulting detection maximum
likelihood of {\tt\string DET\_ML=}71 and 12 ($\gtrsim 3\,\sigma$, Gaussian
equivalent) for the first and second observation, respectively. The total net counts
in each observation from the combined detector images are 155$\pm$18 and
81$\pm$18, respectively (0.2--12\,keV energy band). No source is detected in the third observation and we infer a $3\sigma$ limit of $<0.001\,\rm{c/s}$ in the 0.3-10 keV from the EPIC-pn exposure. 
For the first observation, we extracted three source spectra (i.e.,
one each for the EPIC-pn, MOS1 and MOS2, respectively) using a 30$\arcsec$
radius region, and we estimated the background from a 
source-free region on the same chip. The limited statistics of the second observation do not allow spectral modeling, and we thus  inferred the observed flux from the count rate, assuming a photon index $\Gamma=1.6$ as the limited number statistics did not allow us to constrain the spectrum. The same model is assumed for the count-to-flux conversion of the upper limit of the third and fourth epochs. 

\subsection{NuSTAR (3--79 keV)}\label{SubSec:NuSTAR}
We acquired a total of five epochs of hard X-ray observations of \at\, with the Nuclear Spectroscopic Telescope Array (\textit{NuSTAR}). The first three epochs were acquired starting on 2024-09-30, 19:21:09\,UT ($\delta t=6.0$\,days) under a DDT program (PI: Margutti), with the remaining two observations acquired under a joint \xmm-\nustar\, Guest Observer program (\#090332, PI: Margutti). A complete log of \nustar\, observations is reported in Table \ref{Tab:Xraylog}. 

We used \texttt{nupipeline} and \texttt{nuproducts} to extract spectra and response files using the \textit{NuSTAR} Data Analysis Software (v2.1.2) and calibration files (version 20240104). The source extraction region has a radius of $50\arcsec$, and we estimated the local background with a nearby source-free region. We checked for the presence of solar flares and significant radiation belt backgrounds using standard background plots and custom \texttt{python} scripts. When present, we removed the intervals of time affected by the enhanced background by redefining the Good Time Intervals (GTIs) of extraction of our products. A source is blindly detected at the location of the transient  until $\delta t\approx 20$\,d, which represents the latest \nustar\, detection of an FBOT to date. \at{}\, is weakly detected at $<10$\,keV in the \nustar\, A-module at $\approx 50$\,d and it is not significantly detected at $\approx 75$\,d. These hard X-ray detections, weak detections, and non-detections are critical to anchor the broad-band  spectral fits of \S\ref{Sec:Xraymodeling}. 

\section{Broad-band Radio Observations} \label{Sec:Radio}
\subsection{ALMA} \label{SubSec:ALMAdata}
We observed \at{} with the Atacama Large Millimeter/submillimeter Array (ALMA) in cycle 11 as part of DDT programs 2024.A.00003.T and 2024.A.00009.T (PI: Nayana A.J.). The observations were acquired on 2024 October 14 ($\delta t \approx 19\,\rm days$) and October 31 ($\delta t \approx 36\, \rm days$) in bands-3 (97.5 GHz) and 5 (203 GHz). The Array was in its C3 configuration with 45$-$48 working antennae providing baselines ranging from 14 to 499 meters. J0334-4008 was used as the flux density and bandpass calibrator at both bands on the 2024 October 14 observations and J0006-0623 was used on the 2024 October 31 observations. J0246-1236 and J0241-0815 were used as the phase calibrators in bands 3 and 5, respectively at both epochs. The on-source integration time was $\approx 9$ minutes in band 3 and $\approx 20$ minutes in band 5. We downloaded the pipeline generated images from ALMA archive and estimated the flux density of the source using Common Astronomy Software Applications \citep[CASA;][]{CASAteam2022}. The ALMA flux densities are reported in Table \ref{Tab:radio-ALMA}.
\subsection{ATCA} \label{SubSec:ATCAdata}
The Australian Telescope Compact Array (ATCA) observed \at{} at seven epochs from 2024 October 10 ($\delta t \approx 15$\,days) to 2025 March 19 ($\delta t \approx 175$\,days) in the 15 mm, 7 mm, and 4 cm bands under the Non A-Priori Assignable (NAPA) project C3419 (PI: Nayana A.J.). At each frequency band, the data were recorded in two intermediate frequencies (IFs) each split into 2048 channels. PKS B1934$-$638 and PKS B0237$-$233 were used as the flux density and phase calibrators, respectively, in all three bands. The flux calibrator was also used to calibrate the bandpass in the 15 mm and 4 cm bands whereas PKS B1921$-$293 was used as the bandpass calibrator in the 7\,mm band. The total exposure was $\approx 3$\,hours in the 7 and 15 mm bands and 2.5 hours in the 4 cm band, resulting in $\approx$ 1.5$-$2 hours on-source after overheads, respectively for each band. The data were reduced using CASA following standard flagging and calibration procedure treating each IF and epoch separately. Calibrated data were imaged using two Taylor terms adopting Briggs weighting. ATCA flux density measurements are presented in Table \ref{tab:radio-atca}.
\subsection{ATA} \label{SubSec:Allen-Telescopedata}
We observed the field of AT\,2024wpp with the Allen Telescope Array (ATA; Farah et al. in prep; Pollak et al. in prep) on 2024 October 9 ($\delta t \approx 14$\,days), and November 1 ($\delta t \approx 37$\,days). The ATA is a radio inferometer that comprises 42 dishes, each with a diameter of 6.1 m and can utilize up to four independent frequency tunings in the range of 1--10 GHz, each with $\approx 700$ MHz bandwidth \citep{Bright2023}. Our observations were centered at $3$ and $8$ GHz. We used 3C147 to calibrate the absolute flux scale and the bandpass response and J0241-082 to calibrate the time-dependent complex gains. We used a customized pipeline\footnote{https://github.com/joesbright/ATARI/} utilizing CASA to reduce the data. Imaging was done using CASA task \texttt{TCLEAN} \citep{Offringa2017} adopting Briggs weighting. The source was not detected in our ATA observations \citep{Sfaradi2024}. We report $3\sigma$ image rms as flux density upper limits in Table \ref{Tab:radio-ATA}.
\subsection{MeerKAT} \label{SubSec:MeerKATdata}
We observed \at{} with MeerKAT at three epochs: 2024 October 31 ($\delta t \approx 36$\,days), November 15 ($\delta t \approx 51$\,days), and Dec 11 ($\delta t \approx 77$\,days) at L band under project code SCI-20230907-NA-01 (PI: Nayana A.J.).  J0408-6545 was used as the flux density and bandpass calibrator, and J0240-2309 was used as the phase calibrator. The data were recorded using a correlator bandwidth of 856 MHz split into 4000 channels with an integration time of 8 seconds. We used the calibrated images produced by the SARAO Science Data Processor pipeline (SDP)\footnote{https://skaafrica.atlassian.net/wiki/spaces/ESDKB/pages/338723406/}. The source was not detected in any of the MeerKAT images and 3$\sigma$ flux density limits are reported in Table \ref{Tab:radio-MeerKAT}.
\subsection{GMRT} \label{SubSec:GMRTdata}
Giant Metrewave Radio Telescope (GMRT) observations of \at{} were acquired from 2024 October 10 to 2025 Jan 26 ($\delta t \approx 37-123$\,days) in bands 3 ($0.25-0.50$\,GHz), 4 ($0.55-0.85$\,GHz), and 5 ($1.00-1.46$\,GHz). We observed 3C48 to calibrate the absolute flux densities and bandpass and J0240-231 to calibrate the atmospheric phase fluctuations. The data were recorded using a correlator bandwidth of 200 MHz in bands 3 and 4 and 400 MHz in band 5, split into 2048 channels. The observations were done in standard continuum full polar mode with an integration time of 10 seconds. The GMRT data were reduced using Astronomical Image Processing Software \citep[AIPS;][]{greisen2003} following standard procedure. The data were initially inspected for non-working antennas and RFI-prone channels and flagged using AIPS task UVFLAG. Single-channel calibration was done using a central channel, and the solutions were applied to the entire band. The fully calibrated target data was imaged using IMAGR. A few rounds of phase-only self-calibration were performed to improve the image quality. Radio emission associated with \at{} was not detected in any of the GMRT maps. We quote the details of GMRT observations and the 3 sigma flux density limit at the source position in Table \ref{Tab:radio-gmrt}.

\section{X-ray modeling and inferences}\label{Sec:Xraymodeling}
\begin{figure*} 
 	\centering
 	\includegraphics[width=0.6\textwidth]{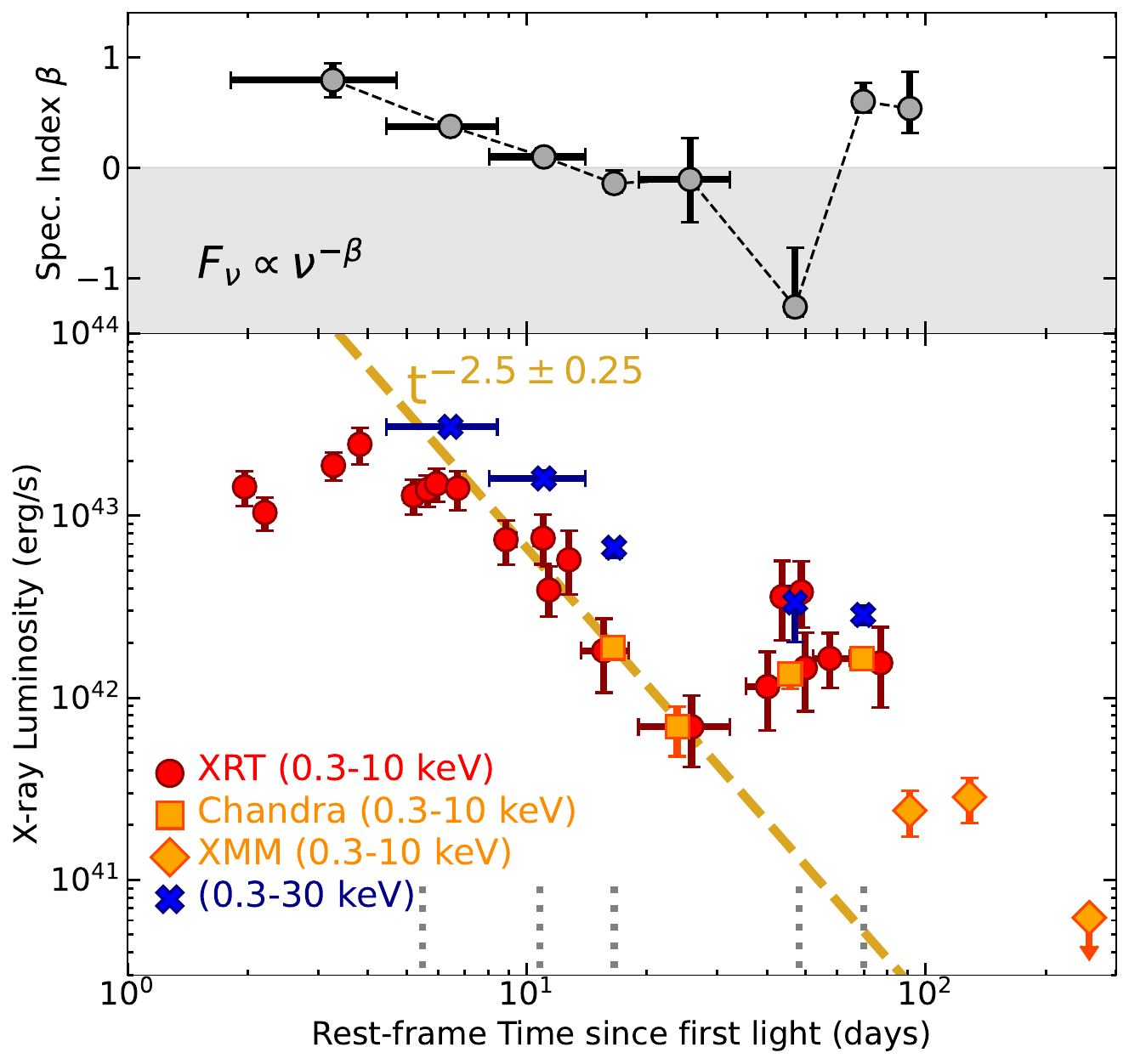} 
  \caption{ \emph{Upper Panel:} Evolution of the spectral photon index $\beta$ with time (where $F_{\nu}\propto \nu^{-\beta}$), showing clear evidence for spectral hardening until the time of the flare peak at $\gtrsim$50\,days, followed by softening of the emission.  Aside from the flare peak, the plotted $\beta$ values apply to the broad-band soft+hard X-ray spectral range; at the time of the flare peak, when there is evidence for a broken power-law spectrum, the plotted value represents the index below the spectral break. The grey-shaded area marks the region of the parameter space associated with a \emph{rising} $F_{\nu}$ spectrum. \emph{Lower Panel}: 0.3--10 keV unabsorbed luminosity light-curve (red filled circles, orange squares and diamonds for XRT, CXO, and XMM observations, respectively) and 0.3--30 keV light-curve (blue crosses) derived from a self-consistent time-dependent flux calibration. Vertical grey dotted lines mark the time of broad-band X-ray spectra acquisition.  Dashed gold line: best-fitting power-law luminosity decay  in the time period 10--50\,d. }
 \label{Fig:SoftXrays}
 \end{figure*}

\begin{figure*}[t!]
 	\centering
 	\includegraphics[width=0.8\textwidth]{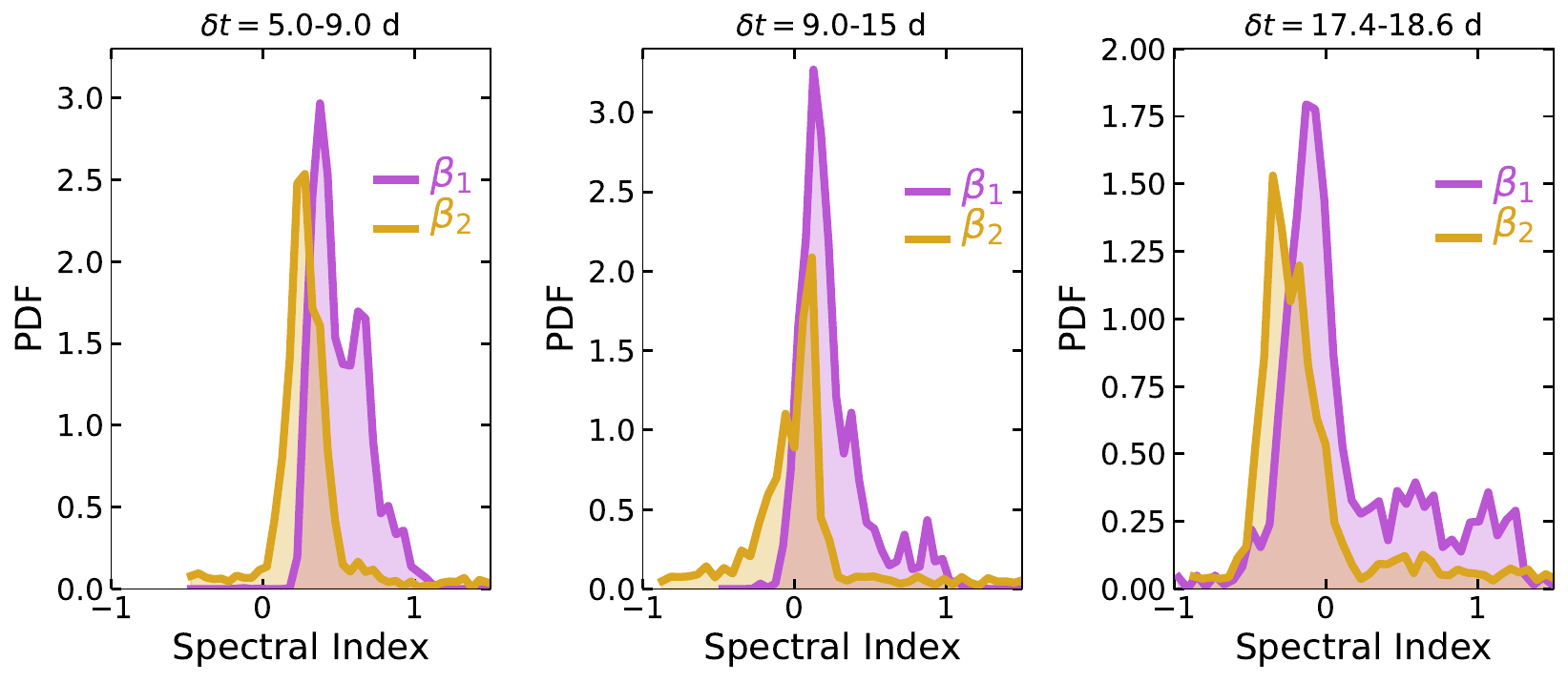} 
  \caption{Probability density distributions of the spectral photon indices derived from the broad-band X-ray spectral fitting of \S\ref{Sec:Xraymodeling} with a broken power-law model. At the time of the flare peak at $\approx 50$\,days, we find  evidence for a broken power-law spectrum with a rising spectrum $F_{\nu}\propto \nu^{1.25}$ at $h\nu \lesssim 8$\,keV.  There is a hint for a harder spectral index at softer energies at earlier times (i.e., $\beta_{2}<\beta_{1}$), which suggests the presence of multiple spectral components, with the relative strength of the harder component increasing with time until the time of the flare peak (see \S\ref{SubSec:ComptonHump}).     }
 \label{Fig:SpecIndex}
 \end{figure*}

\begin{figure*}[t!]
 	\centering
 	\includegraphics[width=1\textwidth]{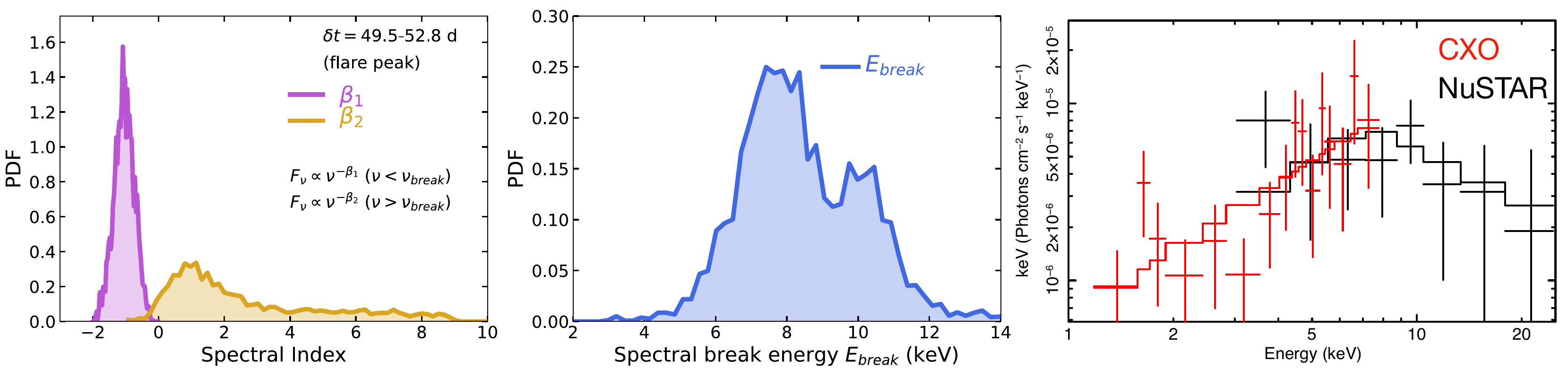} 
  \caption{ Probability density distributions of the spectral photon indices (left panel) and spectral break energy (central panel), for the  broken power-law model that best fits the broad-band X-ray SED at the time of the flare peak (right panel, unfolded spectrum). The CXO data (red) indicate a rising spectrum with extreme properties: $F_{\nu}\propto \nu^{1.25}$. The lack of bright hard X-ray emission at the same time, as constrained by \nustar, demands the presence of a spectral break and significantly softer emission above $E_{\rm{break}}$, with the spectrum bending to $F_{\nu}\propto \nu^{-1}$.  }
 \label{Fig:FourthXraySpec}
 \end{figure*}

\subsection{Joint soft and hard X-ray spectral modeling and X-ray flux calibration}\label{SubSec:Xrayspecmodeling}
We jointly fit the epochs with  soft  (\emph{Swift}-XRT, \xmm\, or \chandra) and hard (\nustar) X-ray data to constrain the broad-band spectral properties of \at{}\, and their evolution. There is no evidence for thermal X-ray emission at any time, which is consistent with all the other FBOTs to date. We use an absorbed simple power-law (SPL) and a broken power-law (BPL) model within \texttt{Xspec}. We adopt Solar abundances and \texttt{tbabs} cross sections with Galactic neutral hydrogen column density $\rm{NH_{MW}}=2.6\times 10^{20}\,\rm{cm^{-2}}$ \citep{HI4PI}. For each epoch and source spectral model, we take two approaches to properly account for the low-number statistics of source counts in \nustar.\footnote{See e.g., \url{https://giacomov.github.io/Bias-in-profile-poisson-likelihood/}} In the first method, we perform full background and source spectral modeling  adopting Cash statistics \citep{Cash79} after ensuring that each bin in each spectrum contains at least one count.  The source and background files are simultaneously fit with a background model for the background spectrum, and a combination of source plus background model for the source file, with the background model parameters tied to the same values. This is the ideal approach to avoid statistically biased parameter inference, but it assumes that it is possible to approximate the background with an analytical function. As a second approach, we use the W-statistics after having ensured that every bin in the background spectrum contains enough counts. We find that the two methods lead to statistically consistent results; in the following, we report and show the results from the W-stat approach. 

We find that a SPL spectrum is a statistically acceptable description at all times with the exception of the broad-band X-ray spectrum acquired at the flare peak at $\delta t \approx 50$\,days, when instead a BPL model is preferred. We list the best-fitting parameters, their uncertainties, and inferred fluxes in Table \ref{Tab:Xrayspecparameters}. Uncertainties are self-consistently derived with MCMC simulations. The spectral parameters of the favored model from this broad-band analysis (identified with a ``$\star$'' symbol in Table \ref{Tab:Xrayspecparameters}) are then used to anchor the time-dependent flux calibration of the intermediate epochs with soft X-ray data only. For soft X-ray data acquired before the first \nustar\, epoch, the flux calibration is based on the best-fitting parameters from a \emph{Swift}-XRT spectrum extracted at $\delta t =2-5$\,d in Table \ref{Tab:Xrayspecparameters}. The resulting unabsorbed 0.3--10 keV X-ray light curve of \at{} is shown in Fig. \ref{Fig:SoftXrays}. 

\at{}\, shows luminous, roughly constant emission at the level of $L_{x}\approx 1.5\times 10^{43}\,\rm{erg\,s^{-1}}$ in the first $\approx 7$\,d, followed by a phase of rapid decay with $L_x\propto t^{-2.5\pm0.25}$ until $\delta t\approx30$\,d. Initially displaying a spectrum $F_{\nu}\propto \nu^{-\beta}$ with $\beta=0.80^{+0.15}_{-0.16}$, the  source spectrum later hardens with time and transitions into a \emph{rising} $F_{\nu}\propto \nu^{-\beta}$ with $\beta<0$ at $\approx 15$\,days (Fig. \ref{Fig:SoftXrays}, upper panel). 
The source displays an episode of major rebrightening of X-rays starting at $\delta t\approx 35$\,days, peaking at $\delta t\approx 50$\,days accompanied by extreme spectral hardening (soft X-ray spectrum $F_{\nu}\propto \nu^{1.25}$). At this time, the broadband X-ray spectrum is bell-shaped, and it is best fit by a BPL model with break energy $\approx 8$\,keV and $F_{\nu}\propto \nu^{-1}$ above $E_{break}$ (Fig. \ref{Fig:FourthXraySpec}).
Remarkably, at $\approx 75$\,days, the soft X-ray spectrum is back to its initial, much softer state of $\beta=0.60^{+0.17}_{-0.10}$ and remains consistent with this softer value until we can constrain the spectrum with \xmm. This phenomenology is unprecedented among FBOTs but has clear physical connections with the Compton hump of AT\,2018cow  \citep{Margutti2019-cow}, and possibly AT\,2020mrf \citep{Yao2021-mrf}, as we detail in the following \S\ref{SubSec:ComptonHump}.

Although an SPL model provides a statistically acceptable fit, a closer inspection of the best fit parameters of the BPL model at $\delta t<20$\,days reveals that the soft X-rays are preferentially best fit by softer spectral indices than the hard X-rays (that is, $\beta_{2}<\beta_{1}$ in Fig. \ref{Fig:SpecIndex}, where $F_{\nu}\propto \nu^{-\beta_{1}}$ for $\nu<\nu_{break}$ and $F_{\nu}\propto \nu^{-\beta_{2}}$ above the break frequency $\nu_{break}$). This observation opens the possibility that two emission components are contributing to the overall shape of the broad-band X-ray spectrum, with the relative strength of the hard component increasing with time until the flare peak. We explore possible physical scenarios consistent with this possibility in \S\ref{SubSec:ComptonHump}.

 \startlongtable
\begin{deluxetable*}{ccccccccccc}
\tablecaption{X-ray spectral parameters and inferred fluxes. The SPL model is parametrized as $F_{\nu}\propto \nu^{-\beta_1}$ where $\beta_1=\Gamma_1-1$ and $\Gamma_1$ is the photon index. The BPL model is $F_{\nu}\propto \nu^{-\beta_1}$ for $\nu<\nu_{break}$ and $F_{\nu}\propto \nu^{-\beta_2}$ for $\nu>\nu_{break}$, $\beta_2=\Gamma_2-1$ and $\Gamma_2$ is the photon index above the spectral break.   }
\tablehead{
\colhead{$\delta t^{\rm{a}}$} & \colhead{model} & \colhead{$\Gamma_1$} & \colhead{$\Gamma_2$}  & \colhead{$E_{break}^{b}$} & \colhead{$F_x$} & \colhead{$F_x$} & \colhead{$F_x^{d}$} & \colhead{Instrument}\\
(d)  &  &  & & (keV) & ($10^{-13}\,\rm{cgs}$) & ($10^{-13}\,\rm{cgs}$)& ($10^{-13}\,\rm{cgs}$) &  \\
     &  &  & &       & [0.3--10 keV] & [0.3--30 keV]& [20--200 keV] & 
 }
\startdata
2.1--5.0  & SPL & $1.79^{+0.15}_{-0.16}$ & -- & -- &  $7.9^{+1.0}_{-1.0}$ & $11.8^{+3.0}_{-2.0}$ & $10.8^{+9.2}_{-5.5}$ & XRT \\
5.0--9.0  & SPL & $1.37^{+0.06}_{-0.04}$ & -- & -- & $7.2^{+0.5}_{-0.3}$ & $15.2^{+0.5}_{-0.5}$ & $40.5^{4.3+}_{-6.4}$ & XRT+NuSTAR \\
9.0--15.0 & SPL & $1.10^{+0.08}_{-0.06}$ & -- & -- &$2.9^{+0.2}_{-0.2}$  & $7.9^{+0.5}_{-0.4}$ & $39.1^{+6.9}_{-8.2}$ & XRT+NuSTAR \\
17.4--18.6 & SPL & $0.86^{+0.11}_{-0.08}$ & -- & -- & $0.93^{+0.08}_{-0.07}$  & $3.3^{+0.3}_{-0.4}$ & $26.9^{+8.3}_{-8.1}$ & CXO+NuSTAR \\
21.3--34.5 & SPL & $0.89^{+0.37}_{-0.39}$ & -- & -- & $0.34^{+0.01}_{-0.01}$ & $1.2^{+0.7}_{-0.6}$ & $8.9^{+13.5}_{-7.4}$ & CXO+XRT \\
49.5--52.8& BPL & $-0.26^{+0.53}_{-0.09}$ & $1.98^{+4.57}_{-0.42}$ & $7.8^{+3.49}_{-0.33}$ & $0.67^{+0.01}_{-0.01}$ & $1.7^{+0.4}_{-0.7}$ & $2.1^{+1.9}_{-2.0}$ & CXO+NuSTAR \\
75.0--76.6& SPL & $1.60^{+0.17}_{-0.10}$ & -- & -- & $0.81^{+0.07}_{-0.05}$ & $1.4^{+0.2}_{-0.2}$ & $2.2^{+1.2}_{-0.9}$ & CXO+NuSTAR \\
98.9--99.5 & SPL &  $1.53^{+0.33}_{-0.22}$ & -- & -- & $0.12^{+0.03}_{-0.03}$ &$0.22^{+0.13}_{-0.09}$ &$0.39^{+0.65}_{-0.31}$ &XMM\\
139.7--140.2& SPL & 1.53& --& --& $0.14^{+0.04}_{-0.04}$ &$0.30^{+0.07}_{-0.07}$ & $0.47^{+0.13}_{-0.13}$ & XMM\\
\enddata
\tablecomments{$^{\rm{a}}$ Observer frame, with respect to time of first light. \\
$^{\rm{b}}$ $E_{break}\equiv h\nu_{break}$.\\
$^{\rm{c}}$ Fluxes are unabsorbed and in units of $\rm{erg\,s^{-1}cm^{-2}}$. \\
$^{\rm{d}}$ Based on the extrapolation of the SPL or BPL spectral model at higher energies not sampled by observations (i.e., it assumes no additional spectral break).
\label{Tab:Xrayspecparameters}}
\end{deluxetable*}

\begin{figure*} 
 	\centering
 	\includegraphics[width=0.33\textwidth]{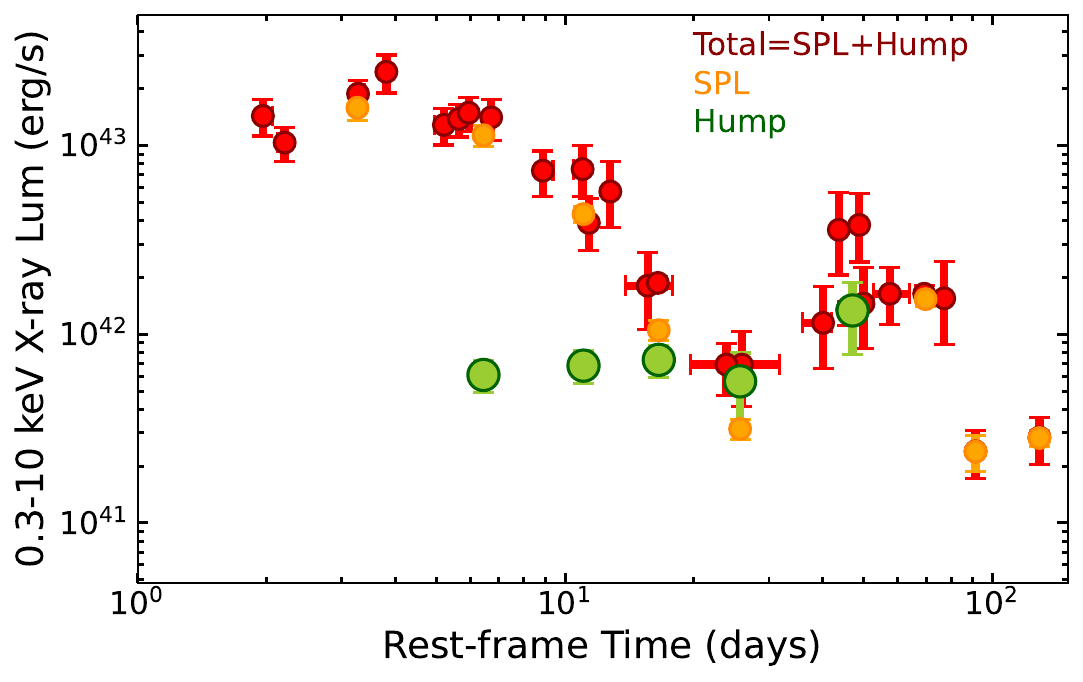} 
 	\includegraphics[width=0.33\textwidth]{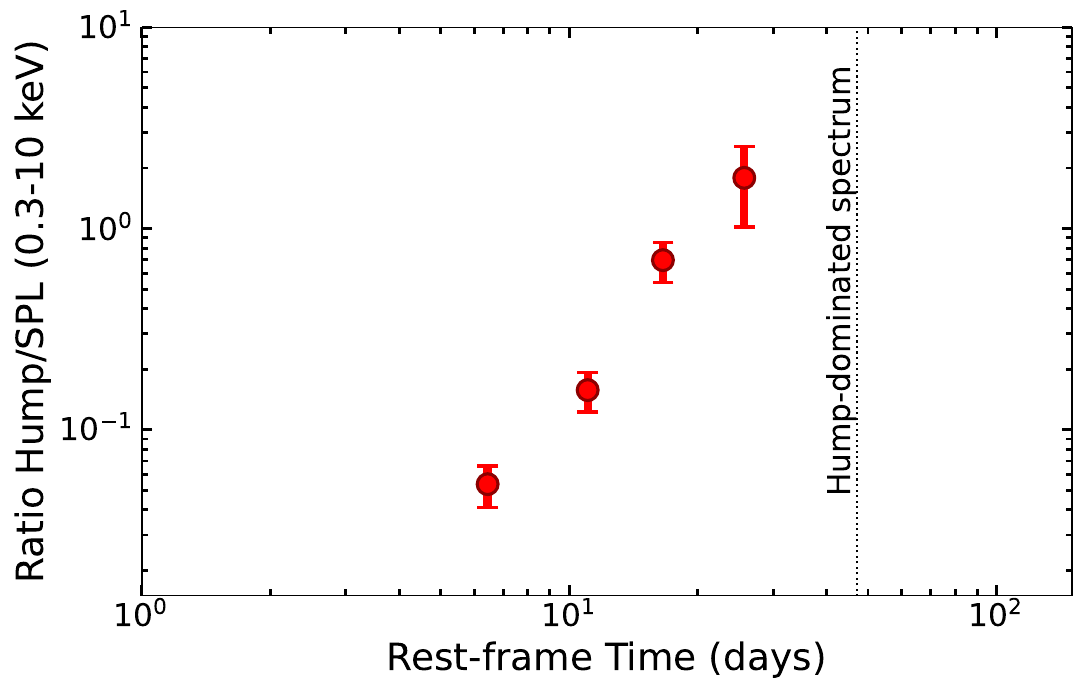} 
   \includegraphics[width=0.3\textwidth]{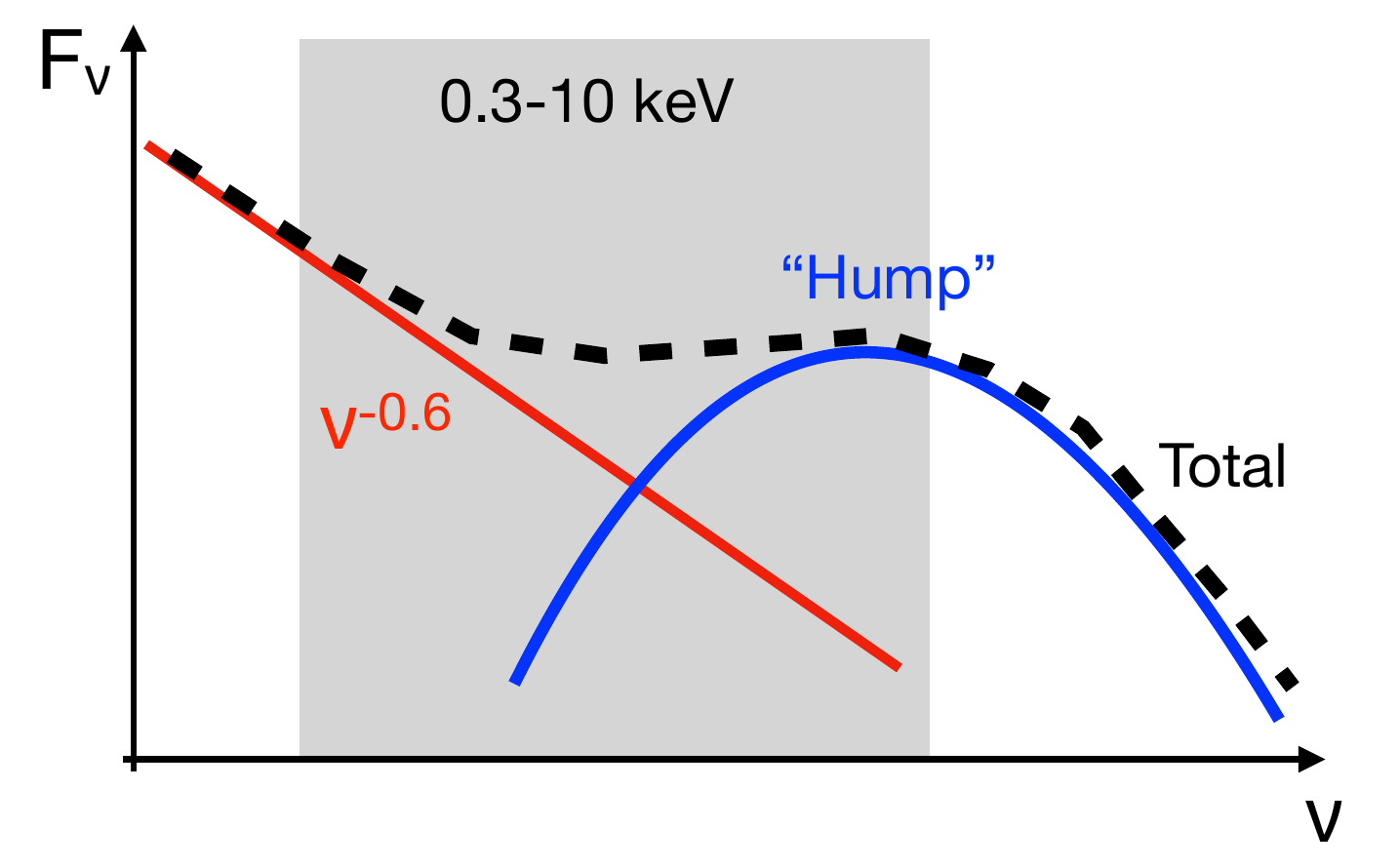} 
  \caption{Evolution of the SPL (orange)  and hump (lime green) component luminosities (\emph{left}) and their ratio (\emph{central}) with time in the fixed observer frame energy band 0.3--10 keV as constrained by our analysis of \S\ref{SubSec:ComptonHump}. The Compton hump contribution increases with time as the optical depth of the reprocessing layer decreases, causing the peak of the hump of emission (initially at $>10$\,keV) to cascade to lower energies and enter the spectral window of interest (Fig. \ref{Fig:IndrekSimulations}). The hump dominates the 0.3--10 keV  emission at $\delta t\approx 50$\,d. \emph{Right:} Cartoon of the SPL+hump model showing how the progressive ``emergence'' of the hump  of emission in the 0.3-10 keV band at  $\delta t\le 50$\,d leads to hardening with time. }
 \label{Fig:hump}
 \end{figure*}

\subsection{A Transient Compton Hump of Emission and Ionization Break Out} \label{SubSec:ComptonHump}
\at{}\, is the second LFBOT after AT\,2018cow with clear evidence for a transient hump of emission. 
Compared with AT\,2018cow \citep{Margutti2019-cow}, the hump of X-ray emission in \at{}\, appears at a later time ($\delta t\approx 50$\,days vs. $\approx 8\,$days) and with a lower peak energy ($E_{\rm{break}}\approx 8\,$keV vs. $\approx50$\,keV).\footnote{We note that a blue-shifted Fe $K\alpha$ feature as the one detected in AT\,2018cow would not be detectable against the continuum here because of the more limited statistics.} In both cases, after the hump disappears the soft $<10$\,keV X-ray spectrum  reverts back to a spectral index value similar to the ``pre-hump'' phase  $\beta \approx 0.5-0.7$. Motivated by the findings of \S\ref{SubSec:Xrayspecmodeling}, we explore an alternative set of fits for which we hypothesize that the observed X-ray spectrum is the superposition of a SPL ($F_{\nu}\propto \nu^{-\beta}$) and a hump component (here modeled with a BPL with slopes frozen to the values inferred for the $\delta t \approx 50\,$days spectrum).   For spectra  at 5$<\delta t < 50$\,days where hardening is apparent in the 0.3-10 keV band, we assume an SPL index $\beta =0.6$.  We fit for the normalization of each component (SPL, BPL) and for the free break energy.  This approach allows us to explore the evolving contribution of a putative hump throughout this initial phase. Although this  model is purely phenomenological and simplified, it is expected to capture the main properties of the evolution of the two components (displayed in Fig. \ref{Fig:hump} in the fixed observer-frame  0.3--10 keV band). Figure \ref{Fig:hump} shows   that the in-band contribution of the hump component grows with time, as it is expected for a hump of emission with decreasing peak energy with time.

\begin{figure} 
 	\centering
 	\includegraphics[width=0.4\textwidth]{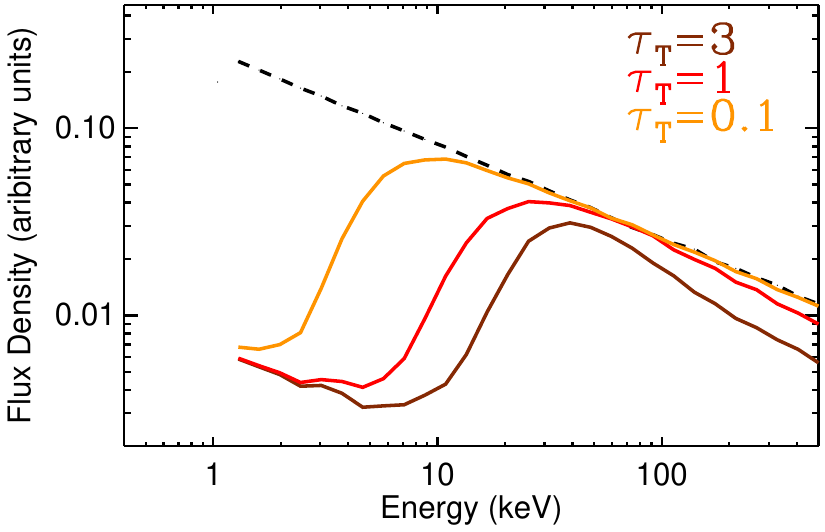} 
  \caption{Selection of X-ray transmission spectra from a central source with intrinsic spectrum $F_{\nu}\propto \nu^{-0.5}$ (black dashed line) from the simulations presented in \cite{Margutti2019-cow} for a range of Thomson optical depth values $\tau_T$ showing the increasing dominance of the Compton hump flux in the 0.3--10 keV band as $\tau_T$ decreases. These simulations do not account for time-dependent ionization effects in the transmission layer. See main text and \cite{Margutti2019-cow} for more details on these simulations.  }
 \label{Fig:IndrekSimulations}
 \end{figure}

Physically, broad-band X-ray spectra with similar properties and evolution are observed and expected in the case of transmission of radiation from a high-energy source through ejecta with time-variable optical depth to Compton scattering and photoelectric absorption, as was suggested for AT\,2018cow \citep{Margutti2019-cow}. In this scenario, the SPL represents the fraction of flux that is transmitted and reaches the observer unmodified, while the BPL mimics the combined results of Compton (down-)scattering at the high photon-energy end, and photoelectric absorption at lower energies. The BPL component dominates the 0.3--10 keV energy range as Thompson optical depth ($\tau_T$) decreases (Fig. \ref{Fig:IndrekSimulations}).  For $\tau_T\ll1$ this model predicts that the 0.3--10 keV spectrum will eventually go back to its initial slope, as observed in both FBOTs.

\begin{figure} 
 	\centering
 	\includegraphics[width=0.4\textwidth]{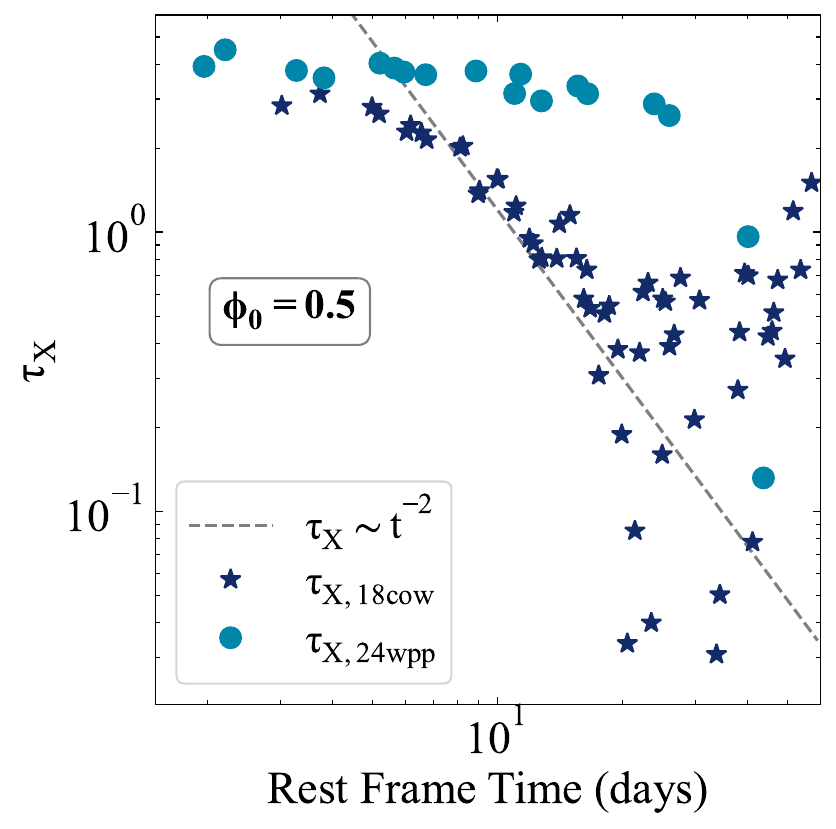} 
  \caption{Effective optical depth to X-rays derived following the reprocessing model of \cite{Metzger22FBOT} and the observed X-ray and UVOIR emission for \at{}\, (filled circles) and AT\,2018cow (stars). \at{}\, maintains higher $\tau_{\rm{X}}$ for a significantly longer time, a behavior that is consistent with the delayed appearance of the Compton hump (\S\ref{SubSec:ComptonHump}). Grey dashed line: $\tau_{\rm{X}}\propto t^{-2}$ scaling expected for radiation shining through a medium expanding with constant velocity. We assume an arbitrary slow-moving ejecta covering fraction $\phi_0=0.5$. }
 \label{Fig:tauX}
 \end{figure}

For AT\,2018cow, at optical peak  $\tau_T\sim (c/v_{\rm{ej}})(\kappa_{\rm{es}}/\kappa)\approx 20-40$, where $\kappa_{\rm{es}}$ is the electron-scattering opacity, $v_{\rm{ej}}$ is the (optically emitting) ejecta velocity, and $\kappa_{\rm{es}} \approx 4 \kappa$  was assumed \citep{Margutti2019-cow}. In the absence of any other effect (e.g., continuous deposition of ejecta, or a change in the ionization state of the ejecta), the expansion of the ejecta leads to  $\tau_T\propto t^{-2}$, implying $\tau_T\approx 3$ at the time of the prominent Compton hump at $\approx 8$\,days with an expected $F_{\nu}$ peak at $\approx 50\,$keV (Fig.\ \ref{Fig:IndrekSimulations}), consistent with observations (Fig. 6 in \citealt{Margutti2019-cow}). However, as discussed next, this reasoning does not apply to \at{}\, in this simple form, fundamentally because the observed $L_{\rm{X}}/(L_{\rm{X}}+L_{\rm{UVOIR}})$ ratio remains constant  and $L_{\rm{X}}/(L_{\rm{X}}+L_{\rm{UVOIR}})\approx L_{\rm{X}}/L_{\rm{UVOIR}}\approx 10^{-2}$ during the first $\lesssim 20$\,days, as opposed to steadily increasing with time as it was observed in AT\,2018cow  (Paper I, Fig. 9). 

To quantitatively explore the implications of the observed evolution of the $L_{\rm{X}}/(L_{\rm{X}}+L_{\rm{UVOIR}})$ ratio, we adopt the parametrization of \cite{Metzger22FBOT}. In this framework, the detected UVOIR luminosity is the result of reprocessing of a centrally located source of energy with luminosity $L_{\rm{engine}}$ by a two-component medium composed of slow-moving ejecta ($v\approx6500\,\rm{km\,s^{-1}}$ inferred in Paper I) covering a fraction $\phi_0$ of solid angle; and fast-moving ejecta  ($v\approx0.2$c) subtending a solid angle $\Omega_{\rm{fast}}=4\pi(1-\phi_0)$. If the slow-moving ejecta is completely opaque to X-rays, we expect $L_{\rm{X}}=L_{\rm{engine}}(1-\phi_0)e^{-\tau_X}$, where we estimate $L_{\rm{engine}}\approx (L_{\rm{X}}+L_{\rm{UVOIR}})$, $\tau_X$ is the effective X-ray optical depth that we expect to roughly trace $\tau_{T}$ in an electron-scattering dominated medium (as the probability of thermalization rapidly increases with the number of scatterings). From this reasoning, $L_{\rm{X}}/(L_{\rm{X}}+L_{\rm{UVOIR}})\propto (1-\phi_0)e^{-\tau_X}$, or $L_{\rm{X}}/L_{\rm{UVOIR}}\propto (1-\phi_0)e^{-\tau_X}$ for $L_{\rm{X}}\ll L_{\rm{UVOIR}}$. The small and constant $L_{\rm{X}}/L_{\rm{UVOIR}}$ ratio of \at{}\, thus implies large $\tau_X$ sustained for a long time (significantly longer than in AT\,2018cow), as opposed to the expected $\propto t^{-2}$ behavior. We quantify this statement in Fig. \ref{Fig:tauX} for an arbitrary choice of $\phi_0=0.5$. Interestingly, $\tau_{X}\approx 0.1$ at $\delta t\sim 50$\,days, for which Fig. \ref{Fig:IndrekSimulations} predicts a $F_{\nu}$ peak energy of $\approx8$\,keV as observed in \at{}\, (Fig. \ref{Fig:hump}).

Two considerations follow: (i) The delayed appearance of a Compton hump in \at{}\, is in line with (and should be expected based on) the large $\tau_{X}$ maintained  until late times.  We also note that the lower peak energy of the hump of emission in \at{}\, compared with AT\,2018cow presumably originates from transmission through a medium with  lower $\tau_{T}$ (Figs. \ref{Fig:IndrekSimulations} and \ref{Fig:tauX}), which happens at a later epoch. (ii)\,$\tau_{X}$ clearly deviates from the $\propto t^{-2}$ evolution. (i)+(ii) can result from a variety of effects, including a time-variable level of ionization of the ejecta, which changes the opacity to X-rays (see below); continuous deposition of mass (as opposed to a one-time mass ejection); or a time-varying covering fraction $\phi_{0}$ of the slow vs. fast moving ejecta (for example as a consequence of different physical conditions of a super-Eddington accretion disk providing the source of ejecta mass).\footnote{We note that at face value, the smaller $L_{\rm{X}}/(L_{\rm{X}}+L_{\rm{UVOIR}})$ ratio of \at{}\, compared with AT\,2018cow can be interpreted as a larger covering fraction $\phi_{0}$ leading to a larger fraction of engine luminosity being reprocessed in UVOIR emission.}

In the rest of this section we explore the possibility that the sudden drop of $\tau_{X}(t)$ at $\approx 40$\,d  is due to an ``ionization break out'', i.e., a reduction of the photoelectric absorption cross section resulting from the ionization of the ejecta by the inner source of energy. \cite{Metzger14} and \cite{MetzgerPiro14} developed the theoretical framework in the context of magnetars formed by either stellar explosions or binary neutron star mergers, ionizing the surrounding ejecta with luminosity $L_{\rm{ion}}$; while  \cite{Tsuna25} invoked a ionization breakout scenario in the context of FBOTs powered by accretion on newly formed NSs and BHs. Here we remain agnostic about the astrophysical nature of the ionizing luminosity $L_{\rm{ion}}$, and we consider a  central engine ionizing the ejecta of mass $M_{ej}$ expanding with velocity $v_{\rm{ej}}$ on a timescale $t$. Following  \cite{Metzger14}, their  Eq. A7 and A11,  we expect the radiation to ionize its way through the ejecta on a timescale:
\begin{eqnarray}
t_{\rm ion} \approx  \left\{
\begin{array}{lr}
30\,{\rm d}\,M_2^{3/4} \left(\frac{v_{ej}}{0.2c}\right )^{-5/4}T_{5}^{-0.2}\left(\frac{X_{Z}}{0.5}\right)^{1/4}\left(\frac{L t}{5\times 10^{50}}\right)^{-1/4}Z_8^{3/4}\\
(\eta_{\rm thr} \ll 1),\\
50\,{\rm d}\,M_2 \left ( \frac{v_{ej}}{0.2c}\right )^{-3/2}T_{5}^{-0.4}\left(\frac{X_{Z}}{0.5}\right)^{1/2}\left(\frac{L t}{5\times 10^{50}}\right)^{-1/2}Z_8^{3/2}\\      
(\eta_{\rm thr} \gg 1), \\
\end{array}
\right.
\label{eq:tbo}
\end{eqnarray}
where $M_2 \equiv M_{\rm ej}/(2M_{\odot})$, $T_{5} \equiv T/10^{5}$ K is the temperature of electrons in the recombination layer; $X_Z$ is the mass fraction  of elements with atomic number $Z = 8Z_8$ in the ejecta and
\begin{eqnarray}
\eta_{\rm thr} \approx 7.5\left(\frac{L t}{5\times 10^{50}{\rm erg}}\right)^{-1}M_{2}\left ( \frac{v_{ej}}{0.2c}\right )^{-1}\left(\frac{X_{Z}}{0.5}\right)T_{5}^{-0.8}Z_8^{3}
\label{eq:etaA}
\end{eqnarray}
is the ratio of ratio of absorptive to scattering opacity in the ejecta.  We have renormalized Eq. \ref{eq:tbo} and \ref{eq:etaA} using parameter values that are relevant to the fast ejecta component of \at{} as constrained by optical and X-ray observations. For oxygen-dominated ejecta,\footnote{We note that lighter elements would be ionized significantly earlier, consistent with the lack of spectral features at early times; instead  the slower moving ejecta would be opaque to X-ray radiation for much longer. We also note that Eq. \ref{eq:tbo} does not account for the possible ongoing deposition of mass, which would delay the emergence of the ionization front compared to the estimate presented above.} we find $t_{\rm{ion}}\approx 30-50$\,d, which compares well with when the ejecta becomes transparent to the X-rays (and the hump dominates in the 0.3--10 keV energy range). While our  analytical arguments require confirmation by detailed simulations in future work, based on the calculations above we consider it plausible that the $\tau_{X}(t)$ drop is at least partially driven by ionization effects.

To conclude, in close analogy with AT\,2018cow \citep{Margutti2019-cow}, our broad-band X-ray analysis favors the presence of a highly variable, centrally located high-energy source shining through expanding aspherical ejecta material with time-dependent ionization (and potentially covering fraction). We address the astrophysical nature of the high-energy source in the next section.
\subsection{The soft X-ray Spectral Index and the Origin of the X-ray emission}\label{SubSec:softXrayindex}
AT\,2018cow-like FBOTs  display similar values of the soft X-ray spectral index of the persistent component (i.e., the component \emph{not} associated with the Compton hump) $F_{\nu}\propto \nu^{-\beta}$ with $\beta\approx 0.6$. As was noted for AT\,2018cow \citep{Margutti2019-cow}, this hard spectrum maintained over tens of days is not compatible with fast cooling of the radiating electrons for any reasonable values of the electron energy distribution index $p\gtrsim 2$ (where $N(E)\propto E^{-p}$). Together with the initial X-ray light-curve plateau followed by a steep temporal decay, prominent X-ray variability with increasing variance with time, and the Compton hump spectrum, this hard spectral index might be a defining trait of this class of transients. From an observational perspective we note that the $F_{\nu} \propto \nu^{1.26}$ emission component is harder than the typical PL component of XRBs and AGNs, which is often attributed to Comptonization of soft disk photons (e.g., \citealt{Titarchuk21} and references therein). Additionally, similar to AT\,2018cow, the X-ray and  radio emission are not part of the same synchrotron spectrum at any time (the X-rays being always brighter than the extrapolated radio spectrum). Following the same line of reasoning as in \cite{Margutti2019-cow} (their Section 3) for AT\,2018cow  that we do not repeat here, we find that these observations imply the presence of an inner, highly variable source capable of continuously ``heating'' the radiating electrons to maintain the slow-cooling spectrum. 

Magnetic reconnection (e.g., in a magnetar nebula or accreting black-hole corona), or the dissipation of outflow kinetic energy (e.g., via internal shocks between multiple episodes of accretion-disk wind or bulk Comptonization) could in principle satisfy these requirements. 

GMRHD simulations of super-Eddington accreting disks \citep{Sadowski15,Sadowski16} reveal the generation of powerful outflows reaching trans-relativistic velocities along the polar direction and carrying a total (i.e., radiative and kinetic) luminosity $L\sim \eta \dot M_{\bullet}c^2$ with $\eta\sim 0.03$ for a non-rotating ($a=0$) stellar-mass BH (for $a=0.7$ $\eta \sim0.08)$.  Simulations by \cite{Sadowski16} extended to accretion rates up to a few hundred $\dot M_{\rm{Edd}}$, which is $\approx 10^3$ times smaller than what is needed to power FBOTs at peak, finding that at high $\dot M$ most of the outflow luminosity is in the kinetic form.\footnote{Recent results from a  wider range of non-MHD simulations by \cite{Yoshioka24} confirm these findings.}  Building on the results from these simulations and extrapolating to significantly higher accretion rates, \cite{Metzger22FBOT} and \cite{Tsuna25} demonstrated how stellar-mass BHs accreting at highly super-Eddington rates produce outflows that can carry enough kinetic energy $E_k>10^{51}\,\rm{erg}$ to match the (extreme) energetic requirements and timescales of FBOTs like \at{} (see their Eq. 18 and Eq. 37, respectively). It is important to note that the astrophysical context of the two models is different: a  tidal disruption and hyper-accretion of a Wolf-Rayet (WR) star by a BH or NS binary companion is invoked by \cite{Metzger22FBOT} vs. the collision of a newly-formed NS or BH from a core-collapse SN explosion with its main sequence companion  by \cite{Tsuna25}. However, both models share the common ingredient of super-Eddington accretion on a compact object, and directly connect the FBOT phenomenology to the dissipation of energy carried by the resulting outflows. 

To conclude, we thus consider likely the possible origins for the central X-ray source: (i) a Pulsar Wind Nebula (PWN)-like system, i.e., a magnetized nebula energized by a compact object (e.g., \citealt{Vurm21}); (ii) emission related to super-Eddington accretion disks around compact objects. Both systems can power collimated jets (for which we have no direct observational evidence in FBOTs) as proposed by \cite{Gottlieb2022}. More generally, irrespective of the details of the astrophysical origin of the X-ray source, the  X-ray emission in FBOTs likely escapes from a lower-density polar region, which in all likelihood implies  geometrical beaming. This fact has two observational consequences: first, the ``true'' X-ray luminosity from the system is lower than what is estimated assuming isotropic emission ($L_{\rm{true}}=L_{iso}\times \Delta \Omega/4\pi$).\footnote{Similar arguments have been used to explain the super-Eddington X-ray luminosity of Ultra-Luminous X-ray sources, ULXs, see e.g., \cite{King23}. } Second,  geometrical beaming  implies a viewing angle dependency of the observed $L_x$, (with observing angles aligned with the polar direction being associated with the brighter displays at early times), which might be at the core of the range of X-ray luminosity behaviors observed in LFBOTs (discussed in detail in \S\ref{SubSec:Xraycomparison}) as well as the appearance and prominence of the Compton hump. 

\section{Radio Modeling and Inferences}
\label{Sec:radio-modeling}
\subsection{General Considerations}
 Radio emission from FBOTs is understood to originate from the shock interaction with the surrounding medium \citep{Ho2019-cow,Margutti2019-cow,coppejans2020,Bright2021-xnd,Ho2022-xnd,Yao2021-mrf,Chrimes2024b-fhn}. The resulting radio SEDs are bell shaped, where the peak is due to synchrotron self-absorption (SSA) of relativistic electrons accelerated at the shock front to a power-law distribution of the form $N(E) \propto E^{-p}$ down to a minimum Lorentz factor $\gamma_{\rm m}$ \citep{chevalier1998}. In a standard SSA scenario, the optically thick spectral index is $\alpha_{\rm 2}=5/2$ (with $F_{\nu} \propto \nu^{\alpha_{2}}$) and the optically thin spectral index is $\alpha_{1}=-(p-1)/2$ (with $F_{\nu} \propto \nu^{\alpha_{1}}$). If synchrotron emitting electrons are efficiently cooled via synchrotron or inverse Compton emission, the optically-thin spectral index above the cooling frequency $\nu_c$ becomes $\alpha_{1}=-p/2$. As the shock propagates in the surrounding medium of density profile $\rho \propto r^{-k}$, the SSA spectral peak shifts to lower frequencies. For a wind-like CSM density profile (i.e., $k=2$), $\nu_{\rm pk} \propto t^{-1}$ while $F_{\rm pk}$ remains constant.
 
\begin{figure*} 
 	\centering	\includegraphics[width=0.90\textwidth]{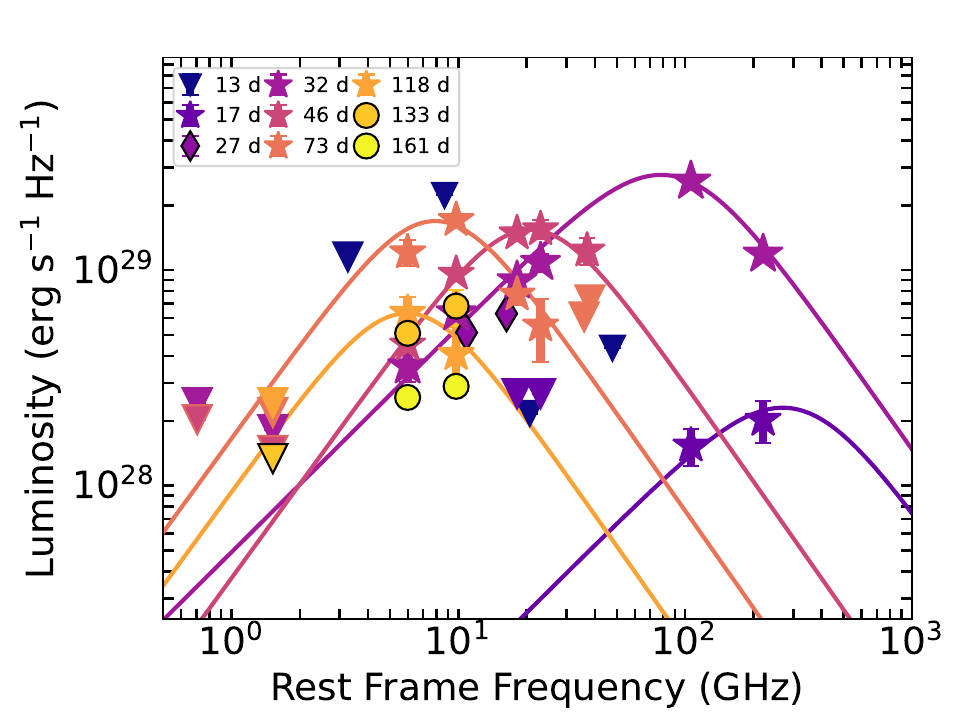} 
  \caption{Radio spectra of \at{} in the time range $\delta t_{\rm rest} \approx 13-161$\,days acquired with MeerKAT, GMRT, ATCA, ATA, and ALMA. Inverted triangles mark the 3$\sigma$ flux density upper limits. Solid lines represent best-fit broken power-law models with smoothing parameter $s = -1$ and optically-thin spectral index $\alpha_{1} = -1.5$. \at{}\, shows a complex evolution with two unprecedented elements: (i) an extremely rapid and delayed rise of the mm emission captured by ALMA; (ii) radio spectra inversion in the last two ATCA epochs (emphasized with circles).}
 \label{Fig:radioSED-fit-single-epoch}
 \end{figure*}

Figure \ref{Fig:radioSED-fit-single-epoch} shows the radio SEDs of AT\,2024wpp at $\delta t_{\rm rest} \approx 13-161$\,days. Although the shapes of the SEDs are similar to the one expected from SSA emission, the evolution of the SEDs is non-standard. The peak flux density of the SEDs brightens by approximately a factor of ten between $\delta t_{\rm rest} \approx 17-32$\,days. Subsequently, at $\delta t_{\rm rest} \approx 46-73$\,days, the peak flux remains roughly constant with the peak frequency moving to lower frequencies followed by a decrease in $\nu_{\rm pk}$ and $F_{\rm pk}$ at $\delta t_{\rm rest} \approx 118$\,days. 
The optically thick phase of the spectrum is best sampled at $\delta t_{\rm rest} \approx 32$\,days, and the spectral index is $\alpha_{\rm 2} \approx 0.7$. The slope is slightly steeper at $\delta t_{\rm rest} \approx 46$\,days;  $\alpha_{\rm 2} \approx 1.5$. In either case, the optically thick spectral indices are flatter than that expected from SSA ($\alpha_{\rm 2} = 2.5$). This deviation is commonly observed in LFBOTs \citep{Ho2019-cow,Margutti2019-cow,Nayana2021,Bright2021-xnd,Ho2022-xnd,Ho2023-at2022tsd} and is often attributed to inhomogeneities in the emitting region \citep{Bjornsson2017,Bjornsson2024,Weiler2002}.  The optically thin slope is $\alpha_{1} \approx -1$ at $\delta t \approx 73$\,d, implying $p\approx 3$.  We thus adopt $p\approx 3$ for synchrotron spectral modeling and parameter estimation.

The spectral behavior at $\delta t_{\rm rest} > 118$\,days is particularly striking. The spectral peak is $\nu_{\rm pk} \approx 6$\, GHz with a 5--9\,GHz spectral slope of $\alpha_{1}=-0.90\pm0.78$. In later observations, we find evidence for a spectral inversion at these frequencies (5--9\,GHz), with $\alpha_{1}=0.59\pm0.31$ and $0.24\pm0.50$ at 133 and 161 days, respectively \citep{Nayana2025-TNS}. This kind of spectral inversion is unprecedented in FBOTs, and may signal the emergence of a new emission component. 
 
\subsection{Spectral modeling and shock parameters}
We model the single epoch SEDs of \at{} as a broken power-law of the form:

\begin{equation}
    F_{\nu} = F_{\rm pk} \left[ \left( \frac{\nu}{\nu_{\rm pk}} \right)^{\alpha_{1}/s} + \left( \frac{\nu}{\nu_{\rm pk}} \right)^{\alpha_{2}/s} \right]^{s} 
\end{equation}

 where $\alpha_{1}$ and $\alpha_{2}$ denote the optically thin and thick spectral slopes, respectively, and $s$ defines the smoothness of the broken power-laws. We fix $\alpha_1=-1.5$ and $s=-1$, while keeping $F_{\rm pk}$, and $\nu_{\rm pk}$ as free parameters. We choose $\alpha_{\rm 1}=-1.5$ based on the optically thin spectral index at $\delta t_{\rm rest} \approx 73$\,days and the position of cooling frequencies which are self-consistently calculated later in this section. At $\delta t_{\rm rest} \approx 17$ and 32\,days, the cooling frequencies are below the SED peak, while at $\delta t_{\rm rest} \approx 46$\,days, it is closer to the peak. This would result in a steepening in the optically thin spectral slope by $\Delta \alpha=0.5$ and the optically thin slope at $\delta t_{\rm rest} \approx 73$\,days is $\approx -1$. In addition, we keep $\alpha_{2}$ as a free parameter at $\delta t_{\rm rest} \approx 32$ and 46\,days. At other epochs, we fix $\alpha_{2}$ to the best fit values obtained in the nearest epoch SED.

Following \citet{chevalier1998}, we calculate the shock radius ($R$) and magnetic field ($B$) for $p=3$ using the following equations:

\startlongtable
\begin{deluxetable*}{lcccccccccc}
\tablecaption{Shock parameters of \at{} estimated from single-epoch radio spectral modeling}
\tablehead{
\colhead{Time$^{\rm{a}}$} & \colhead{$\alpha_{2}$} & \colhead{$\nu_{pk}^{\rm{b}}$} & \colhead{$F_{pk}^{\rm{b}}$}  & \colhead{$R$} & \colhead{$(\Gamma \beta)c$} & \colhead{$B$} & \colhead{$n$} & \colhead{$U$}\\
(d)  &  & (GHz) & (mJy) & ($\times 10^{16}$cm) & ($c$) & (Gauss) & (cm$^{-3}$) & ($\times 10^{48}$erg) \\
 }
\startdata
17.5$^{\rm{d}}$ & 1.05 & 310$^{+123}_{-98}$ & 0.224$^{+0.058}_{-0.045}$ 
& $0.06^{+0.02}_{-0.01}$ & $0.013^{+0.004}_{-0.002}$ & $36.9^{+24.4}_{-16.6}$ & $6.42^{+1.9}_{-0.5}\times10^{8}$ & $0.07^{+0.04}_{-0.03}$ \\
32.4 & $1.05^{+0.05}_{-0.05}$ & $90^{+5}_{-5}$ & $2.697^{+0.077}_{-0.069}$ 
& $0.56^{+0.03}_{-0.03}$ & $0.067^{+0.004}_{-0.004}$ & $4.1^{+0.4}_{-0.4}$ & $0.31^{+0.09}_{-0.07}\times10^{6}$ & 0.76$^{+0.01}_{-0.01}$ \\
46.1 & $1.45^{+0.18}_{-0.16}$ & $21^{+3}_{-2}$  &  $1.524^{+0.062}_{-0.052}$
& $2.91^{+0.36}_{-0.36}$ & $0.244^{+0.03}_{-0.03}$ & $1.3^{+0.2}_{0.2}$ & $2.5^{+1.8}_{-0.9}\times10^{3}$ & 11.44$^{+1.33}_{1.33}$ \\
72.7 & 1.45 & $8^{+0.5}_{-0.5}$  & $1.673^{+0.074}_{-0.068}$ 
& $7.88^{+0.65}_{-0.56}$ & $0.418^{+0.035}_{-0.030}$ & $0.5^{+0.03}_{-0.03}$ & $1.23^{+0.38}_{-0.31}\times10^{2}$ & 33.12$^{+3.77}_{-3.10}$ \\
117.6 & 1.45 & 6$^{+4}_{-2}$ & 0.624$^{+0.159}_{-0.112}$ & $6.04^{+2.87}_{-2.23}$ & $0.198^{+0.094}_{-0.073}$ & $0.4^{+0.3}_{-0.1}$ & $4.2^{+4.6}_{-0.3}\times10^{2}$ & 11.68$^{+7.6}_{-3.9}$ \\
\enddata
\tablecomments{$^{\rm{a}}$ With respect to date of first light in rest frame $\delta t_{\rm rest}=\delta t_{\rm obs}/(1+z)$. \\ $^{\rm{b}}$ $\nu_{\rm pk}$ and $F_{\rm pk}$ are the intersection of the optically thick and thin power laws of synchrotron spectrum. The parameters are estimated assuming equipartition ($\epsilon_{\rm e}=\epsilon_{\rm B}=0.33$). \\ $^{\rm{c}}$ Mean shock velocity ($\Gamma \beta)c=Rc/t$.\\
$^{\rm{d}}$ The parameters at $\delta t_{\rm rest} \approx 17.5$\,days are not physical as the SED is free-free absorbed due to the surrounding medium up to radius $R \approx 10^{16}$\,cm (see \S \ref{SubSec:shock-breakout from dense CSM shell}).
\label{Tab:radio-shock-parameters}}
\end{deluxetable*}

\begin{eqnarray}
\label{eqn:R-C98}
    R=8.8 \times 10^{15} f_{\rm eB}^{-1/19} \left( \frac{f}{0.5}\right)^{-1/19} \left(  \frac{F_{\rm pk}}{\rm Jy}\right)^{9/19} \\
    \times \left( \frac{D}{\rm Mpc} \right)^{18/19} \left( \frac{\nu_{\rm pk}}{\rm GHz} \right)^{-1} \, \rm cm
\end{eqnarray}

\begin{eqnarray}
\label{eqn:B-C98}
    B=0.58 f_{\rm eB}^{-4/19} \left( \frac{f}{0.5} \right)^{-4/19} \left( \frac{F_{\rm pk}}{\rm Jy} \right)^{-2/19} \\ \nonumber \times \left( \frac{D}{\rm Mpc}\right)^{-4/19} \left( \frac{\nu_{\rm pk}}{5\,\rm GHz}\right) \, \rm G
\end{eqnarray}

The shock internal energy ($U$) is given by
\begin{equation}
\label{eqn:U-C98}
    U=\frac{1}{\epsilon_{\rm B}} \frac{4}{3} \pi f R^{3} \frac{B^{2}}{8\pi}
\end{equation}
\noindent
Here $f_{\rm eB}\equiv\frac{\epsilon_{\rm e}}{\epsilon_{\rm B}}$, where $\epsilon_{\rm e}$ and $\epsilon_{\rm B}$ are the fractions of post-shock energy in the relativistic electrons and magnetic fields, respectively. We assume equipartition of energy, i.e., $f_{\rm eB} =1$ (for $\epsilon_{\rm e}=\epsilon_{\rm B} = 0.3$). $f$ is the volume filling factor of the synchrotron emitting region, and is taken to be $f=0.5$ \citep{chevalier1998}. 

The \cite{chevalier1998} model assumes that the cooling frequency ($\nu_{\rm c}$) is above the SSA frequency ($\nu_{\rm a}$ ) and the synchrotron characteristic frequency emitted by minimum energy electrons ($\nu_{\rm m}$) is below $\nu_{a}$ (i.e., $\nu_{\rm m}<\nu_{a}<\nu_{\rm c}$). This order of characteristic frequencies may not be valid at all times. For e.g., in AT\,2018cow, $\nu_{a}>\nu_{\rm c}$ at early times due to the presence of a dense medium in the immediate environment and shock energy was dissipated in a small volume \citep{Ho2019-cow}.

We calculate $\nu_{\rm m}$ and cooling frequencies (both synchrotron cooling frequency, $\nu_{\rm c,sync}$,  and Inverse Compton (IC) cooling frequency $\nu_{\rm c,IC}$) at different epochs to check the validity of this model. The minimum Lorentz factor ($\gamma_{\rm m}$) of a distribution of accelerated electrons of energy power-law index $p$ is given by $ \gamma_{\rm m} = \frac{p-2}{p-1} \frac{m_{\rm p}}{m_{\rm e}} \epsilon_{\rm e} \beta^{2}$, where $\beta$ is the shock velocity in units of $c$.
The corresponding minimum characteristic synchrotron frequency is $\nu_{\rm m} = \gamma_{\rm m}^{2} \nu_{\rm g}$, where $\nu_{\rm g} =  \frac{qB}{2\pi m_{\rm e}c}$ is the gyro frequency. Here, $q$ and $m_{\rm e}$ are the charge and mass of electron, respectively. We find $\nu_{\rm m} \ll \nu_{\rm pk} \equiv \nu_{\rm a}$ during the observed epochs. The synchrotron and IC cooling frequencies are $\nu_{\rm c,sync}=\gamma_{\rm sync}^{2} \nu_{\rm g}$ and $\nu_{\rm c, IC}=\gamma_{\rm IC}^{2} \nu_{\rm g}$, where electrons with $\gamma > \gamma_{\rm sync}$ or $\gamma_{IC}$ cool at time $t$. $\gamma_{\rm sync}$ and $\gamma_{\rm IC}$ are given by \citep{Rybicki1979}
\begin{equation}
\label{eqn:Gamma_cool_sync}
    \gamma_{\rm sync} = \frac{6\pi m c^{2}}{\sigma_{\rm T} c B^{2}t}
\end{equation}
 
\begin{equation}
\label{eqn:Gamma_cool_IC}
    \gamma_{\rm IC} =  \frac{3\pi m c^{2} R^{2}}{\sigma_{\rm T} L_{\rm bol}  t}
\end{equation}
Here $\sigma_{\rm T}$ is Thomson scattering cross-section and $L_{\rm bol}$ is the bolometric luminosity at time $t$. We calculate $\nu_{\rm c, sync}$ and $\nu_{\rm c, IC}$ using $R$ and $B$ from Table \ref{Tab:radio-shock-parameters}. We use $L_{\rm bol} \approx (8, 0.9, 0.3, 0.07, 0.01) \times 10^{43}\,\rm erg\,s^{-1}$ at $\delta t_{\rm rest} =$ 17.5, 32.4, 46.1, 72.7, and 117.6\,days, respectively (Paper I). The cooling frequencies are lower than $\nu_{\rm pk}$ at $\delta t_{\rm rest} =$ 17.5 and 32.4\,days, which implies that \cite{chevalier1998} is not self-consistent at these epochs. \cite{Ho2022-xnd} present formulas to estimate source properties in the regime $\nu_{\rm a} > \nu_{\rm c}$ (see their Appendix C). For $p=3$, the expressions for $R$ and $B$ are the following

\begin{eqnarray}
\label{eqn:R-Ho2022}
    R=4.2 \times 10^{15} f_{\rm eB}^{-1/13} \left(  \frac{F_{\rm pk}}{\rm Jy}\right)^{6/13} 
     \left( \frac{D}{\rm Mpc} \right)^{12/13} \\ \times \nonumber \left( \frac{\nu_{\rm pk}}{\rm GHz} \right)^{-11/13} \left( \frac{t}{100\, \rm days}\right)^{1/13} \, \rm cm
\end{eqnarray}

\begin{eqnarray}
\label{eqn:B-Ho2022}
    B=0.14 f_{\rm eB}^{-4/13} \left( \frac{F_{\rm pk}}{\rm Jy} \right)^{-2/13} \left( \frac{D}{\rm Mpc}\right)^{-4/13} \\ \nonumber \times \left( \frac{\nu_{\rm pk}}{5\,\rm GHz}\right)^{21/13} \left( \frac{t}{100\, \rm days}\right)^{4/13} \, \rm G
\end{eqnarray}

We estimate shock parameters at $\delta t_{\rm rest} \approx 17$ and 32\,days using these equations  and report the self-consistent estimates of $R$, $v$, $B$, $U$, and $n$ at all epochs in Table \ref{Tab:radio-shock-parameters}. We note that the parameters at $\delta t_{\rm rest} \approx 17.5$\,days are not physical as the SED at $\delta t_{\rm rest} \approx 17$\,days is free-free absorbed due to the surrounding medium up to radius $R \approx 10^{16}$\,cm. The SED evolution between $\delta t_{\rm rest} \approx 17-32$\,days is best explained in a scenario where the shock is propagating through a dense shell (see \S \ref{SubSec:shock-breakout from dense CSM shell}).

\subsection{Synchrotron emission from thermal electrons}
\label{subsec:synchrotron-thermal}
Collisionless strong shocks are commonly believed to accelerate electrons into a non-thermal power-law distribution via diffusive shock acceleration \citep{Bell1978,Blandford1978,Blandford1987,Spitkovsky2008,Sironi2009,Sironi2011,Caprioli2015}. Numerical models of collisionless shocks suggest that the electron distribution consists of both thermal and non-thermal populations. \cite{Margalit2021} discuss the contribution of thermal electrons to the emergent synchrotron flux in transrelativistic shocks and find a significant contribution from thermal electrons towards the peak emission for shocks of velocity $v \gtrsim 0.2$c. The key observational signatures of synchrotron emission from thermal electrons are a steep optically thin spectral index and a $F_{\nu} \propto \nu^{2}$ spectral slope in the optically thick regime. 

\cite{Ho2022-xnd} reported a steep optically-thin spectral index $\alpha_{1} \approx -2$ in the case of AT\,2020xnd and modeled the SEDs as synchrotron emission from a thermal population of electrons. The observed optically-thin spectral index for \at{} is $\alpha_{1} = -1.29\pm0.17$ with an optically-thick spectral slope of $\alpha_{2} = 0.66\pm0.02$ ($\delta t_{\rm rest} \approx 32$\,days) and $\alpha_{2} = 1.45\pm0.18$ ($\delta t_{\rm rest} \approx 46$\,days). The shock velocities from single epoch spectral modeling are $v > 0.2$c at $\delta t_{\rm rest} \approx 46$ and 73\,days (see Table \ref{Tab:radio-shock-parameters}). Even though the spectral slope is not as steep as that expected from a thermal population, motivated by the large shock velocities, we fit individual SEDs at $\delta t_{\rm rest} \approx $ 32, 46, and 73\,days with the ``thermal + non-thermal" synchrotron emission model of \cite{Margalit2021}, and explore the parameter space using MCMC. We keep the number density, shock radius, and the electron thermalization efficiency ($\epsilon_{\rm T}$) as free parameters and run the fit keeping the ratio of energy in non-thermal electrons to thermal electrons ($\delta=\epsilon_{\rm e}/\epsilon_{\rm T}$) and $\epsilon_{\rm B}$ fixed. We use $\epsilon_{\rm B}=0.1$ and run the model for a range of $\delta$ values ($\delta =$0.001, 0.01, 0.1, and 1). The model failed to reproduce the SEDs at $\delta t_{\rm rest} \approx$ 32 and 46\,days, but resulted in reasonable fits for $\delta t_{\rm rest} \approx 73$\,days. The model resulted in optically thick spectral slopes that were too steep and could not match the observed values at $\delta t_{\rm rest} \approx 32$\,days. At $\delta t_{\rm rest} \approx 46$\,days, the peak and decay were not reproduced by the models. The best-fit values of parameters at $\delta t_{\rm rest} \approx 73$\,days are $R=7.13^{+0.91}_{-0.67} \times 10^{16}$\,cm, $n=129^{+32}_{-27}\,\rm{cm^{-3}}$, and $\epsilon_{\rm T}=0.56^{+0.28}_{-0.21}$ for $\epsilon_{\rm B}=0.1$ and $\delta=0.1$ (i.e., $\epsilon_{\rm e}=0.005$). The shock radius and density derived from this model are very similar to the ones estimated from single-epoch spectral modeling as expected \citep[see Fig. 2 of][\S 4]{Margalit2021} and also seen in the case AT\,2020xnd \citep{Ho2022-xnd}. Although there may be some contribution to the synchrotron flux of thermal electrons at $\delta t_{\rm rest} \approx 73$\,days, we conclude that it is not necessary to invoke this model to interpret the observed SEDs. 
\section{The origin and evolution of the radio emitting outflow}
\label{sec:origin of mm emission}
The spectral evolution of \at{} is characterized by distinct phases in the time evolution of both the peak flux density and peak frequency. The peak flux density increases as $F_{\rm pk} \propto t^{4.04 \pm 0.42}$ during $\delta t_{\rm rest} \approx 17-32$\,days. This is followed by a slight decline between 32$-$73 days ($F_{\rm pk} \propto t^{-0.67 \pm 0.06}$) and a steeper decay as $F_{\rm pk} \propto t^{-2.05 \pm 0.54}$ at $\delta t_{\rm rest} > 73$\,days. The corresponding evolution of the peak frequency is $\nu_{\rm pk} \propto t^{-2.01 \pm 0.65}$ at $\delta t_{\rm rest} \approx 17-32$\,days, $\nu_{\rm pk} \propto t^{-2.99 \pm 0.10}$ at $\delta t_{\rm rest} \approx 32-73$\,days, and $\nu_{\rm pk} \propto t^{-0.60 \pm 1.05}$ at $\delta t_{\rm rest} \approx 73-118$\,days.  The rapid brightening of SEDs between $\delta t_{\rm rest} \approx 17-32$\,days may be attributed to suppression of the intrinsic synchrotron emission due to radiative cooling processes (IC and/or synchrotron cooling). In particular, depending on the position of the cooling frequency, the cooling process can reduce the observable flux at early epochs. We investigate the relative importance of different cooling processes to explain the observed SED evolution in \S \ref{SubSec:Cooling}.

The physical parameters inferred from single epoch spectral analysis imply a rapidly growing radio emitting region, with a temporal evolution of $R\propto t^{3.12\pm0.12}$, consistent with an accelerating outflow. The outflow velocity increases from $\Gamma \beta c \approx 0.07c$ to $0.42c$ during $\delta_{\rm rest} t=32-73$\,days, with an increasing amount of internal energy coupled to this outflow (see Table \ref{Tab:radio-shock-parameters}). The increase in the blast-wave energy with time and its acceleration can be (i) intrinsic or (ii) apparent. If intrinsic, more energy is being deposited into the shock with time (e.g., via continuous winds from an accretion disk). The shock then breaks out from this dense region and accelerates in a very steep density profile, a process similar to that of shocks expanding in the outer layers of exploding massive stars. Alternatively, if the effect is apparent, such evolution can arise due to geometric effects. In off-axis jet models, the observer sees more energy progressively coming into their line of sight as a consequence of the deceleration of the outflow. We explore both scenarios in detail in \S \ref{SubSec:offaxis models} and \S \ref{SubSec:shock-breakout from dense CSM shell}.

\subsection{Energy losses of synchrotron emitting electrons}
\label{SubSec:Cooling}
\begin{figure*}
 	\centering   
     \includegraphics[width=0.49\textwidth]{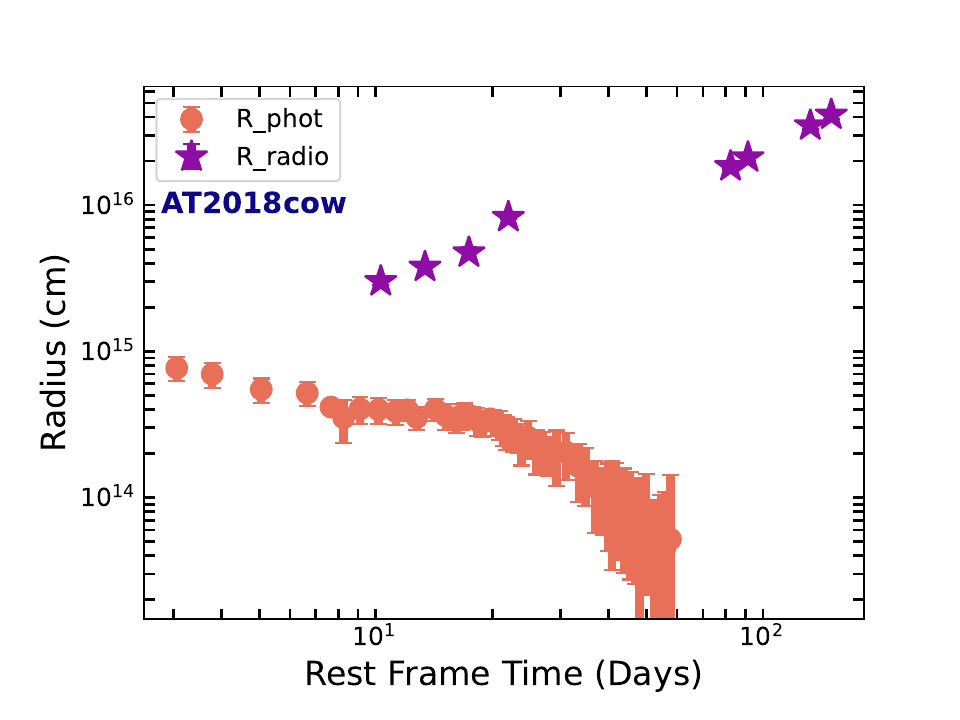}   	\includegraphics[width=0.49\textwidth]{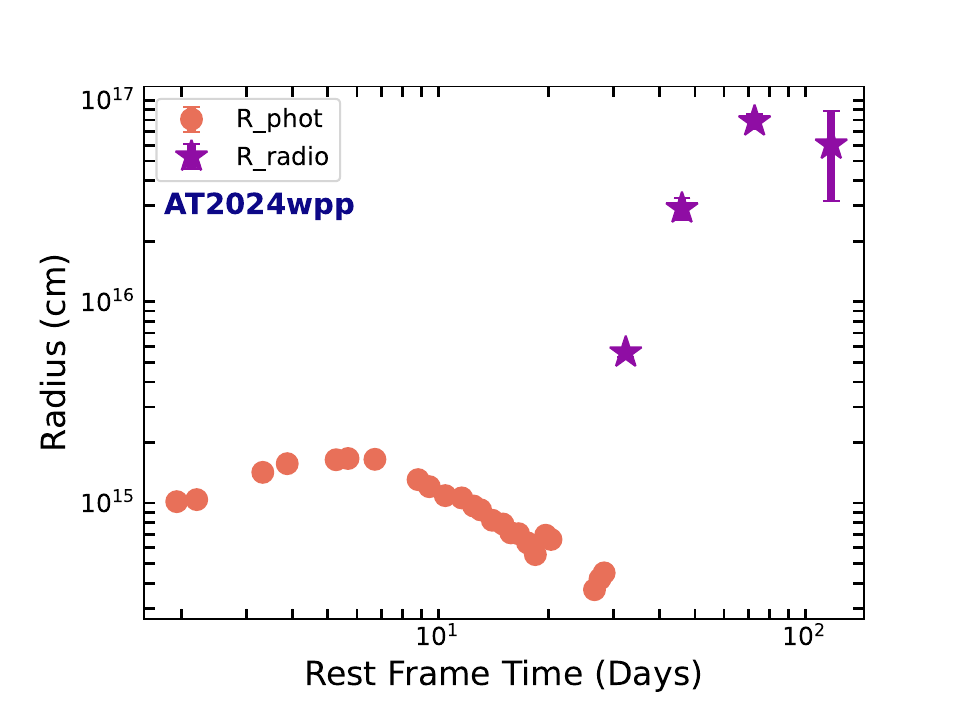}  
  \caption{Evolution of the radius of optical photosphere and radio photosphere of AT\,2018cow in left panel \citep{Margutti2019-cow} and \at{} in right panel (Paper I). The size of optical photosphere is derived by fitting a blackbody function to the bolometric luminosities. The size of radio photosphere (rest frame) is estimated by modeling single epoch radio SEDs adopting SSA formalism \citep{chevalier1998}.  }
 \label{Fig:RadioPhot}
 \end{figure*}

Synchrotron-emitting electrons can lose energy via various cooling processes: synchrotron cooling and/or IC cooling. In synchrotron cooling, relativistic electrons lose energy by emitting radiation in the presence of magnetic fields, with the cooling efficiency increasing at higher magnetic field strengths and electron energies. In the presence of a strong radiation field, relativistic electrons can upscatter low-energy optical photons to higher energies and IC cooling can be important. The mm emission from \at{} at $\delta t_{\rm rest} \approx 17$\,days is an order of magnitude lower than the mm emission at $\delta t_{\rm rest} \approx 32$\,days. We explore the possibility of flux suppression via electron cooling to account for this. In the case of an IC cooling scenario, the radio outflow needs to always be in front of the optical photosphere. We plot the optical and radio photosphere radii of \at{}  at multiple epochs ($\delta t_{\rm rest} \approx 32-118$\,days) in Fig \ref{Fig:RadioPhot} (right panel) and find that this is indeed the case.

We extrapolate the radio SED from $\delta t_{\rm rest} \approx 46.1$\,days to $\delta t_{\rm rest} \approx 17.5$\,days assuming a density profile $\rho_{\rm CSM} \propto r^{-3}$ and a constant shock velocity of $\Gamma \beta c = 0.24$\,c. The expected flux densities at 100 and 200 GHz on this SED are $\approx$ 3.1 and 1.6 mJy, respectively. At this epoch, the Lorentz factor of electrons cooling via IC cooling is $\gamma_{\rm IC} = 12$ and via synchrotron cooling is $\gamma_{\rm sync} = 14$ (for $\epsilon_{\rm e}=\epsilon_{\rm B}=0.3$), indicating that IC cooling is marginally dominant. The Lorentz factor of electrons emitting at 100 and 200 GHz are $\approx$ 117 and 166 at $\delta t_{\rm rest} \approx 17.5$\,days. Thus, cooling can suppress the flux at 100 GHz by a factor of $\approx 10$ and at 200 GHz by a factor of $\approx 14$. This will result in observed flux densities of $F_{\rm 100\,GHz} \approx 0.3$ mJy and $F_{\rm 200\,GHz} \approx 0.1$ mJy. The observed ALMA flux densities at $\delta t_{\rm rest} \approx 17.5$\,days are $F_{100\, \rm GHz}=0.08$ mJy and $F_{200\, \rm GHz}=0.10$ mJy, which are $\approx 4$ times lower at 100 GHz and similar at 200 GHz. So, the ALMA flux densities are suppressed beyond what can be accounted for by electron cooling alone at 100 GHz whereas at 200 GHz, the flux density is consistent with that expected from IC suppression.

If we carry out a similar exercise by extrapolating the synchrotron SED from $\delta t_{\rm rest} \approx 46.1$\,days to $\delta t_{\rm rest} \approx 32.4$\,days, the Lorentz factor of electrons experiencing IC cooling is $\approx 203$ and synchrotron cooling is $\approx 49$, indicating synchrotron cooling is the dominant cooling process. 
The effect of cooling should suppress the flux to $F_{\rm 100 \, GHz} \approx 0.15$ mJy and $F_{\rm 200 \, GHz} \approx 0.04$ mJy. However, the observed flux densities at $\delta t_{\rm rest} \approx 32.4$\,days are $F=1.3$ mJy at 100 GHz and $F=0.6$ mJy at 200 GHz, which are 9$-$15 times higher.    

We caution that the above estimates based on cooling timescales depends strongly on the CSM density profile and shock velocities as the IC cooling time scale is a strong function of $R$ and synchrotron cooling timescale is a strong function of $B$. It is clear that (not trying to match the absolute numbers), the fluxes at 100 and 200 GHz will decrease from $\delta t \approx 17$ to 32\,days in a scenario where fluxes are suppressed due to cooling whereas the observed ALMA fluxes are increasing by approximately an order of magnitude. Thus, cooling effects cannot account for the observed sharp rise in millimeter fluxes.
\subsection{Off-axis models}
\label{SubSec:offaxis models}  
Alternatively, the steep rise of the radio emission can be due to emission from an off-axis relativistic jet, where the emission is initially beamed away from the observer's line of sight. In this scenario, the outflow needs to be a relativistic jet with an initial off-axis viewing angle. The radio emission at early times will be suppressed by relativistic beaming and then increase rapidly as the jet decelerates and relativistic beaming becomes less severe. 
The millimeter flux density of \at{} rises approximately an order of magnitude between 17 and 32\,days with a temporal slope of $F_{\rm 97.5\,GHz} \propto t^{4.44\pm0.39}$. The temporal indices of synchrotron flux from a non-spreading jet viewed off-axis can be as steep as $F \propto t^{10}$ for a flat CSM density profile \citep[][see their Table A1]{Beniamini2023}. Thus the observed fast rise in flux densities can be accommodated by off-axis jet models \citep{Sfaradi2024}. The millimeter band flux densities drop to $F_{\rm 97.5\, GHz} \approx 150\,\mu$Jy and $F_{\rm 203\,GHz} \approx 50\,\mu$Jy at $\delta t_{\rm rest} \approx 46$\,days indicating that the time of peak of the millimeter band light curve is at $\delta t_{\rm rest} \approx 32$\,days. Note that the above-mentioned flux densities at $\delta t_{\rm rest} \approx 46$\,days are from the best-fit SED. After the peak time, the observer should see most of the outflow and the inferred kinetic energy can be considered to be representative of the actual kinetic energy of the relativistic outflow implying $E_{k,iso}\approx 3.3 \times 10^{49}\,$erg (from the SED analysis at $\delta t_{\rm rest} \approx 73$\,days). We note that this estimate of $E_{\rm k, iso}$ is a lower limit due to the equipartition assumption.

Consider a top-hat jet in which $dE_{\rm k}/d\Omega=E_{\rm k,iso}/4\pi$ is constant up to a certain opening angle $\theta_{0}$ with initial Lorentz factor $\Gamma_{0}$ (where $\Gamma_{0} \gg \theta_{0}^{-1}$) propagating in a medium of density profile $\rho = A r^{-k}$. The jet decelerates as it propagates through the medium and the deceleration radius is given by \citep{Beniamini2023}:
\begin{equation}
\label{eqn:Rdec-Beniamini2023}
    R_{\rm dec}=\left[ \frac{(3-k) E_{\rm k,iso}}{4\pi A c^{2} \Gamma_{0}^{2}} \right]^{\frac{1}{3-k}} \, \rm cm
\end{equation}
The corresponding deceleration timescale is $t_{\rm dec} = (1+z)R_{\rm dec}/2c\Gamma_{0}^{2}$ for an on-axis observer. After $t_{\rm dec}$, the jet bulk Lorentz factor evolves as $\Gamma(t) \propto t^{-\left[ \frac{3-k}{2(4-k)} \right]}$.  For an observer located at an angle $\theta_{\rm obs}$ from the initial direction of the jet, the peak of the light curve would be at $t_{\rm pk} = t_{\rm dec} (\theta_{\rm obs} \Gamma_{0})^{\frac{8-2k}{3-k}}$. For $E_{\rm K, iso} \approx 3.3 \times 10^{49}$\,erg and $n=0.3 \times 10^{6}\, \rm cm^{-3}$ from equipartition analysis (see Table \ref{Tab:radio-shock-parameters}), the deceleration radius is $R_{\rm dec} \approx 1.2 \times 10^{14}$\,cm for $\Gamma_{\rm 0}=100$ and $\theta_{\rm obs}=30^{\circ}$. This translates to $t_{\rm pk} \approx$ 2.3 hours (for $k=0$) and $t_{\rm pk} \approx$ 0.07 hours (for $k=2$), which is significantly smaller than the time of peak of the mm emission. This argues against the off-axis jet model for the observed increase in the millimeter emission at $\delta t_{\rm rest} \approx 17-32$\,days. 
Furthermore, the temporal decay of the mm component from $\delta t_{\rm rest} \approx 32$\,days appears to be $F_{97.5\, \rm GHz} \propto t^{-5.9}$ and $F_{203\, \rm GHz} \propto t^{-6.8}$ based on the extrapolation to the best-fit SED model at $\delta t_{\rm rest} \approx 46$\,days, which is difficult to explain in an off-axis model. The maximum temporal decay index post jet break is expected to be $F_{\nu} \propto t^{-p}$ \citep{Sari1999}. We also note the extremely fast decay of AT\,2018cow at $\approx 230$\,GHz with a decay index of $F \propto t^{-4.7}$ at $\delta t_{\rm rest} > 44$\,days (see Fig. \ref{Fig:10GHz-mm-lightcurves-FBOTs}), possibly indicates a similar physical origin. 

We further explore the off-axis model following the generalized equipartition analysis presented in \cite{Matsumoto23equipartition}. It is assumed that the observed emission is dominated from a small region of the order $\pi/\Gamma$, where $\Gamma$ is the bulk Lorentz factor. A critical parameter in this setup is the apparent velocity in Newtonian limits ($\beta_{\rm eq,N}$) defined by \citep{Matsumoto23equipartition}
\begin{equation}
    \beta_{\rm eq, N} = \frac{(1+z)R_{\rm eq,N}}{ct}
\end{equation}
Here $R_{\rm eq,N}$ is the Newtonian equipartition radius (listed in Table \ref{Tab:radio-shock-parameters} for AT\,2024wpp). $\beta_{\rm eq,N} = 0.23$ is the critical value above which the relativistic off-axis solution transitions into the Newtonian on-axis branch \citep{Matsumoto23equipartition}. Although $\beta_{\rm eq,N}=0.23$ was estimated for a maximum viewing angle of $\pi$, \cite{Beniamini2023} suggest the critical value to be $\beta_{\rm eq,N} = 0.44$ for a more realistic maximum viewing angle of $\pi/2$. The apparent velocity for AT\,2024wpp appears to increase to $\beta_{\rm eq, N} \approx 0.4$ by $\delta t_{\rm rest} \approx 73$\,days (see Table \ref{Tab:radio-shock-parameters}) approaching this critical value and then drops to $\beta_{\rm eq, N} \approx 0.2$ by $\delta t_{\rm rest} \approx 118$\,days. The shock energy also increases from $U \approx 0.8\times10^{48}$\,erg to $\approx 33 \times 10^{48}$\,erg during $\delta t_{\rm rest} \approx 32-73$\,days. We calculate the radius that minimizes the energy at each epoch for $p=3$ assuming equipartition ($\epsilon_{\rm e}=\epsilon_{\rm B}=0.33$). We consider four different observer viewing angles $\theta_{\rm obs} =30^{\circ},\,45^{\circ},\,60^{\circ},\,90^{\circ}$. At $\delta t_{\rm rest} \approx 118$\,days, for an off-axis angle $\theta=90^{\circ}$, the model requires $\Gamma \approx 6$ and the corresponding energy is $U \approx 5 \times 10^{49}$\,erg. For other viewing angles ($\theta_{\rm obs}=30^{\circ},\,45^{\circ},\,60^{\circ}$), the inferred velocities at these late times are even higher ($\Gamma>7$). These high $\Gamma$ values represent strongly collimated outflows and will not result in emission that peaks at $\delta t_{\rm rest} \approx 32$\,days for off-axis observers. Thus, this model does not provide a natural explanation for the observations.

\begin{figure*} 
 	\centering 	\includegraphics[width=0.90\textwidth]{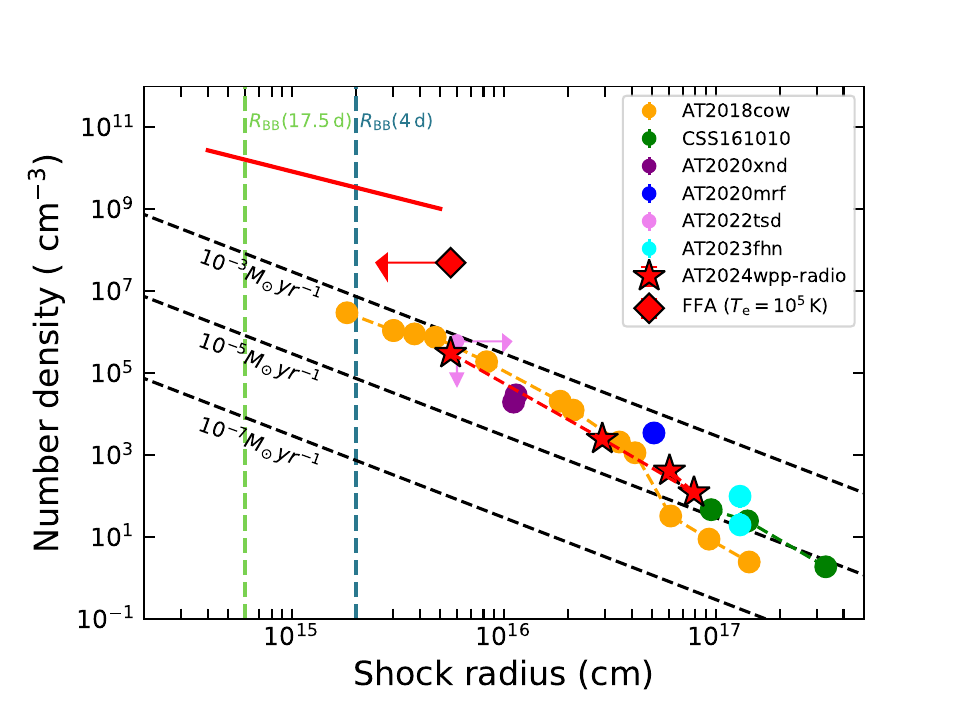} 
  \caption{Density profile of the medium around LFBOTs. \at{} (this work), AT\,2018cow \citep{Ho2019-cow,Margutti2019-cow}, CSS161010 \citep{coppejans2020}, AT2020xnd \citep{Bright2021-xnd,Ho2022-xnd}, AT\,2020mrf \citep{Yao2021-mrf}, AT\,2022tsd \citep{Ho2023-at2022tsd}, and AT\,2023fhn \citep{Chrimes2024b-fhn}. Dashed black lines denote number density profiles corresponding to constant mass-loss rates for an arbitrary wind velocity of $v_{\rm w} \approx 1000 \rm \,km\,s^{-1}$. The vertical blue and green lines denote the position of optical photosphere at optical peak ($\delta t \approx 4$\,days) and at $\delta t \approx 17.5$\,days, respectively (Paper I). The solid red line denotes the CSM density profile $\rho_{\rm CSM}(r) \propto r^{-1.3}$ inferred from interpreting the NIR excess as related to effects of free-free emission (see \S6.2 in Paper I). The inferred CSM density profiles of different events at $>10^{16}\,\rm{cm}$ are remarkably similar, which suggests a simple physical mechanism or a self-regulating process (\S\ref{subsec:csm-density-profile}).  }
 \label{Fig:density-FBOTs}
 \end{figure*}

 \begin{figure*} [t!]
 	\centering       	\includegraphics[width=0.45\textwidth]{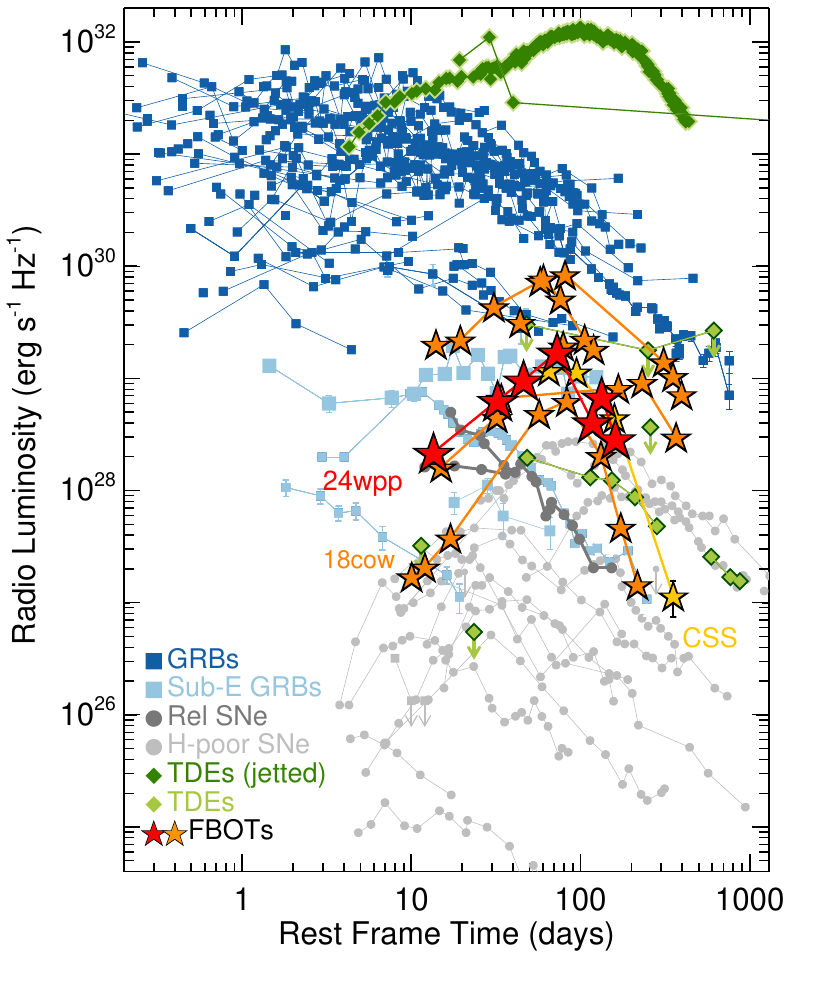}  
 	\includegraphics[width=0.475\textwidth]{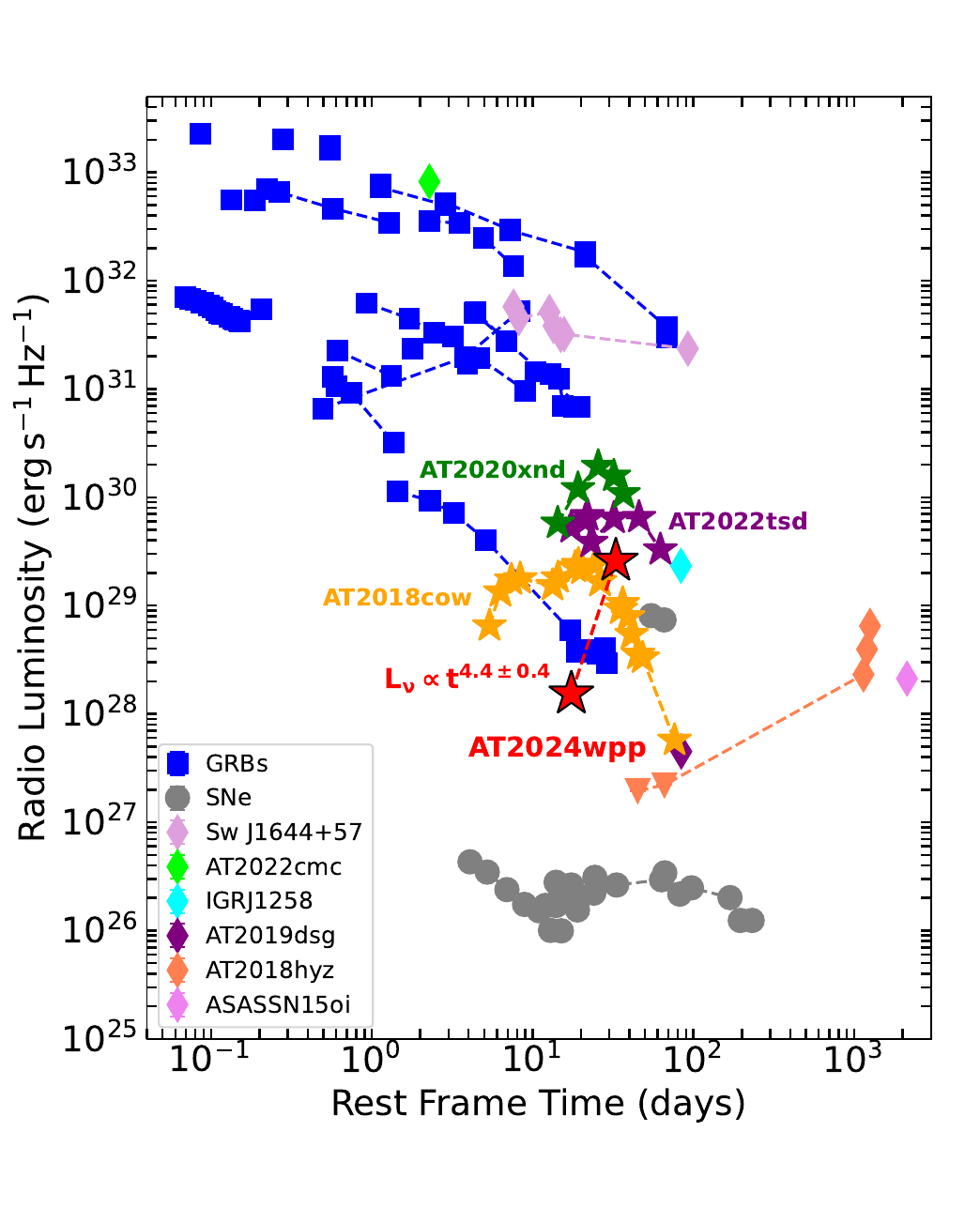}  
    \vskip -0.5 cm
  \caption{\emph{Left Panel:} Radio luminosity light-curve of \at{}\, at $\approx$ 10\,GHz in the context of other transients. FBOTs stand out for their luminosities that are intermediate between ultra-relativistic GRBs and  SNe, while also showing a characteristic bell-shaped light-curve peaking at $\approx80$\,d  followed by a very steep decay. This shared radio behavior is at the core of the ``universal density profile'' that we discuss in \S\ref{subsec:csm-density-profile}.  \emph{Right Panel:} The (sparsely populated) millimeter band (80-200 GHz) phase space of light curves of different extragalactic transients  GRBs \citep[][and references therein]{Eftekhari2022}, TDEs \citep{Berger2012-swJ1644,Perley2022-at2022cmc-gcn,Andreoni2022,Yuan2016,Cendes2021}, CCSNe \citep{Weiler2007,Horesh2013-sn2011dh} and FBOTs \citep{Ho2019-cow,Ho2022-xnd,Ho2023-at2022tsd}.  In stark contrast with the well-behaved 10\,GHz emission,  the mm-emission from even this small sample of FBOTs displays remarkable diversity, likely mirroring the diverse properties of the innermost material the shock emission is emerging from.}
 \label{Fig:10GHz-mm-lightcurves-FBOTs}
 \end{figure*}

\subsection{Radio emitting outflow propagating through a dense CSM}
\label{SubSec:shock-breakout from dense CSM shell}
 The SED evolution of \at{} can be interpreted in a scenario in which the shock interacts with a dense and compact CSM shell at early times ($\delta t_{\rm rest} \approx 17-32$\,days). 
In this case, the radio emission can be initially suppressed by free-free absorption (FFA),
and as the shock propagates and emerges out from the dense shell, there is an increase in flux density as a result of the lower optical depth. 
If the mm-flux rise is due to the different optical depths, then $F_1/F_2\propto e^{-\tau_{\rm FFA}}\approx 0.06$, where  $F_1$ and $F_2$ are the 97.5 GHz flux densities at 17 and 32\,days. We derive $\tau_{\rm FFA}\approx3$ with $\tau_{\rm FFA}\approx \kappa_{\rm FFA}\hat{n}(R_2-R_1)$. Here, $\kappa_{\rm FFA}$ is the free-free opacity and $\hat{n}$ is the average density between an inner $R_1$ and outer $R_2$ radius of the dense CSM. 
The radial extent of this dense medium is also limited, as we observe a steep temporal decay of light curves and an SSA evolution of the SED at $\delta t_{\rm rest} > 32$\,days. 
Assuming the shock emergence to be happening at the edge of the dense shell \citep{Khatami2024}, we approximate $R_{2}\approx 0.56 \times 10^{16}$\,cm with  the equipartition radius at $\delta t_{\rm rest} \approx 32$\,days.  
We assume the radio emitting region to be above the optical photosphere (estimated in Paper I) at 17\,days, and we use $R_{\rm 1} \gtrsim 0.06 \times 10^{16}$ cm. The equipartition shock radius at $\delta t_{\rm rest} \approx 17$\,days ($\approx 0.05 \times 10^{16}$ cm) is \emph{not} physical as it does not account for FFA. Free-free opacity ($\kappa_{\rm FFA}$) is defined as \citep{Rybicki1979} 
\begin{eqnarray}
    \kappa_{\rm FFA} = 0.018 \times T_{\rm e}^{-3/2} Z^{2} \hat{n} \nu^{-2} g_{\rm ff}
\end{eqnarray}

We infer $\hat{n}\approx 0.5 (T/10^5\,\rm{K})^{3/4} \times 10^{8} \rm cm^{-3}$ assuming constant density between $R_{\rm 1}$ and $R_{\rm 2}$. The corresponding CSM mass would be $M_{\rm CSM} \approx 0.07 (T/10^5\,\rm{K})^{3/4} M_{\odot}$. A similar physical scenario was invoked to explain the early bright millimeter emission from AT\,2018cow: the size of the CSM shell was inferred to be $\approx 1.7 \times 10^{16}$\,cm with a mass $\approx 0.002 M_{\odot}$ \citep{Ho2019-cow}.

The CSM density profile derived from later ($\delta t_{\rm rest} \approx 46-118$\, days) radio SEDs is $\rho_{\rm CSM} \propto r^{-3.10 \pm 0.16}$ at $R \gtrsim 3 \times 10^{16}\, \rm cm$ (Fig. \ref{Fig:density-FBOTs}). The shock is expected to accelerate at the outer edge of the dense shell above which the density profile is steeper than $s>3$, where $\rho(r) \propto r^{-s}$ \citep{Matzner99,Waxman1993}. The shock velocities derived from the equipartition analysis increase from $\Gamma \beta c = 0.07$ to 0.42 from $\delta t_{\rm rest} \approx 32$ to 73\,days, indicating an accelerating shock wave. The shock internal energy increases from $U \approx 0.8 \times 10^{48}$\, erg to $U \approx 33 \times 10^{48}$\, erg between these epochs. A high-density medium can efficiently convert the kinetic energy to thermal energy leading to large radio luminosities \citep{Khatami2024}. The actual scaling of luminosity would be with the thermal energy per unit radius ($U/R$). From equations \ref{eqn:U-C98}, \ref{eqn:R-C98}, and \ref{eqn:B-Ho2022}, one can write $U/R \propto L_{\rm pk}^{8/13}$. Thus, the high millimeter luminosities and SED evolution of \at{} at early times are consistent with a scenario where the shock is propagating through a dense shell at small radii ($\lesssim 10^{16}\, \rm cm$). 

In Paper\,I, we estimate a density profile of $\rho_{\rm CSM}(r) \propto r^{-1.3}$ for the medium above the optical photosphere to account for the NIR excess observed at $\delta t_{\rm rest} \approx 30$\,days, under the assumption that the NIR excess is due to free-free optical depth effects in a scattering dominated medium (see section 6.2 in Paper\,I).  The optical photosphere is $R_{\rm BB}(30\,\rm{days}) \approx 4 \times 10^{14}$\,cm. In this scenario, the observed NIR luminosity implies a density $n \approx 4 \times 10^{8}\, \rm cm^{-3}$ at $R \approx 10^{16}\, \rm cm$. While there are caveats to the NIR excess interpretation, the combined inferences on CSM densities from NIR analysis and radio modeling that we show in Fig. \ref{Fig:density-FBOTs} paint a picture of a dense shell at $R \lesssim 10^{16}$\, cm, with a $\rho_{\rm CSM} \propto r^{-3}$ at larger radii.

\section{Comparison with other LFBOTs}
\label{Sec:comparisonRadio-Xray}

\subsection{LFBOTs at Radio and millimeter wavelengths}

\begin{figure} [t!]
 	\centering   
 	\includegraphics[width=0.5\textwidth]{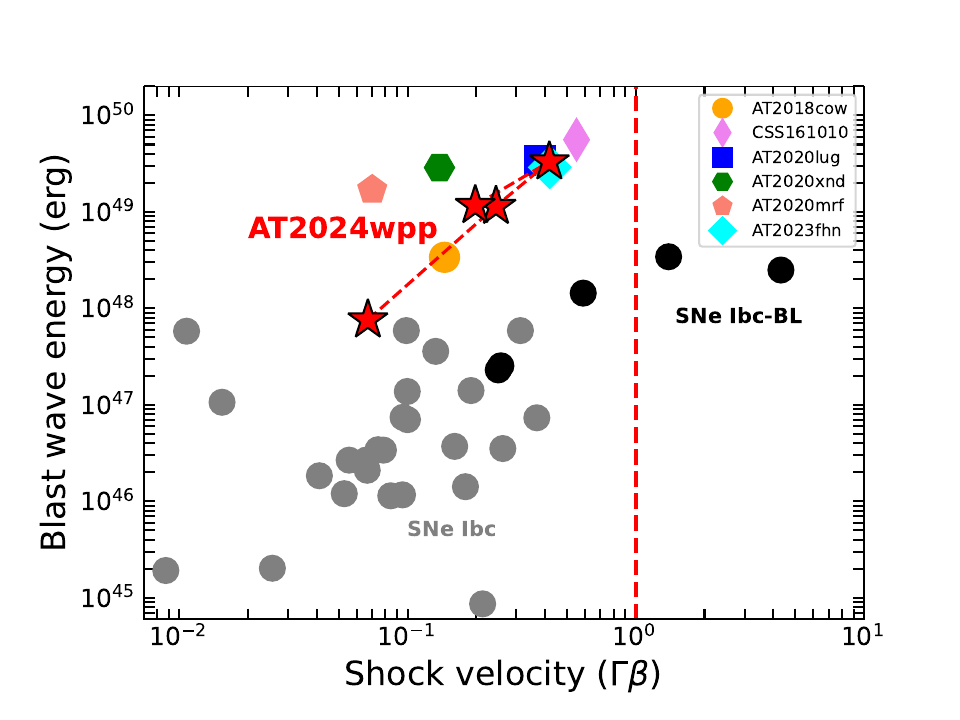}    
  \caption{Blast wave energy versus shock velocity of FBOTs: AT\,2024wpp ($\delta t \approx 32-118$\,days) AT\,2018cow ($\delta t \approx 22$\,d), CSS161010 ($\delta t \approx 99$\,d), AT\,2020xnd ($\delta t \approx 38$\,d), AT2022mrf ($\delta t \approx 261$\,d, and AT\,2023fhn ($\delta t \approx 138$\,d). References: \cite{Margutti2019-cow,Ho2019-cow,coppejans2020,Bright2021-xnd,Ho2022-xnd,Yao2021-mrf,Chrimes2024b-fhn}. }
 \label{Fig:shock-energy-FBOTs}
 \end{figure}
 
 \begin{figure} [b!]
 	\centering   
        \includegraphics[width=0.43\textwidth]{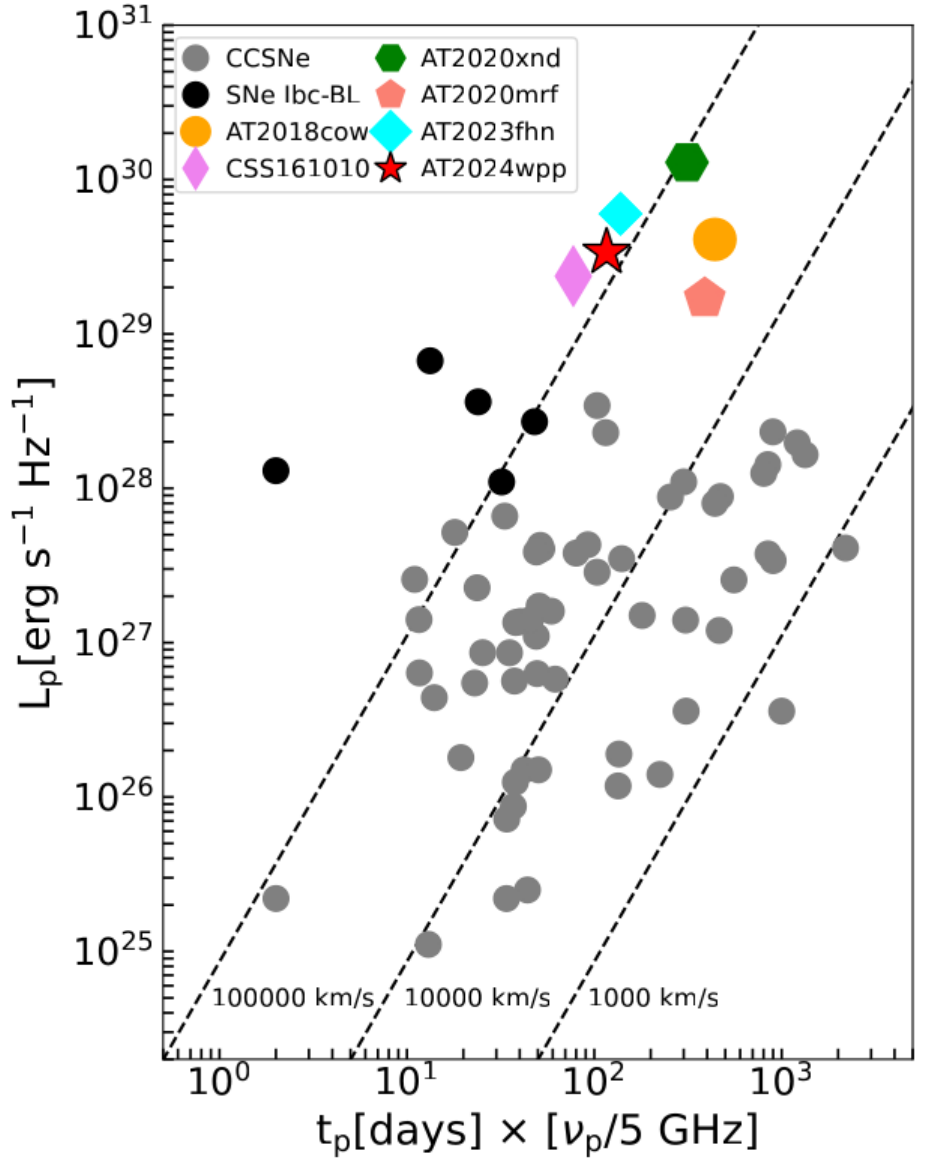}   \caption{Radio spectral luminosities of different astrophysical transients. $L_{\rm p}$ denotes the peak spectral luminosity in the $8-10$ GHz band. $\nu_{\rm p}$ and $t_{\rm p}$ represent the peak frequency and peak time of radio SED, respectively. The dashed lines denote the mean shock velocity in a synchrotron self-absorption  scenario \citep{chevalier1998}. References: \citep{Margutti2019-cow,Ho2019-cow,coppejans2020,Bright2021-xnd,Ho2022-xnd,Yao2021-mrf,Chrimes2024b-fhn}. }
 \label{Fig:Lp-nuptp diagram}
 \end{figure}

Other than \at{}, there are seven LFBOTs with long-term X-ray and radio observations: AT\,2018cow \citep{Ho2019-cow,Margutti2019-cow}, CSS161010 \citep{coppejans2020}, AT\,2018lug \citep{Ho2020-AT2018lug}, AT\,2020mrf \citep{Yao2021-mrf}, AT\,2020xnd \citep{Bright2021-xnd,Ho2022-xnd} AT\,2022tsd \citep{Ho2023-at2022tsd}, and AT\,2023fhn \citep{Chrimes2024b-fhn}. Out of these seven, only three are detected in the millimeter bands: AT\,2018cow\citep{Ho2019-cow}, AT\,2020xnd \citep{Ho2022-xnd,Bright2021-xnd}, and AT\,2022tsd \citep{Ho2023-at2022tsd,Matthews23}. In Figure \ref{Fig:10GHz-mm-lightcurves-FBOTs}, we show the 8--10 GHz light curves of all detected LFBOTs along with other transients and a compilation of light curves of all millimeter transients with the four millimeter-bright LFBOTs. 

One of the most striking features is the similarity in rise timescales and the rapid flux decay across all radio-bright FBOTs in the radio bands, as highlighted in Fig \ref{Fig:10GHz-mm-lightcurves-FBOTs} (left panel). This is also reflected in Fig \ref{Fig:Lp-nuptp diagram}, where FBOTs occupy a distinct and compact region of the parameter space. Instead, the mm emission from even a very small sample of LFBOTs shows an impressive range of behaviors and luminosities (for example, there is a factor $>100$ difference between the mm luminosity of AT\,2022tsd and \at{}\, at $\approx$\,20\,days). This phenomenology likely primarily reflects the diversity of the innermost medium around LFBOTs vs.\ the more ``universal'' CSM at $>10^{16}\,\rm{cm}$, in addition to possible differences in their central engines.

Broadly speaking, the properties of the radio emission from \at{} are in line with those of other LFBOTs. Specifically, \at{}  shows the steep rise and very steep decay that are  hallmark observational features of LFBOTs compared to other transients at cm wavelengths (Fig. \ref{Fig:10GHz-mm-lightcurves-FBOTs}, left).
The peak radio spectral luminosity of \at{} is $L_{\rm 9 GHz} \approx 1.7 \times 10^{29}\, \rm erg\,s^{-1}\,Hz^{-1}$ at $\delta t \approx 73$\,days, similar to that of other LFBOTs (see Figure \ref{Fig:10GHz-mm-lightcurves-FBOTs}, left panel) with a steep post-peak temporal decay index of $F_{\rm 9\,GHz} \propto t^{-2.05 \pm 0.54}$. 

However, the SED evolution of \at{} is unique and distinct compared to that of other LFBOTs (i) at early times $\delta t_{\rm rest} <32$\,days; (ii) at $\delta t_{\rm rest} \approx 133$\,days when we observe a spectral radio inversion. The evolution of millimeter band flux densities during $\delta t_{\rm rest} \approx 17-32$\,days is unprecedented, with $\nu_{\rm pk} \propto t^{-2.01\pm0.65}$ and $F_{\rm pk} \propto t^{4.04 \pm 0.42}$. The closest analog is AT\,2018cow, though the mm rise is not as extreme (Fig \ref{Fig:10GHz-mm-lightcurves-FBOTs}, right panel). In both events, the rapid mm brightening  can be attributed to the shock propagating through a dense and radially-confined medium, and efficiently converting the ejecta kinetic energy into thermal energy \citep{Ho2019-cow,Khatami2024}.

The average blast-wave velocity of \at{} increases from $\Gamma \beta c \approx 0.07c$ to $0.42c$ between $\delta t_{\rm rest} \approx 32$ and 73\,days and then decreases to $0.2c$ by $\delta t_{\rm rest} \approx 118$\,days. This acceleration is consistent with the shock breaking out of the dense CSM shell and entering a lower-density medium. While not as evident as in \at{}, AT\,2018cow also shows indication of an accelerating blast wave with velocities going from $v \approx 0.1c$ at early times \citep[$\delta t \approx 22$\,days;][]{Ho2019-cow} to $v \approx 0.2c$ at later times \citep[see Fig 13 of][]{Margutti2019-cow,Nayana2021}. Radio spectral information on AT\,2018lug is limited; CSS161010 shows evidence for a decelerating outflow between $\delta t \approx 69-357$\,days and doesn't have observations at early times. 

For \at{}, the SEDs evolve as $\nu_{\rm pk} \propto t^{-2.99\pm0.10}$ and $F_{\rm pk} \propto t^{-0.67\pm0.06}$ during $\delta t_{\rm rest} \approx 32-73$\,days. At later times ($\delta t_{\rm rest} > 118$\,days), the SED evolution is characterized by $\nu_{\rm pk} \propto t^{-0.60\pm1.05}$ and $F_{\rm pk} \propto t^{-2.05\pm0.54}$. In the case of AT\,2018cow, the SED evolution at $\delta t >80$\,days followed $\nu_{\rm pk} \propto t^{-2.2 \pm 0.1}$ and $F_{\rm pk} \propto t^{-1.7 \pm 0.1}$ \citep{Ho2019-cow,Margutti2019-cow}. CSS161010 showed an evolution of $\nu_{\rm pk} \propto t^{-1.26 \pm 0.07}$ and $F_{\rm pk} \propto t^{-1.79\pm0.09}$ at $\delta t >99$\,days \citep{coppejans2020}. 
In terms of  $\nu_{\rm pk}(t)$ and $F_{\rm pk}(t)$, the SED evolution of \at{} at $\delta t \gtrsim 32$\,days closely resembles that of other radio-bright FBOTs. The spectral peak frequency ($\nu_{\rm pk}$) cascading to lower values over time is broadly consistent with expectations of an expanding shock wave. However, the $\nu_{\rm pk}(t)$ and  $F_{\rm pk}(t)$ of LFBOTs is significantly different from that seen in typical CCSNe interacting with a wind-like CSM (where we expect $\nu_{\rm pk} \propto t^{-1}$ and $F_{\rm pk} \sim$ constant). The relatively faster evolution of $\nu_{\rm pk}$ and $F_{\rm pk}$ seen in LFBOTs is indicative of a steeper CSM density profile than that of a canonical wind-like profile ($\rho_{\rm CSM}(r) \propto r^{-2}$). 

LFBOTs are also clearly distinct from ordinary CCSNe in terms of shock velocity. We plot \at{} along with other LFBOTs and SNe in the velocity-energy phase space in Figure \ref{Fig:shock-energy-FBOTs}. 
\at{} has an inferred outflow velocity of $\Gamma \beta c \approx 0.4c$ at $\delta t \approx 73$\,days, and belongs to the class of LFBOTs that show mildly relativistic outflows similar to CSS161010 \citep{coppejans2020}, AT\,2018lug \citep{Ho2020-AT2018lug}, and AT\,2023fhn \citep{Chrimes2024b-fhn}. The outflow velocity of AT\,2018lug is $\Gamma \beta c \geq 0.3c$ at $\delta t \approx 100$\,days and of CSS161010 is $\Gamma \beta c \geq 0.55c$ at a similar epoch. These outflow velocities are higher compared to the non-relativistic velocities seen in AT\,2018cow ($v \sim 0.1c$) \citep{Ho2019-cow,Margutti2019-cow}. At $\delta t \approx 73$\,days, the kinetic energy coupled to the fast-moving radio-emitting shock of velocity $\Gamma \beta  \approx 0.4c$ is $E_{\rm k} \approx 3.3 \times 10^{49}\, \rm erg$. 
For a standard spherical hydrodynamical collapse of a star, this would imply  $E_{\rm k} > 10^{55}$\,erg coupled with the slow-moving material at $v \approx 10,000$\,km\,s$^{-1}$,  where $E_{\rm k} \propto (\Gamma \beta)^{-5.2}$ for a polytropic index of 3 \citep{Tan2001}. This energy largely exceeds the limit ($E_{\rm k} \approx 10^{51}$\,erg) of typical neutrino-driven stellar explosions (and challenges most stellar explosion models) and argues against a spherical stellar collapse as the astrophysical origin of \at{}.

\begin{figure*}[t!]
 	\centering	\includegraphics[width=0.85\textwidth]{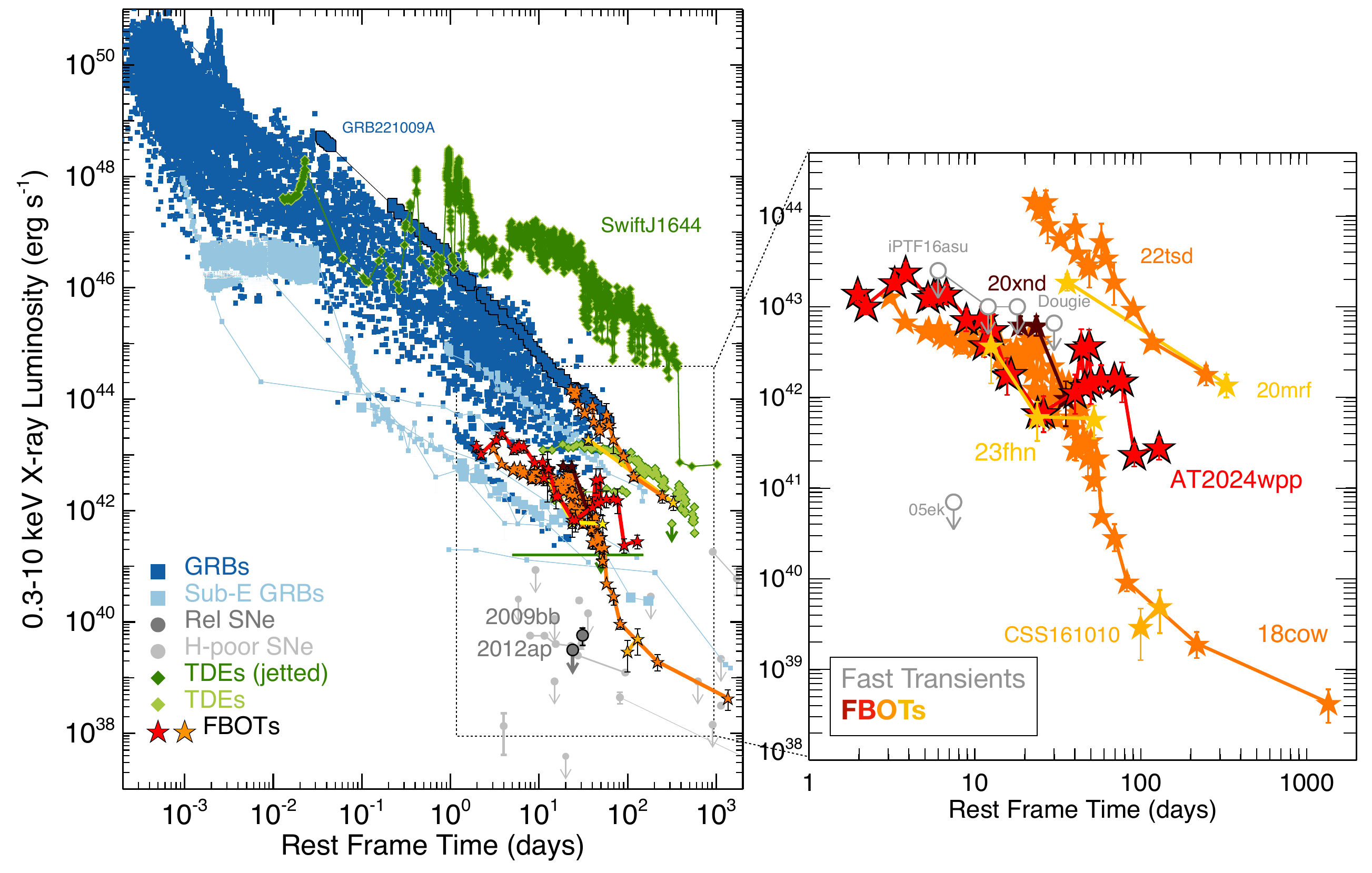} 
  \caption{\emph{Left Panel:} Soft X-ray luminosity evolution of \at{}\, in the context of  explosive transients capable of launching relativistic ejecta (long GRBs, TDEs, H-poor SNe). FBOTs span the entire dynamical range of X-ray luminosity observed for long GRBs to date.
  \emph{Right Panel:} Zoom-in of the region of luminous FBOTs with X-ray emission (stars) and fast transients (empty gray circles). \at{}\, is only the second FBOT for which we were able to sample the X-ray emission at $\delta t<10$\,days. While the initial $L_x\approx $ a few $10^{43}\,\rm{erg\,s^{-1}}$ is not dissimilar from ATs 2018cow, 2020xnd, and 2023fhn, \at{}\, stands out at later times $\delta t \gtrsim 30$\,days for its longer-lived and spectrally harder X-ray emission. Similar to AT\,2018cow is the remarkably fast late-time decay. We add to this panel other fast transients observed in the X-rays: of these the very luminous and fast-evolving optical emission from ``Dougie'' bears close similarities with luminous FBOTs.  References:  \cite{Margutti13,Margutti2019-cow,coppejans2020,Bright2021-xnd,Yao2021-mrf,Matthews23,Ho2023-at2022tsd,Chrimes2024a-fhn,Migliori24} and references there in. }
 \label{Fig:XrayFBOTcomparison}
 \end{figure*}

Finally, a unique radio aspect of \at{} is the evidence for radio spectral inversion at $\delta t_{\rm rest} \approx 133$ and 161\,days (see Fig \ref{Fig:radioSED-fit-single-epoch}). The spectral behavior indicates that $\nu_{\rm pk} \gtrsim 9$\,GHz, implying $R < 6 \times 10^{16}$\,cm at these epochs, which is smaller than the shock radius at $\delta t_{\rm rest} \approx 118$\,days. This could be indicative of another outflow, possibly associated with slow disk winds and/or due to a complex CSM density profile.

To summarize, even though the sample is limited, LFBOTs present a combination of radio properties that make them clearly distinct from other transients (like the rapid rise and decay of their radio light-curves), while at the same time showing significant diversity in terms of spectral evolution and outflow velocities. The properties of \at{} are particularly striking due to the accelerating outflow with increasing shock internal energy. We present a detailed comparison of the inferred CSM densities of LFBOTs and their astrophysical implication in \S \ref{subsec:csm-density-profile}.

\subsection{LFBOTs in the X-rays}
\label{SubSec:Xraycomparison}

Only LFBOTs have been detected in the X-rays (Fig. \ref{Fig:XrayFBOTcomparison}), showing very luminous displays $L_x>10^{43}\,\rm{erg\,s^{-1}}$ 
in line with those of long GRBs: the FBOT AT\,2022tsd  \citep{Matthews23,Ho2023-at2022tsd} even rivals GRB\,221009A, the brightest GRB detected so far. We note that our independent spectral extraction and re-analysis of the FBOT AT\,2023fhn does not confirm the claim of sub-luminous X-ray emission by \cite{Chrimes2024a-fhn} and points instead to a harder spectrum (and hence more luminous emission) than what was assumed by those authors, in line with other FBOTs. While LFBOTs share with GRBs\footnote{TDEs are also known to show rapid X-ray variability, while no known SN displayed rapid soft X-ray variability weeks after explosion (e.g., \citealt{Dwarkadas25}) for a recent review. Rapid X-ray variability is a hallmark feature of engine-driven transients.} rapid X-ray variability time scales ($\Delta t /t< 1$) and the non-thermal nature of their X-ray emission, their spectral properties are markedly different: GRB X-ray afterglow emission is typically consistent with a $F_{\nu}\propto \nu^{-1}$ spectrum (e.g., \citealt{Margutti13}), while LFBOTs have harder spectra even before the emergence of the Compton hump for years after the FBOT \citep{Migliori24}.

Among FBOTs, only \at{}\, and AT\,2018cow have shown clear evidence for a Compton hump of emission. However, it is interesting to note that the hard 0.3--10 keV spectrum $F_{\nu}\propto \nu^0$ of the very X-ray luminous AT\,2020mrf at $\approx$330\,d, compared to its significantly softer spectrum $F_{\nu}\propto \nu^{-0.8}$ at early times $\approx 36$\,days \citep{Yao2021-mrf},  is suggestive of a similar phenomenology and evolution as in \at{}\, and AT\,2018cow, albeit on a significantly longer timescale. At the time of writing there are only seven other LFBOTs with published X-ray light-curves: at $\delta t\approx 20$\,days when most of the sample has observations, the sample covers a $\approx10^2$ dynamic range of luminosities with two groups of LFBOTs: ``18cow-like'' FBOTs with a plateau+steep decay light-curve morphology; and ``22tsd-like'' FBOTs displaying the most luminous X-ray emission. Given the likely geometrically beamed nature of the emission (\S\ref{SubSec:softXrayindex}), we consider it possible that the observed diversity is in part due to viewing angle effects, with pole-on views being associated with more luminous displays.
 \begin{figure*} 
 	\centering   
        \includegraphics[width=0.95\textwidth]{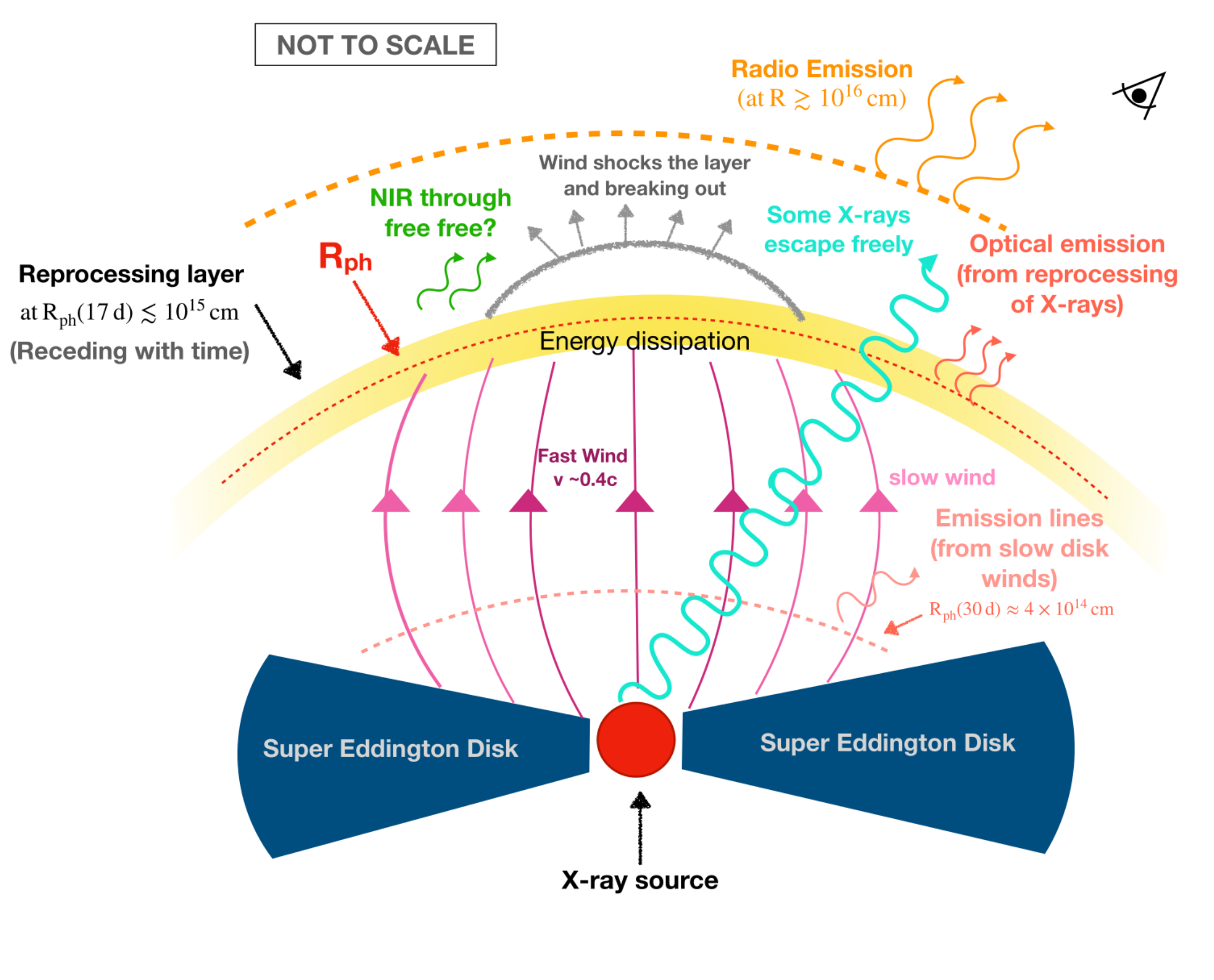}   \caption{A cartoon diagram (not to scale) showing the geometry of \at{} and various emission components in the context of an engine-driven progenitor model. The physical picture is motivated by the models presented in \cite{Tsuna25} and \cite{Metzger22FBOT}. Both these models invoke super-Eddington accretion onto a compact object (NS or BH) in different astrophysical contexts. The disk-wind outflows from the accretion disk generate $^{56}\rm Ni$-poor asymmetric ejecta with a range of velocities$-$fast outflow ($v \approx 0.4c$) from the interior of the accretion disk and slow winds from the outer radii of the disk. In this framework, the X-ray emission originates from the central engine, while the optical emission arises from reprocessing of X-rays. Here, we are agnostic about the reprocessing layer whereas \cite{Tsuna25} assume it to be an SN ejecta and \cite{Metzger22FBOT} assumes it to be the fast disk-wind ejecta. The low-velocity emission lines are from the slow disk-winds from the outer radii of the accretion disk. Radio emission arises due to the interaction between fastest disk-wind and CSM at $R \gtrsim 10^{16}$\,cm. }
 \label{Fig:cartoon-geometry of AT2024wpp}
 \end{figure*}

\section{Plausible Physical Models}
\label{Sec:physical models}
Various progenitor models have been proposed to explain the observed properties of FBOTs. Here, we explore different physical models that could explain the X-ray and radio properties of \at{}. Based on the observations presented in this work, the progenitor system of \at{} should be able to produce: (1) luminous and variable X-ray emission with a non-thermal spectrum and a transient Compton hump appearing at $\delta t \approx 50$\,days.  (2) Shock carrying $E_{\rm k} \approx 3.3 \times 10^{49}\, \rm erg$ with velocities as fast as $v \approx 0.4$c at $\delta t_{\rm rest} \approx 73$\,days (3) A dense CSM of $n \approx 0.3 \times 10^{6}\,\rm cm^{-3}$ at a distance of $R \approx 0.6 \times 10^{16}$\,cm from the explosion center and significantly larger densities inwards. (4) A radial density profile of $\rho_{\rm CSM} \propto r^{-3.1}$ extending up to $R \approx 10^{17}\, \rm cm$.   

Progenitor models that exclusively rely  on CSM interaction \citep{Fox2019,Leung2021,Pellegrino2022} cannot explain the presence of variable, non-thermal X-ray emission  and do not provide a natural explanation for the mildly-relativistic outflows that characterize LFBOTs: hence, CSM-interaction-only models can be ruled out. The relativistic outflows imply the presence of compact objects. Indeed, a  class of viable LFBOTs models involves the presence of a central engine: a failed SN that produces an accreting BH and small ejecta mass ejected via accretion disc winds \citep{Quataert2019,Antoni2022}, a successful CCSN from a rotating massive star that gives birth to a BH or NS \citep{Margutti2019-cow,Perley2019,Gottlieb2022}, pulsational pair instability SNe \citep{Leung2020}, tidal disruption of a star by intermediate mass BH or stellar mass BH \citep{Perley2019,Kuin2019,Kremer2021,Gutierrez2024}, merger-induced tidal disruption and hyperaccretion of a WR star by  NS or BH \citep{Metzger22FBOT}, collision of a newly-born NS or BH from a core-collapse explosion with a companion star leading to tidal disruption and hyper-accretion \citep{Tsuna25}. Although these models can drive non-relativistic to mildly relativistic outflows since they invoke an accretion disk around a BH/NS in some regime of hyper accretion \citep{Sadowski15,Sadowski16}, reproducing the quasi-universal density profile of Fig. \ref{Fig:density-FBOTs} is more challenging. We expand on this aspect in the next section.

\subsection{A universal CSM density profile in LFBOTs}
\label{subsec:csm-density-profile}
The density profile of the environment of \at{} is $\rho_{\rm CSM}(r) \propto r^{-3.10\pm0.16}$, shown in Fig. \ref{Fig:density-FBOTs} at $R \gtrsim 6 \times 10^{15}$\,cm along with that of other LFBOTs. The environmental densities of LFBOTs are strikingly similar, with $n \approx 10^{6}\, \rm cm^{-3}$ at $R \approx 10^{16}\, \rm cm$, and an approximate profile $\rho_{\rm CSM} \propto r^{-3}$ at distances $\approx$ a few times $10^{15}$ to $10^{17}\, \rm cm$. The innermost density profile at $<3\times 10^{15}\,\rm{cm}$ has only been sampled with radio observations for AT\,2018cow \citep{Ho2019-cow,Margutti2019-cow} and \at{}. In both cases, there is some evidence for a flatter inner density profile (see the first four points in orange in Fig.\ \ref{Fig:density-FBOTs} for AT\,2018cow). For \at{} the evidence comes from two angles: first, the rapid mm-band rise of \S\ref{SubSec:shock-breakout from dense CSM shell}; second, the NIR excess observed at $\delta t \approx 30\, \rm days$ being consistent with free-free opacity effects occurring in an extended medium of shallow density profile of $\rho \propto r^{-1.3}$ above the optical photosphere at $R_{\rm BB} \approx 4 \times 10^{14}$\,cm (see Sec 6.2 of Paper I).

Based on these observations, we find that the LFBOT environment likely consists of a high-density component with a flat profile up to distances of a few $10^{15}\,\rm{cm}$ to $10^{16}\,\rm{cm}$, and a steep density profile $\rho_{\rm{CSM}} \propto r^{-3}$ outwards. These broadly similar LFBOTs CSM density profiles are likely manifestations of similar stellar evolution processes. We discuss the astrophysical implications below. 

While clearly not consistent with  wind-like density profiles $\rho\propto r^{-2}$ (which are the result of constant mass loss to wind velocity ratio), the environmental densities of LFBOTs correspond to \emph{effective} mass-loss rates up to $\dot{M} \approx 10^{-3}\,M_{\odot}\,\rm yr^{-1}$ for an arbitrary wind velocity of $v_{\rm w} = 1000\,\rm km\,s^{-1}$ (Fig. \ref{Fig:density-FBOTs}). These effective $\dot M$ are significantly higher than those of H-stripped CCSNe (e.g., \citealt{chevalier2006}) and long Gamma Ray Bursts (GRBs, e.g., \citealt{Gompertz2018}).\footnote{We limit our comparison to H-poor stellar explosions as the LFBOT ejecta is H depleted (see Paper\,I)} Such dense environments can be formed during the brief evolutionary phases of intense mass loss from some massive stars \citep{Smith14}. 
However, the presence of dense material at such close distances from the LFBOTs requires some form of timing between the mass ejection event(s) and the onset of the LFBOT emission. Some LFBOT models struggle to provide a natural explanation for this timing (and hence explain the quasi-universal CSM density profile of LFBOTs at $\gtrsim10^{16}\,\rm{cm}$). Among these is the Pulsation Pair Instability SN model (PPISN, \citealt{Woosley2017}), for which the CSM density is set by previous PPI events \citep{Renzo2020,Leung2021}. Even for (single-star) models that involve the presence of a BH/NS and can in principle create large densities at $R\lesssim 10^{16}\,\rm cm$, (like the failed explosion of a single massive star ejecting a small amount of ejecta via disk winds, or the successful core-collapse of a massive star forming a NS or BH) some fine tuning is required between the mass loss and progenitor's evolutionary phase to reproduce observations.

More natural explanations of the universal CSM-density profile observed in LFBOTs are offered by models like a (i) merger-initiated tidal disruption and hyper-accretion of a WR star onto a NS or BH binary companion \citep{Metzger22FBOT}; (ii) the collision of a newly-born NS or BH from a core-collapse explosion with a companion star \citep{Tsuna25}. The appeal of these models is that they can reproduce other key observational LFBOT properties like the asymmetric ejecta, variable non-thermal X-ray emission (potentially with a Compton hump), and mildly relativistic outflow velocities (this is fundamentally because both models involve super-Eddington accretion on a compact object). 

In the first model (i) by \cite{Metzger22FBOT}, dense and confined CSM is established by the pre-merger WR star mass loss \citep{Pejcha2016a,Pejcha2016b,Pejcha2017,MacLeod2017,MacLeod2020} on radial scales $\approx 10^{14}-10^{15}\, \rm cm$ \citep{Matsumoto2022} which will be similar in all WR/BH-NS mergers. The extended CSM ($R \gtrsim 10^{15}\, \rm cm$) results from circumbinary disk outflows \citep{Keto2007,Hollenbach1994}, and the circumbinary relic disk from the common envelope phase \citep{Kashi2011}. The properties of this extended CSM can depend on the binary parameters and mass transfer. Probing the extended CSM of LFBOTs at radial scales $R \gg 10^{18}\, \rm cm$ in the future might help reveal their diversity in this regard. In the model proposed by \cite{Tsuna25}, the observational characteristics of FBOTs are best reproduced in a scenario where the accreting NS/BH is formed from the explosion of low mass ($\lesssim 3 M_{\odot}$) helium star. Mass transfer at a rate of $\approx 10^{-4}\, \rm M_{\odot}\,yr^{-1}$ onto the companion can occur from the helium star at 0.1$-$1 kyr before the core collapse \citep{Tauris2015,Wu2022,Ercolino2025}. For typical equatorial mass-loss speeds ($10-100\, \rm km\,s^{-1}$) from the binary, this translates to distances $R \gtrsim 10^{15}-10^{17}\,\rm cm$. In the decades before the explosion, extreme mass-loss rates $\gtrsim 10^{-2}\, \rm M_{\odot}\,yr^{-1}$ can occur as a result of the rapid expansion of the outer layer of the helium star \citep{Wu2022}. This will lead to dense CSM at radius $R \lesssim 10^{15}\, \rm cm$. 

To conclude this section, at the time of writing, the most promising LFBOTs models are those that involve binary systems, where outflows launched by a hyper-accreting compact object interact with an environment that was shaped by the previous evolution of the binary. 

\subsection{The geometry of \at{} and various emission components}
\label{Sec:geometry and emission from wpp}
Figure \ref{Fig:cartoon-geometry of AT2024wpp} represents a schematic illustration (not to scale) of the geometry of \at{} and various emission components in the context of a progenitor model involving a central engine. This physical picture is motivated by the models presented in  \cite{Metzger22FBOT} and \cite{Tsuna25}. While these models are different in their astrophysical context, both of them invoke super-Eddington accretion onto a compact object (NS or BH).  
The disk-wind outflows powered by the release of gravitational energy \citep{Narayan1995,Blandford1999,Kitaki2021} generate the $^{56}\rm Ni$-poor aspherical ejecta in a velocity range observed in FBOTs. The outflow speed of disk winds from the outer radii of the accretion disc is low with mean velocities $\approx 3000-4000\, \rm km\,s^{-1}$ \citep{Margalit2016,Metzger22FBOT}, which can explain the low-velocity emission lines seen in \at{} (Paper I). The disk wind outflow velocities from the interior part of the accretion disk are much higher; results from GRMHD simulations of super-Eddington accretion disks indicate trans-relativistic outflow velocities \citep{Sadowski15,Sadowski16}. 

X-ray emission is from the central engine. Depending on the viewing angle and X-ray covering fraction, some X-rays can escape early on and be detected. X-ray emission from \at{} is detected from $\delta t \approx 2$\,days. The transient Compton hump seen in the X-rays naturally fits in this scenario as the X-ray source is embedded in an expanding ejecta with time-variable optical depth to Compton scattering. Optical emission is powered by the reprocessing of X-ray emission, where the origin and nature of the reprocessing layer are different in these two models. In the \cite{Tsuna25} model, the collision between a newly born NS/BH and the companion star leads to tidal disruption and super-Eddington accretion. The fast disk winds collide with the SN ejecta, effectively converting wind kinetic energy to radiation, resulting in a luminous optical transient with peak luminosity $L_{\rm pk} \approx 10^{44}\, \rm erg\,s^{-1}$. In the \cite{Metzger22FBOT} model, a WR in a binary system is tidally disrupted and accreted onto the NS/BH. The optical emission is due to a combination of reprocessing of X-rays from the inner accretion disk/jets by fast disk wind ejecta and shock interaction between the disk wind outflow and premerger CSM of the WR star. Based on our multi-wavelength observations and analysis of \at{}  (paper I and this work), we remain agnostic about the origin and nature of the reprocessing layer and hypothesize that the optical radiation is reprocessed X-rays from the central engine. From the NIR analysis, the mass of the reprocessing layer is $M \approx 2M_{\odot}$ (paper I) and is likely to be pre-existing material. However, we emphasize that the pre-existing material need not be the only medium that reprocesses the X-rays, as the fast disk winds could also contribute to the process \citep{Metzger22FBOT}. In any case, it is less likely that the reprocessing layer is SN ejecta, as SN Ib/c ejecta would develop some CNO lines in the optical spectra at later times, which we do not see in the case of \at{}, unless the nondetection is due to the low CNO mass fraction in low-mass helium stars \citep{Dessart2020,Dessart2021}. Radio emission arises as a result of shock interaction between the fastest disk wind outflow with the CSM at $r \approx 10^{16}-10^{18}$\,cm. The blast-wave velocities inferred from radio SED modeling of \at{} go up to $\Gamma \beta c \approx 0.42c$, in line with the prediction of mildly-relativistic disk wind outflows from super-Eddington disks from simulations \citep{Sadowski15,Sadowski16}. This physical picture is consistent with the results presented in \cite{Pursiainen25} as at early times the optical photosphere is inside the fast-moving outflow. 

The high disk-wind velocities can be used to put rough constraints on the mass and size of the central compact source. For typical mass ($M \approx 1.1M_{\odot}$) and size ($R \approx 12$\,km) of a NS, the escape velocity will be $v_{\rm esc} \approx 0.5c$. Thus, the observed outflow velocities can be barely achieved in the disk winds from an NS, whereas they can be easily achieved from a BH \citep{Sadowski15,Sadowski16}. Radiation hydrodynamic simulations of supercritical accretion onto NS predicts outflow velocities $v \approx 0.2-0.3c$ \citep{Ohsuga2007} while GRMHD simulations predict velocities $v \approx 0.4c$ \citep{Takahashi2017}. The large energy budget of \at{} ($E_{\rm rad}=10^{51}\, \rm erg$ from Paper I) also favors a BH over NS  \cite[see \S 2.4 of][]{Tsuna25}.  

\section{Summary and Conclusions} 
\label{Sec:conclusions}
We present extensive X-ray (0.3--79 keV) and radio (0.25--203\,GHz) observations of the FBOT \at{} spanning $\delta t \approx 2-280$\,days after first light. Major findings from the combined X-ray and radio analysis are the following:
\begin{enumerate}
    \item \at{} shows luminous and variable X-ray emission, being only the third FBOT with a hard X-ray detection. The X-ray luminosity ($L_{\rm x} \approx 1.5 \times 10^{43}\, \rm erg\,s^{-1}$) remains roughly constant in the first 7\,days, and then decays with an index of $L_{\rm x} \propto t^{-2.5\pm0.25}$, followed by a re-brightening (flaring) starting from $\delta t \approx 35$\,days that peaks at $\delta t \approx 50$\,days (see Fig \ref{Fig:SoftXrays}). 
    \item The X-ray spectra are initially soft ($F_{\nu} \propto \nu^{-0.8}$) and gradually transition to a harder state over time with an extremely hard spectrum ($F_{\nu} \propto \nu^{1.25}$) at the peak of the re-brightening ($\delta t \approx 50$\,days). The spectrum becomes soft ($F_{\nu} \propto \nu^{-0.6}$) again at $\delta t \approx 75$\,days after the flare peak  (Figs. \ref{Fig:SoftXrays}, \ref{Fig:SpecIndex}, and \ref{Fig:hump}).
    \item The X-ray emission from \at{} shows clear evidence for a transient Compton hump at $\delta t \gtrsim 50$\,days, similar to that of AT\,2018cow in which the Compton hump was present at much earlier times ($\delta t \approx 8$\,days). Compton humps are unprecedented in the field of stellar explosions, but are commonly observed in accretion-powered systems like AGNs and XRBs.
    \item The spectral and temporal evolution of X-ray emission from \at{} favors the presence of a high-energy source embedded inside expanding aspherical ejecta, similar to the picture invoked to explain AT\,2018cow \citep{Margutti2019-cow}. The delayed appearance of the Compton hump in \at{} relative to AT\,2018cow can be attributed to a variety of effects, including time-dependent ionization of the ejecta and larger ejecta mass.
    \item \at{} displays luminous radio emission with a peak spectral luminosity of $L_{\rm 9 GHz} \approx 1.7 \times 10^{29}\, \rm erg\,s^{-1}\,Hz^{-1}$ at $\delta t_{\rm rest} \approx 73$\,days, significantly larger than SNe and comparable to other LFBOTs (see Fig \ref{Fig:10GHz-mm-lightcurves-FBOTs}). Radio emission is also detected in the millimeter bands (97.5 and 203\,GHz) at $\delta t_{\rm rest} \approx 17\, \rm and \,32$\,days, marking \at{} as the fourth millimeter-bright FBOT.
    \item The radio spectral evolution is unprecedented with an extremely rapid rise in the millimeter flux densities at early times $\delta t_{\rm rest} \approx 17-32$\,days. Subsequently, the spectral peak flux slowly declines with peak frequency cascading to lower bands as expected in a shock-driven synchrotron emission model. At a later time ($\delta t_{\rm rest} > 118$\,days), we find first evidence for a spectral inversion, possibly indicating the emergence of a new emission component (see Fig \ref{Fig:radioSED-fit-single-epoch}).        
    \item The shock velocities from radio SED modeling indicate an accelerating outflow with velocities evolving from $\Gamma \beta c \approx 0.07c$ to $0.42c$ during $\delta t_{\rm rest} \approx 32-73$\,days with an increasing amount of energy ($U \approx 0.8-33 \times 10^{48}$\,erg) coupled to this outflow (see Table \ref{Tab:radio-shock-parameters}).
    \item We interpret the radio emission from \at{} in a scenario in which the radio-emitting shock is propagating through a dense CSM shell of outer radius $\approx 10^{16}$\,cm. The shock then accelerates at the edge of this shell through a medium of density profile $\rho_{\rm CSM}(r) \propto r^{-3}$ ($\dot{M} \approx 10^{-3}\,\rm M_{\odot}\,yr^{-1}$ for $v_{\rm w} = 1000\, \rm km\,s^{-1}$). 
    \item We compile the CSM densities of all radio-bright FBOTs from the literature and note that the environmental densities are strikingly similar with $n \approx 10^{6}\, \rm cm^{-3}$ at $R \approx 10^{16}$\,cm with an approximate profile of $\rho_{\rm CSM}(r) \propto r^{-3}$ over $R \approx 10^{16}-10^{18}$\,cm (Fig \ref{Fig:density-FBOTs}). This indicates that similar evolutionary processes and mass-loss mechanisms of the progenitor system are setting up these environments.
    \item Our extensive X-ray and radio monitoring of \at{} and combined inferences from these observations favor a progenitor model that involves super-Eddington accretion onto a compact object capable of producing disk-wind outflows of velocities up to $\sim 0.4c$. 
\end{enumerate}

FBOTs remain one of the least understood classes of transients, with detailed multiwavelength data available only for a handful of events. Future wide-field time-domain surveys and rapid-response follow-up capabilities will be key to expanding this sample. Upcoming missions such as UVEX \citep{Kulkarni2021-UVEX} and ULTRASAT \citep{Shvartzvald2024-ULTRASAT} will enable the prompt discovery and early characterization of many more FBOTs. Coupled with coordinated multi-wavelength campaigns, this will allow us to systematically probe the diversity of FBOT progenitors, their environments, and central engines.
\vspace{5mm}

\facilities{Swift(XRT and UVOT), AAVSO, CTIO:1.3m,
CTIO:1.5m, CXO}
\software{astropy \citep{2013A&A...558A..33A,2018AJ....156..123A},  
          }
\section*{Acknowledgments}
This paper makes use of the following ALMA data: $ADS/JAO.ALMA\#2024.A.00003.T$ and $ADS/JAO.ALMA\#2024.A.00009.T$. ALMA is a partnership of ESO (representing its member states), NSF (USA) and NINS (Japan), together with NRC (Canada), MOST and ASIAA (Taiwan), and KASI (Republic of Korea), in cooperation with the Republic of Chile. The Joint ALMA Observatory is operated by ESO, AUI/NRAO and NAOJ.
The Allen Telescope Array refurbishment program and its ongoing operations are being substantially funded through the Franklin Antonio Bequest. Additional contributions from Frank Levinson, Greg Papadopoulos, the Breakthrough Listen Initiative and other private donors have been instrumental in the renewal of the ATA. Breakthrough Listen is managed by the Breakthrough Initiatives, sponsored by the Breakthrough Prize Foundation. The Paul G. Allen Family Foundation provided major support for the design and construction of the ATA, alongside contributions from Nathan Myhrvold, Xilinx Corporation, Sun Microsystems, and other private donors. The ATA has also been supported by contributions from the US Naval Observatory and the US National Science Foundation.
We thank the GMRT staff for making these observations possible. The GMRT is run by the National Centre for Radio Astrophysics of the Tata Institute of Fundamental Research.
This research has made use of the NuSTAR Data Analysis Software (NuSTARDAS) jointly developed by the ASI Space Science Data Center (SSDC, Italy) and the California Institute of Technology (Caltech, USA). 
This research has made use of data obtained 
from the Chandra Data Archive provided by the Chandra X-ray Center (CXC).
This research has made use of the XRT Data Analysis Software (XRTDAS) developed under the responsibility
of the ASI Science Data Center (ASDC), Italy.
This work made use of data supplied by the UK Swift Science Data Centre at the University of Leicester.
This research has made use of data and software provided by the High Energy Astrophysics Science Archive Research Center (HEASARC), which is a service of the Astrophysics Science Division at NASA/GSFC.
\par R.M. acknowledges support by the National Science
Foundation under award No. AST-2224255, and by NASA under grants 80NSSC22K1587, 80NSSC25K7591 and 80NSSC22K0898. 
G.M. acknowledges financial support from the INAF mini-grant "The high-energy view of jets and transient" (Bando Ricerca Fondamentale INAF 2022).
B.D.M. acknowledges support from NASA AAG (grant number 80NSSC22K0807), the Fermi Guest Investigator Program (grant number 80NSSC24K0408) and the Simons Foundation (grant number 727700). The Flatiron Institute is supported by the Simons Foundation.
D.T. is supported by the Sherman Fairchild Postdoctoral Fellowship at Caltech. 
FDC acknowledges support from the DGAPA/PAPIIT grant IN113424. 
DLC acknowledges support from the Science and Technology Facilities Council (STFC) grant ST/X001121/1. 
N.L. thanks the LSST-DA Data Science Fellowship Program, which is funded by LSST-DA, the Brinson Foundation, the WoodNext Foundation, and the Research Corporation for Science Advancement Foundation; her participation in the program has benefited this work. 
CTC and KDA gratefully acknowledge support from NSF under grant AST-2307668 and from the Alfred P. Sloan Foundation. 

\appendix
Tables \ref{Tab:radio-ALMA}, \ref{tab:radio-atca}, \ref{Tab:radio-ATA}, \ref{Tab:radio-MeerKAT}, and \ref{Tab:radio-gmrt} show radio observation logs and flux measurements of AT\,2024wpp with ALMA, ATCA, ATA, MeerKAT, and GMRT, respectively. Table \ref{Tab:Xraylog} shows the details of X-ray observations of AT\,2024wpp with XMM-Newton, CXO, and NuSTAR.

\section{Radio Data Table}\label{AppendixRadio}
\startlongtable
\begin{deluxetable*}{ccccccccc}
\tablecaption{ALMA observations of AT\,2024wpp}
\tablehead{
\colhead{Start Date} & \colhead{Project ID}  & \colhead{Centroid MJD} & \colhead{Phase$^{\rm{a}}$} & \colhead{Frequency}   & \colhead{Flux Density$^{\rm{b}}$} \\
 (dd/mm/yyyy) &  &  & (d) & (GHz) & (mJy) &
 }
\startdata
14/10/2024 & 2024.A.00003.T  & 60597.22 & 18.92 & 97.5  & 0.076$\pm$0.019  \\ 
14/10/2024 & 2024.A.00003.T  & 60597.19 & 18.89 & 203.0 & 0.100$\pm$0.024  \\
31/10/2024 & 2024.A.00009.T  & 60614.06 & 35.76 & 97.5  & 1.282$\pm$0.015  \\
31/10/2024 & 2024.A.00009.T  & 60614.09 & 35.79 & 203.0 & 0.588$\pm$0.031  \\
\enddata
\tablecomments{$^{\rm{a}}$ With respect to first light. $^{\rm{b}}$ The uncertainties on flux measurements are 1$\sigma$.
\label{Tab:radio-ALMA}}
\end{deluxetable*}

\startlongtable
\begin{deluxetable*}{ccccccccc}
\tablecaption{ATCA observations of AT\,2024wpp}
\tablehead{
\colhead{Start Date} & \colhead{Project ID}  & \colhead{Centroid MJD} & \colhead{Phase$^{\rm{a}}$} & \colhead{Frequency} & \colhead{Bandwidth} & \colhead{Flux Density$^{\rm{b}}$} \\
 (dd/mm/yyyy) &    &  & (d) & (GHz) & (GHz) & ($\mu$Jy) &
 }
\startdata
10/10/2024 & C3419 &  60593.00 & 14.70 &   16.7   & 2.0 &  $<$ 108  \\
10/10/2024 & C3419 &  60593.00 & 14.70 &   21.2   & 2.0 &  $<$ 108 \\
10/10/2024 & C3419 &  60593.00 & 14.70  &   43.0   & 2.0 &  $<$ 216 \\
10/10/2024 & C3419 &  60593.00 & 14.70  &   45.0   & 2.0 &  $<$ 216 \\
14/10/2024 & C3419 &  60597.48 & 19.18  &   16.7   & 2.0 &  $<$ 132 \\
14/10/2024 & C3419 &  60597.48 & 19.18  &   21.2  & 2.0 &  $<$ 132 \\
30/10/2024 & C3419 &  60613.47 & 35.17  &  5.5   & 2.0 &  178$\pm$28 \\
30/10/2024 & C3419 &  60613.47 & 35.17  &  9.0   & 2.0 &  314$\pm$20 \\
30/10/2024 & C3419 &  60613.47 & 35.17  &  16.7 & 2.0 & 442$\pm$28 \\
30/10/2024 & C3419 &  60613.47 & 35.17  &  21.2 & 2.0 & 538$\pm$53  \\
14/11/2024 & C3419 &  60628.37 & 50.07  &  5.5  & 2.0 &  224$\pm$20 \\
14/11/2024 & C3419 &  60628.37 & 50.07  &  9.0  & 2.0 &  477$\pm$18 \\
14/11/2024 & C3419 &  60628.37 & 50.07  &  16.7 & 2.0 & 734$\pm$23 \\
14/11/2024 & C3419 &  60628.37 & 50.07  &  21.2 & 2.0 & 767$\pm$76  \\
14/11/2024 & C3419 &  60628.37 & 50.07  &  34.0 & 2.0 &  611$\pm$87 \\
13/12/2024 & C3419 &  60657.27 & 78.97  &  5.5  & 2.0 & 599$\pm$82  \\
13/12/2024 & C3419 &  60657.27 & 78.97  &   9.0 & 2.0 & 847$\pm$30  \\
13/12/2024 & C3419 &  60657.27 & 78.97  &   16.7 & 2.0 & 382$\pm$42  \\
13/12/2024 & C3419 &  60657.27 & 78.97  &   21.2 & 2.0 & 274$\pm$88  \\
13/12/2024 & C3419 &  60657.27 & 78.97  &   33.0 & 2.0 & $<$300 \\
13/12/2024 & C3419 &  60657.27 & 78.97  &   35.0 & 2.0 & $<$360 \\
31/01/2025 & C3419 &  60706.13 & 127.83 &   5.5  & 2.0 & 314$\pm$35 \\
31/01/2025 & C3419 &  60706.13 & 127.83 &   9.0  & 2.0 & 202$\pm$25 \\
17/02/2025 & C3419 &  60723.13 & 144.83 &   5.5  & 2.0 & 252$\pm$25 \\
17/02/2025 & C3419 &  60723.13 & 144.83 &   9.0  & 2.0 & 337$\pm$38 \\
19/03/2025 & C3419 &  60753.04 & 174.74 &   5.5  & 2.0 & 127$\pm$19 \\
19/03/2025 & C3419 &  60753.04 & 174.74 &   9.0  & 2.0 & 143$\pm$28 \\
\enddata
\tablecomments{$^{\rm{a}}$ With respect to first light in observer frame. $^{\rm{b}}$ The uncertainties on flux measurements includes map rms values (1$\sigma$) and a 5\% systematic uncertainty on the flux density added in quadrature. The flux density upper limits are $3\sigma$.
\label{tab:radio-atca}}
\end{deluxetable*}

\startlongtable
\begin{deluxetable*}{ccccccccc}
\tablecaption{ATA observations of AT\,2024wpp}
\tablehead{
\colhead{Start Date} & \colhead{Project ID}  & \colhead{Centroid MJD} & \colhead{Phase$^{\rm{a}}$} & \colhead{Frequency}  & \colhead{Bandwidth} & \colhead{Flux Density$^{\rm{b}}$} \\
 (dd/mm/yyyy) &  &  & (d) & (GHz) & (GHz) & ($\mu$Jy) &
 }
\startdata
09/10/2024 & P053  & 60592.27  & 13.97  & 3.0 & 0.67 & $<570$     \\ 
09/10/2024 & P053  & 60592.27  & 13.97  & 8.0 & 0.67 & $<1100$   \\
01/11/2024 & P053  & 60615.21  & 36.91  & 3.0 & 0.67 & $<810$     \\ 
01/11/2024 & P053  & 60615.21  & 36.91  & 8.0 & 0.67 & $<1440$   \\
\enddata
\tablecomments{$^{\rm{a}}$ With respect to first light. $^{\rm{b}}$ The flux density upper limits are $3\sigma$.
\label{Tab:radio-ATA}}
\end{deluxetable*}

\startlongtable
\begin{deluxetable*}{ccccccccc}
\tablecaption{MeerKAT observations of AT\,2024wpp}
\tablehead{
\colhead{Start Date} & \colhead{Project ID}  & \colhead{Centroid MJD} & \colhead{Phase$^{\rm{a}}$} & \colhead{Frequency}  & \colhead{Bandwidth} & \colhead{Flux Density$^{\rm{b}}$} \\
 (dd/mm/yyyy) &  &  & (d) & (GHz) & (GHz) & ($\mu$Jy) &
 }
\startdata
31/10/2024 & SCI-20230907-NA-01  & 60614.72  & 36.42  & 1.28 & 0.86 & $<66$     \\ 
15/11/2024 & SCI-20230907-NA-01  & 60629.68  & 51.38  & 1.28 & 0.86 & $<60$   \\
11/12/2024 & SCI-20230907-NA-01  & 60655.69  & 77.39  & 1.28 & 0.86 & $<75$   \\
\enddata
\tablecomments{$^{\rm{a}}$ With respect to first light. $^{\rm{b}}$ The flux density upper limits are $3\sigma$.
\label{Tab:radio-MeerKAT}}
\end{deluxetable*}

\startlongtable
\begin{deluxetable*}{ccccccccc}
\tablecaption{GMRT observations of AT\,2024wpp}
\tablehead{
\colhead{Start Date} & \colhead{Project ID}  & \colhead{Centroid MJD} & \colhead{Phase$^{\rm{a}}$} & \colhead{Frequency}  & \colhead{Bandwidth} & \colhead{Flux Density$^{\rm{b}}$} \\
 (dd/mm/yyyy) &  &  & (d) & (GHz) & (GHz) & ($\mu$Jy) &
 }
\startdata
01/11/2024 & 47$_{-}$059  & 60615.00  & 36.70  & 1.25 & 0.40 & $<90$   \\ 
02/11/2024 & 47$_{-}$059  & 60616.00  & 37.70  & 0.65 & 0.20 & $<86$   \\
02/11/2024 & 47$_{-}$059  & 60616.00  & 37.70  & 0.44 & 0.20 & $<210$   \\
15/11/2024 & 47$_{-}$059  & 60629.00  & 50.70  & 0.65 & 0.20 & $<81$ \\
02/12/2024 & 47$_{-}$059  & 60646.00  & 67.70  & 0.44 & 0.20 & $<240$ \\
04/12/2024 & 47$_{-}$059  & 60648.00  & 69.70  & 0.65 & 0.20 & $<75$ \\
26/01/2025 & DDTC414   & 60701.55 & 123.25 & 1.25 & 0.40 & $<$84 \\
\enddata
\tablecomments{$^{\rm{a}}$ With respect to first light. $^{\rm{b}}$ The flux density upper limits are $3\sigma$
\label{Tab:radio-gmrt}}
\end{deluxetable*}

\section{X-ray observations logs and tables}\label{AppendixXray}

\startlongtable
\begin{deluxetable*}{lcccccc}
\tablecaption{X-ray observations of AT\,2024wpp with \nustar\, \chandra, and \xmm, (PI Margutti).}
\tablehead{
\colhead{Instrument} & \colhead{Start date} & \colhead{Mid time$^{\rm{a}}$} & \colhead{Obs ID} & \colhead{Exposure Time$^{\rm{b}}$}  \\
  & (yyyy/mm/dd) & (d) &  & (ks)  &  &
 }
\startdata
NuSTAR & 2024-09-30 &  6.0 & 91001341002 &  41.5/41.0   \\
NuSTAR & 2024-10-06 &  11.0 & 91001341004 &  42.7/42.3   \\
NuSTAR & 2024-10-12 &  18.0 & 91001341006 &  57.1/56.6   \\
NuSTAR & 2024-11-16 &  52.3 & 80802406002  &  43.0/42.6    \\
NuSTAR & 2024-12-09 &  76.1 & 80802406004  &  35.7/37.1    \\
CXO/ACIS-S &   2024-10-13  & 19.9  & 30566  &  19.8   \\
CXO/ACIS-S &   2024-10-21  & 25.9 & 30567 &  19.8   \\
CXO/ACIS-S &   2024-11-14 &  49.7 & 30568 &  36.1   \\
CXO/ACIS-S &   2024-12-09 &  75.2 & 30642 &  36.2   \\
XMM/EPIC-pn &   2025-01-02 &  99.2 & 0903320501& 33.1   \\
XMM/EPIC-MOS1 &   2025-01-02 &  99.2 & 0903320501&  44.8   \\
XMM/EPIC-MOS2 &   2025-01-02 &  99.2 & 0903320501&  44.8    \\
XMM/EPIC-pn &   2025-02-11 &  140.0 & 0903320601 &  13.3   \\
XMM/EPIC-MOS1 &   2025-02-11 &  140.0 & 0903320601 &  40.4   \\
XMM/EPIC-MOS2 &   2025-02-11 &  140.0 & 0903320601 & 40.4   \\
XMM/EPIC-pn &   2025-07-01 &  279.4 & 0903320701 &  42.7  \\
XMM/EPIC-MOS1 &   2025-07-01 & 279.4 & 0903320701 &  44.8   \\
XMM/EPIC-MOS2 &   2025-07-01 &  279.4 & 0903320701 &  44.8   \\
XMM/EPIC-pn &   2025-08-16 & 325.2 & 0903320801 &  26.8   \\
XMM/EPIC-MOS1 &   2025-08-16  & 325.2 & 0903320801 &   40.1   \\
XMM/EPIC-MOS2 &    2025-08-16 & 325.2 & 0903320801 &   40.1   \\
\enddata
\tablecomments{$^{\rm{a}}$ With respect to first light. $^{\rm{b}}$ For NuSTAR we report the exposures for the A and B modules, respectively, after removing the interval of times severely affected by solar flares.
\label{Tab:Xraylog}}
\end{deluxetable*}

\bibliography{AT2024wpp_v1}{}

\begin{thebibliography}{}
\expandafter\ifx\csname natexlab\endcsname\relax\def\natexlab#1{#1}\fi
\providecommand{\url}[1]{\href{#1}{#1}}
\providecommand{\dodoi}[1]{doi:~\href{http://doi.org/#1}{\nolinkurl{#1}}}
\providecommand{\doeprint}[1]{\href{http://ascl.net/#1}{\nolinkurl{http://ascl.net/#1}}}
\providecommand{\doarXiv}[1]{\href{https://arxiv.org/abs/#1}{\nolinkurl{https://arxiv.org/abs/#1}}}

\bibitem[{{Andreoni} {et~al.}(2022){Andreoni}, {Coughlin}, {Perley}, {Yao},
  {Lu}, {Cenko}, {Kumar}, {Anand}, {Ho}, {Kasliwal}, {de Ugarte Postigo},
  {Sagu{\'e}s-Carracedo}, {Schulze}, {Kann}, {Kulkarni}, {Sollerman}, {Tanvir},
  {Rest}, {Izzo}, {Somalwar}, {Kaplan}, {Ahumada}, {Anupama}, {Auchettl},
  {Barway}, {Bellm}, {Bhalerao}, {Bloom}, {Bremer}, {Bulla}, {Burns},
  {Campana}, {Chandra}, {Charalampopoulos}, {Cooke}, {D'Elia}, {Das}, {Dobie},
  {Ag{\"u}{\'\i} Fern{\'a}ndez}, {Freeburn}, {Fremling}, {Gezari}, {Goode},
  {Graham}, {Hammerstein}, {Karambelkar}, {Kilpatrick}, {Kool}, {Krips},
  {Laher}, {Leloudas}, {Levan}, {Lundquist}, {Mahabal}, {Medford}, {Miller},
  {M{\"o}ller}, {Mooley}, {Nayana}, {Nir}, {Pang}, {Paraskeva}, {Perley},
  {Petitpas}, {Pursiainen}, {Ravi}, {Ridden-Harper}, {Riddle}, {Rigault},
  {Rodriguez}, {Rusholme}, {Sharma}, {Smith}, {Stein}, {Th{\"o}ne},
  {Tohuvavohu}, {Valdes}, {van Roestel}, {Vergani}, {Wang}, \&
  {Zhang}}]{Andreoni2022}
{Andreoni}, I., {Coughlin}, M.~W., {Perley}, D.~A., {et~al.} 2022, \nat, 612,
  430, \dodoi{10.1038/s41586-022-05465-8}

\bibitem[{{Antoni} \& {Quataert}(2022)}]{Antoni2022}
{Antoni}, A., \& {Quataert}, E. 2022, \mnras, 511, 176,
  \dodoi{10.1093/mnras/stab3776}

\bibitem[{{Arcavi} {et~al.}(2016){Arcavi}, {Wolf}, {Howell}, {Bildsten},
  {Leloudas}, {Hardin}, {Prajs}, {Perley}, {Svirski}, {Gal-Yam}, {Katz},
  {McCully}, {Cenko}, {Lidman}, {Sullivan}, {Valenti}, {Astier}, {Balland},
  {Carlberg}, {Conley}, {Fouchez}, {Guy}, {Pain}, {Palanque-Delabrouille},
  {Perrett}, {Pritchet}, {Regnault}, {Rich}, \&
  {Ruhlmann-Kleider}}]{Arcavi2016}
{Arcavi}, I., {Wolf}, W.~M., {Howell}, D.~A., {et~al.} 2016, \apj, 819, 35,
  \dodoi{10.3847/0004-637X/819/1/35}

\bibitem[{{Astropy Collaboration} {et~al.}(2013){Astropy Collaboration},
  {Robitaille}, {Tollerud}, {Greenfield}, {Droettboom}, {Bray}, {Aldcroft},
  {Davis}, {Ginsburg}, {Price-Whelan}, {Kerzendorf}, {Conley}, {Crighton},
  {Barbary}, {Muna}, {Ferguson}, {Grollier}, {Parikh}, {Nair}, {Unther},
  {Deil}, {Woillez}, {Conseil}, {Kramer}, {Turner}, {Singer}, {Fox}, {Weaver},
  {Zabalza}, {Edwards}, {Azalee Bostroem}, {Burke}, {Casey}, {Crawford},
  {Dencheva}, {Ely}, {Jenness}, {Labrie}, {Lim}, {Pierfederici}, {Pontzen},
  {Ptak}, {Refsdal}, {Servillat}, \& {Streicher}}]{2013A&A...558A..33A}
{Astropy Collaboration}, {Robitaille}, T.~P., {Tollerud}, E.~J., {et~al.} 2013,
  \aap, 558, A33, \dodoi{10.1051/0004-6361/201322068}

\bibitem[{{Astropy Collaboration} {et~al.}(2018){Astropy Collaboration},
  {Price-Whelan}, {Sip{\H{o}}cz}, {G{\"u}nther}, {Lim}, {Crawford}, {Conseil},
  {Shupe}, {Craig}, {Dencheva}, \& et~al.}]{2018AJ....156..123A}
{Astropy Collaboration}, {Price-Whelan}, A.~M., {Sip{\H{o}}cz}, B.~M., {et~al.}
  2018, \aj, 156, 123, \dodoi{10.3847/1538-3881/aabc4f}

\bibitem[{{Bell}(1978)}]{Bell1978}
{Bell}, A.~R. 1978, \mnras, 182, 147, \dodoi{10.1093/mnras/182.2.147}

\bibitem[{{Beniamini} {et~al.}(2023){Beniamini}, {Piran}, \&
  {Matsumoto}}]{Beniamini2023}
{Beniamini}, P., {Piran}, T., \& {Matsumoto}, T. 2023, \mnras, 524, 1386,
  \dodoi{10.1093/mnras/stad1950}

\bibitem[{{Berger} {et~al.}(2012){Berger}, {Zauderer}, {Pooley}, {Soderberg},
  {Sari}, {Brunthaler}, \& {Bietenholz}}]{Berger2012-swJ1644}
{Berger}, E., {Zauderer}, A., {Pooley}, G.~G., {et~al.} 2012, \apj, 748, 36,
  \dodoi{10.1088/0004-637X/748/1/36}

\bibitem[{{Bj{\"o}rnsson}(2024)}]{Bjornsson2024}
{Bj{\"o}rnsson}, C.~I. 2024, \apj, 963, 93, \dodoi{10.3847/1538-4357/ad1109}

\bibitem[{{Bj{\"o}rnsson} \& {Keshavarzi}(2017)}]{Bjornsson2017}
{Bj{\"o}rnsson}, C.~I., \& {Keshavarzi}, S.~T. 2017, \apj, 841, 12,
  \dodoi{10.3847/1538-4357/aa6cad}

\bibitem[{{Blandford} \& {Eichler}(1987)}]{Blandford1987}
{Blandford}, R., \& {Eichler}, D. 1987, \physrep, 154, 1,
  \dodoi{10.1016/0370-1573(87)90134-7}

\bibitem[{{Blandford} \& {Begelman}(1999)}]{Blandford1999}
{Blandford}, R.~D., \& {Begelman}, M.~C. 1999, \mnras, 303, L1,
  \dodoi{10.1046/j.1365-8711.1999.02358.x}

\bibitem[{{Blandford} \& {Ostriker}(1978)}]{Blandford1978}
{Blandford}, R.~D., \& {Ostriker}, J.~P. 1978, \apjl, 221, L29,
  \dodoi{10.1086/182658}

\bibitem[{{Bright} {et~al.}(2021){Bright}, {Margutti}, {Matthews}, {Brethauer},
  {Coppejans}, {Wieringa}, {Metzger}, {DeMarchi}, {Laskar}, {Romero},
  {Alexander}, {Horesh}, {Migliori}, {Chornock}, {Berger}, {Bietenholz},
  {Devlin}, {Dicker}, {Jacobson-Gal{\'a}n}, {Mason}, {Milisavljevic}, {Motta},
  {Mroczkowski}, {Ramirez-Ruiz}, {Rhodes}, {Sarazin}, {Sfaradi}, \&
  {Sievers}}]{Bright2021-xnd}
{Bright}, J.~S., {Margutti}, R., {Matthews}, D., {et~al.} 2021, arXiv e-prints,
  arXiv:2110.05514.
\newblock \doarXiv{2110.05514}

\bibitem[{{Bright} {et~al.}(2023){Bright}, {Rhodes}, {Farah}, {Fender}, {van
  der Horst}, {Leung}, {Williams}, {Anderson}, {Atri}, {DeBoer}, {Giarratana},
  {Green}, {Heywood}, {Lenc}, {Murphy}, {Pollak}, {Premnath}, {Scott},
  {Sheikh}, {Siemion}, \& {Titterington}}]{Bright2023}
{Bright}, J.~S., {Rhodes}, L., {Farah}, W., {et~al.} 2023, Nature Astronomy, 7,
  986, \dodoi{10.1038/s41550-023-01997-9}

\bibitem[{{Burrows} {et~al.}(2005){Burrows}, {Hill}, {Nousek}, {Kennea},
  {Wells}, {Osborne}, {Abbey}, {Beardmore}, {Mukerjee}, {Short}, {Chincarini},
  {Campana}, {Citterio}, {Moretti}, {Pagani}, {Tagliaferri}, {Giommi},
  {Capalbi}, {Tamburelli}, {Angelini}, {Cusumano}, {Br{\"a}uninger}, {Burkert},
  \& {Hartner}}]{Burrows05}
{Burrows}, D.~N., {Hill}, J.~E., {Nousek}, J.~A., {et~al.} 2005, \ssr, 120,
  165, \dodoi{10.1007/s11214-005-5097-2}

\bibitem[{{Caprioli}(2015)}]{Caprioli2015}
{Caprioli}, D. 2015, in International Cosmic Ray Conference, Vol.~34, 34th
  International Cosmic Ray Conference (ICRC2015), 8,
  \dodoi{10.22323/1.236.0008}

\bibitem[{{CASA Team} {et~al.}(2022){CASA Team}, {Bean}, {Bhatnagar}, {Castro},
  {Donovan Meyer}, {Emonts}, {Garcia}, {Garwood}, {Golap}, {Gonzalez Villalba},
  {Harris}, {Hayashi}, {Hoskins}, {Hsieh}, {Jagannathan}, {Kawasaki},
  {Keimpema}, {Kettenis}, {Lopez}, {Marvil}, {Masters}, {McNichols},
  {Mehringer}, {Miel}, {Moellenbrock}, {Montesino}, {Nakazato}, {Ott}, {Petry},
  {Pokorny}, {Raba}, {Rau}, {Schiebel}, {Schweighart}, {Sekhar}, {Shimada},
  {Small}, {Steeb}, {Sugimoto}, {Suoranta}, {Tsutsumi}, {van Bemmel},
  {Verkouter}, {Wells}, {Xiong}, {Szomoru}, {Griffith}, {Glendenning}, \&
  {Kern}}]{CASAteam2022}
{CASA Team}, {Bean}, B., {Bhatnagar}, S., {et~al.} 2022, \pasp, 134, 114501,
  \dodoi{10.1088/1538-3873/ac9642}

\bibitem[{{Cash}(1979)}]{Cash79}
{Cash}, W. 1979, \apj, 228, 939, \dodoi{10.1086/156922}

\bibitem[{{Cendes} {et~al.}(2021){Cendes}, {Alexander}, {Berger}, {Eftekhari},
  {Williams}, \& {Chornock}}]{Cendes2021}
{Cendes}, Y., {Alexander}, K.~D., {Berger}, E., {et~al.} 2021, \apj, 919, 127,
  \dodoi{10.3847/1538-4357/ac110a}

\bibitem[{{Chevalier}(1998)}]{chevalier1998}
{Chevalier}, R.~A. 1998, \apj, 499, 810, \dodoi{10.1086/305676}

\bibitem[{{Chevalier} {et~al.}(2006){Chevalier}, {Fransson}, \&
  {Nymark}}]{chevalier2006}
{Chevalier}, R.~A., {Fransson}, C., \& {Nymark}, T.~K. 2006, \apj, 641, 1029,
  \dodoi{10.1086/500528}

\bibitem[{{Chrimes} {et~al.}(2024{\natexlab{a}}){Chrimes}, {Coppejans},
  {Jonker}, {Levan}, {Groot}, {Mummery}, \& {Stanway}}]{Chrimes2024b-fhn}
{Chrimes}, A.~A., {Coppejans}, D.~L., {Jonker}, P.~G., {et~al.}
  2024{\natexlab{a}}, \aap, 691, A329, \dodoi{10.1051/0004-6361/202451172}

\bibitem[{{Chrimes} {et~al.}(2024{\natexlab{b}}){Chrimes}, {Jonker}, {Levan},
  {Coppejans}, {Gaspari}, {Gompertz}, {Groot}, {Malesani}, {Mummery},
  {Stanway}, \& {Wiersema}}]{Chrimes2024a-fhn}
{Chrimes}, A.~A., {Jonker}, P.~G., {Levan}, A.~J., {et~al.} 2024{\natexlab{b}},
  \mnras, 527, L47, \dodoi{10.1093/mnrasl/slad145}

\bibitem[{{Coppejans} {et~al.}(2020){Coppejans}, {Margutti}, {Terreran},
  {Nayana}, {Coughlin}, {Laskar}, {Alexander}, {Bietenholz}, {Caprioli},
  {Chandra}, {Drout}, {Frederiks}, {Frohmaier}, {Hurley}, {Kochanek},
  {MacLeod}, {Meisner}, {Nugent}, {Ridnaia}, {Sand}, {Svinkin}, {Ward}, {Yang},
  {Baldeschi}, {Chilingarian}, {Dong}, {Esquivia}, {Fong}, {Guidorzi},
  {Lundqvist}, {Milisavljevic}, {Paterson}, {Reichart}, {Shappee}, {Stroh},
  {Valenti}, {Zauderer}, \& {Zhang}}]{coppejans2020}
{Coppejans}, D.~L., {Margutti}, R., {Terreran}, G., {et~al.} 2020, \apjl, 895,
  L23, \dodoi{10.3847/2041-8213/ab8cc7}

\bibitem[{{Dessart} {et~al.}(2021){Dessart}, {Hillier}, {Sukhbold}, {Woosley},
  \& {Janka}}]{Dessart2021}
{Dessart}, L., {Hillier}, D.~J., {Sukhbold}, T., {Woosley}, S.~E., \& {Janka},
  H.~T. 2021, \aap, 656, A61, \dodoi{10.1051/0004-6361/202141927}

\bibitem[{{Dessart} {et~al.}(2020){Dessart}, {Yoon}, {Aguilera-Dena}, \&
  {Langer}}]{Dessart2020}
{Dessart}, L., {Yoon}, S.-C., {Aguilera-Dena}, D.~R., \& {Langer}, N. 2020,
  \aap, 642, A106, \dodoi{10.1051/0004-6361/202038763}

\bibitem[{{Drout} {et~al.}(2014){Drout}, {Chornock}, {Soderberg}, {Sanders},
  {McKinnon}, {Rest}, {Foley}, {Milisavljevic}, {Margutti}, {Berger},
  {Calkins}, {Fong}, {Gezari}, {Huber}, {Kankare}, {Kirshner}, {Leibler},
  {Lunnan}, {Mattila}, {Marion}, {Narayan}, {Riess}, {Roth}, {Scolnic},
  {Smartt}, {Tonry}, {Burgett}, {Chambers}, {Hodapp}, {Jedicke}, {Kaiser},
  {Magnier}, {Metcalfe}, {Morgan}, {Price}, \& {Waters}}]{Drout2014}
{Drout}, M.~R., {Chornock}, R., {Soderberg}, A.~M., {et~al.} 2014, \apj, 794,
  23, \dodoi{10.1088/0004-637X/794/1/23}

\bibitem[{{Dwarkadas}(2025)}]{Dwarkadas25}
{Dwarkadas}, V.~V. 2025, Universe, 11, 161, \dodoi{10.3390/universe11050161}

\bibitem[{{Eftekhari} {et~al.}(2022){Eftekhari}, {Berger}, {Metzger}, {Laskar},
  {Villar}, {Alexander}, {Holder}, {Vieira}, {Whitehorn}, \&
  {Williams}}]{Eftekhari2022}
{Eftekhari}, T., {Berger}, E., {Metzger}, B.~D., {et~al.} 2022, \apj, 935, 16,
  \dodoi{10.3847/1538-4357/ac7ce8}

\bibitem[{{Ercolino} {et~al.}(2025){Ercolino}, {Jin}, {Langer}, \&
  {Dessart}}]{Ercolino2025}
{Ercolino}, A., {Jin}, H., {Langer}, N., \& {Dessart}, L. 2025, \aap, 696,
  A103, \dodoi{10.1051/0004-6361/202453426}

\bibitem[{{Evans} {et~al.}(2009){Evans}, {Beardmore}, {Page}, {Osborne},
  {O'Brien}, {Willingale}, {Starling}, {Burrows}, {Godet}, {Vetere}, {Racusin},
  {Goad}, {Wiersema}, {Angelini}, {Capalbi}, {Chincarini}, {Gehrels}, {Kennea},
  {Margutti}, {Morris}, {Mountford}, {Pagani}, {Perri}, {Romano}, \&
  {Tanvir}}]{Evans09}
{Evans}, P.~A., {Beardmore}, A.~P., {Page}, K.~L., {et~al.} 2009, \mnras, 397,
  1177, \dodoi{10.1111/j.1365-2966.2009.14913.x}

\bibitem[{{Fox} \& {Smith}(2019)}]{Fox2019}
{Fox}, O.~D., \& {Smith}, N. 2019, \mnras, 488, 3772,
  \dodoi{10.1093/mnras/stz1925}

\bibitem[{{Gehrels} {et~al.}(2004){Gehrels}, {Chincarini}, {Giommi}, {Mason},
  {Nousek}, {Wells}, {White}, {Barthelmy}, {Burrows}, {Cominsky}, {Hurley},
  {Marshall}, {M{\'e}sz{\'a}ros}, {Roming}, {Angelini}, {Barbier}, {Belloni},
  {Campana}, {Caraveo}, {Chester}, {Citterio}, {Cline}, {Cropper}, {Cummings},
  {Dean}, {Feigelson}, {Fenimore}, {Frail}, {Fruchter}, {Garmire}, {Gendreau},
  {Ghisellini}, {Greiner}, {Hill}, {Hunsberger}, {Krimm}, {Kulkarni}, {Kumar},
  {Lebrun}, {Lloyd-Ronning}, {Markwardt}, {Mattson}, {Mushotzky}, {Norris},
  {Osborne}, {Paczynski}, {Palmer}, {Park}, {Parsons}, {Paul}, {Rees},
  {Reynolds}, {Rhoads}, {Sasseen}, {Schaefer}, {Short}, {Smale}, {Smith},
  {Stella}, {Tagliaferri}, {Takahashi}, {Tashiro}, {Townsley}, {Tueller},
  {Turner}, {Vietri}, {Voges}, {Ward}, {Willingale}, {Zerbi}, \&
  {Zhang}}]{Gehrels04}
{Gehrels}, N., {Chincarini}, G., {Giommi}, P., {et~al.} 2004, \apj, 611, 1005,
  \dodoi{10.1086/422091}

\bibitem[{{Gompertz} {et~al.}(2018){Gompertz}, {Fruchter}, \&
  {Pe'er}}]{Gompertz2018}
{Gompertz}, B.~P., {Fruchter}, A.~S., \& {Pe'er}, A. 2018, \apj, 866, 162,
  \dodoi{10.3847/1538-4357/aadba8}

\bibitem[{{Gottlieb} {et~al.}(2022){Gottlieb}, {Tchekhovskoy}, \&
  {Margutti}}]{Gottlieb2022}
{Gottlieb}, O., {Tchekhovskoy}, A., \& {Margutti}, R. 2022, \mnras, 513, 3810,
  \dodoi{10.1093/mnras/stac910}

\bibitem[{{Greisen}(2003)}]{greisen2003}
{Greisen}, E.~W. 2003, in Astrophysics and Space Science Library, Vol. 285,
  Information Handling in Astronomy - Historical Vistas, ed. A.~{Heck}, 109,
  \dodoi{10.1007/0-306-48080-8_7}

\bibitem[{{Guti{\'e}rrez} {et~al.}(2024){Guti{\'e}rrez}, {Mattila},
  {Lundqvist}, {Dessart}, {Gonz{\'a}lez-Gait{\'a}n}, {Jonker}, {Dong},
  {Coppejans}, {Chen}, {Charalampopoulos}, {Elias-Rosa}, {Reynolds},
  {Kochanek}, {Fraser}, {Pastorello}, {Gromadzki}, {Neustadt}, {Benetti},
  {Kankare}, {Kangas}, {Kotak}, {Stritzinger}, {Wevers}, {Zhang}, {Bersier},
  {Bose}, {Buckley}, {Dastidar}, {Gangopadhyay}, {Hamanowicz}, {Kollmeier},
  {Mao}, {Misra}, {Potter}, {Prieto}, {Romero-Colmenero}, {Singh}, {Somero},
  {Terreran}, {Vaisanen}, \& {Wyrzykowski}}]{Gutierrez2024}
{Guti{\'e}rrez}, C.~P., {Mattila}, S., {Lundqvist}, P., {et~al.} 2024, \apj,
  977, 162, \dodoi{10.3847/1538-4357/ad89a5}

\bibitem[{{HI4PI Collaboration} {et~al.}(2016){HI4PI Collaboration}, {Ben
  Bekhti}, {Fl{\"o}er}, {Keller}, {Kerp}, {Lenz}, {Winkel}, {Bailin},
  {Calabretta}, {Dedes}, {Ford}, {Gibson}, {Haud}, {Janowiecki}, {Kalberla},
  {Lockman}, {McClure-Griffiths}, {Murphy}, {Nakanishi}, {Pisano}, \&
  {Staveley-Smith}}]{HI4PI}
{HI4PI Collaboration}, {Ben Bekhti}, N., {Fl{\"o}er}, L., {et~al.} 2016, \aap,
  594, A116, \dodoi{10.1051/0004-6361/201629178}

\bibitem[{{Ho} {et~al.}(2024){Ho}, {Srinivasaragavan}, {Perley}, {Andreoni},
  {Rehentulla}, \& {Qin}}]{Ho2024-AT2024wpp}
{Ho}, A.~Y.~Q., {Srinivasaragavan}, G., {Perley}, D., {et~al.} 2024, Transient
  Name Server AstroNote, 272, 1

\bibitem[{{Ho} {et~al.}(2019){Ho}, {Phinney}, {Ravi}, {Kulkarni}, {Petitpas},
  {Emonts}, {Bhalerao}, {Blundell}, {Cenko}, {Dobie}, {Howie}, {Kamraj},
  {Kasliwal}, {Murphy}, {Perley}, {Sridharan}, \& {Yoon}}]{Ho2019-cow}
{Ho}, A. Y.~Q., {Phinney}, E.~S., {Ravi}, V., {et~al.} 2019, \apj, 871, 73,
  \dodoi{10.3847/1538-4357/aaf473}

\bibitem[{{Ho} {et~al.}(2020){Ho}, {Perley}, {Kulkarni}, {Dong}, {De},
  {Chandra}, {Andreoni}, {Bellm}, {Burdge}, {Coughlin}, {Dekany}, {Feeney},
  {Frederiks}, {Fremling}, {Golkhou}, {Graham}, {Hale}, {Helou}, {Horesh},
  {Kasliwal}, {Laher}, {Masci}, {Miller}, {Porter}, {Ridnaia}, {Rusholme},
  {Shupe}, {Soumagnac}, \& {Svinkin}}]{Ho2020-AT2018lug}
{Ho}, A. Y.~Q., {Perley}, D.~A., {Kulkarni}, S.~R., {et~al.} 2020, \apj, 895,
  49, \dodoi{10.3847/1538-4357/ab8bcf}

\bibitem[{{Ho} {et~al.}(2022){Ho}, {Margalit}, {Bremer}, {Perley}, {Yao},
  {Dobie}, {Kaplan}, {O'Brien}, {Petitpas}, \& {Zic}}]{Ho2022-xnd}
{Ho}, A. Y.~Q., {Margalit}, B., {Bremer}, M., {et~al.} 2022, \apj, 932, 116,
  \dodoi{10.3847/1538-4357/ac4e97}

\bibitem[{{Ho} {et~al.}(2023{\natexlab{a}}){Ho}, {Perley}, {Gal-Yam}, {Lunnan},
  {Sollerman}, {Schulze}, {Das}, {Dobie}, {Yao}, {Fremling}, {Adams}, {Anand},
  {Andreoni}, {Bellm}, {Bruch}, {Burdge}, {Castro-Tirado}, {Dahiwale}, {De},
  {Dekany}, {Drake}, {Duev}, {Graham}, {Helou}, {Kaplan}, {Karambelkar},
  {Kasliwal}, {Kool}, {Kulkarni}, {Mahabal}, {Medford}, {Miller}, {Nordin},
  {Ofek}, {Petitpas}, {Riddle}, {Sharma}, {Smith}, {Stewart}, {Taggart},
  {Tartaglia}, {Tzanidakis}, \& {Winters}}]{Ho2023-FBOT-rate}
{Ho}, A. Y.~Q., {Perley}, D.~A., {Gal-Yam}, A., {et~al.} 2023{\natexlab{a}},
  \apj, 949, 120, \dodoi{10.3847/1538-4357/acc533}

\bibitem[{{Ho} {et~al.}(2023{\natexlab{b}}){Ho}, {Perley}, {Chen}, {Schulze},
  {Dhillon}, {Kumar}, {Suresh}, {Swain}, {Bremer}, {Smartt}, {Anderson},
  {Anupama}, {Awiphan}, {Barway}, {Bellm}, {Ben-Ami}, {Bhalerao}, {de Boer},
  {Brink}, {Burruss}, {Chandra}, {Chen}, {Chen}, {Cooke}, {Coughlin}, {Das},
  {Drake}, {Filippenko}, {Freeburn}, {Fremling}, {Fulton}, {Gal-Yam},
  {Galbany}, {Gao}, {Graham}, {Gromadzki}, {Guti{\'e}rrez}, {Hinds}, {Inserra},
  {A J}, {Karambelkar}, {Kasliwal}, {Kulkarni}, {M{\"u}ller-Bravo}, {Magnier},
  {Mahabal}, {Moore}, {Ngeow}, {Nicholl}, {Ofek}, {Omand}, {Onori}, {Pan},
  {Pessi}, {Petitpas}, {Polishook}, {Poshyachinda}, {Pursiainen}, {Riddle},
  {Rodriguez}, {Rusholme}, {Segre}, {Sharma}, {Smith}, {Sollerman},
  {Srivastav}, {Strotjohann}, {Suhr}, {Svinkin}, {Wang}, {Wiseman}, {Wold},
  {Yang}, {Yang}, {Yao}, {Young}, \& {Zheng}}]{Ho2023-at2022tsd}
{Ho}, A. Y.~Q., {Perley}, D.~A., {Chen}, P., {et~al.} 2023{\natexlab{b}}, \nat,
  623, 927, \dodoi{10.1038/s41586-023-06673-6}

\bibitem[{{Hollenbach} {et~al.}(1994){Hollenbach}, {Johnstone}, {Lizano}, \&
  {Shu}}]{Hollenbach1994}
{Hollenbach}, D., {Johnstone}, D., {Lizano}, S., \& {Shu}, F. 1994, \apj, 428,
  654, \dodoi{10.1086/174276}

\bibitem[{{Horesh} {et~al.}(2013){Horesh}, {Stockdale}, {Fox}, {Frail},
  {Carpenter}, {Kulkarni}, {Ofek}, {Gal-Yam}, {Kasliwal}, {Arcavi}, {Quimby},
  {Cenko}, {Nugent}, {Bloom}, {Law}, {Poznanski}, {Gorbikov}, {Polishook},
  {Yaron}, {Ryder}, {Weiler}, {Bauer}, {Van Dyk}, {Immler}, {Panagia},
  {Pooley}, \& {Kassim}}]{Horesh2013-sn2011dh}
{Horesh}, A., {Stockdale}, C., {Fox}, D.~B., {et~al.} 2013, \mnras, 436, 1258,
  \dodoi{10.1093/mnras/stt1645}

\bibitem[{{Kashi} \& {Soker}(2011)}]{Kashi2011}
{Kashi}, A., \& {Soker}, N. 2011, \mnras, 417, 1466,
  \dodoi{10.1111/j.1365-2966.2011.19361.x}

\bibitem[{{Keto}(2007)}]{Keto2007}
{Keto}, E. 2007, \apj, 666, 976, \dodoi{10.1086/520320}

\bibitem[{{Khatami} \& {Kasen}(2024)}]{Khatami2024}
{Khatami}, D.~K., \& {Kasen}, D.~N. 2024, \apj, 972, 140,
  \dodoi{10.3847/1538-4357/ad60c0}

\bibitem[{{King} {et~al.}(2023){King}, {Lasota}, \& {Middleton}}]{King23}
{King}, A., {Lasota}, J.-P., \& {Middleton}, M. 2023, \nar, 96, 101672,
  \dodoi{10.1016/j.newar.2022.101672}

\bibitem[{{Kitaki} {et~al.}(2021){Kitaki}, {Mineshige}, {Ohsuga}, \&
  {Kawashima}}]{Kitaki2021}
{Kitaki}, T., {Mineshige}, S., {Ohsuga}, K., \& {Kawashima}, T. 2021, \pasj,
  73, 450, \dodoi{10.1093/pasj/psab011}

\bibitem[{{Kremer} {et~al.}(2021){Kremer}, {Lu}, {Piro}, {Chatterjee}, {Rasio},
  \& {Ye}}]{Kremer2021}
{Kremer}, K., {Lu}, W., {Piro}, A.~L., {et~al.} 2021, \apj, 911, 104,
  \dodoi{10.3847/1538-4357/abeb14}

\bibitem[{{Kuin} {et~al.}(2019){Kuin}, {Wu}, {Oates}, {Lien}, {Emery},
  {Kennea}, {de Pasquale}, {Han}, {Brown}, {Tohuvavohu}, {Breeveld}, {Burrows},
  {Cenko}, {Campana}, {Levan}, {Markwardt}, {Osborne}, {Page}, {Page},
  {Sbarufatti}, {Siegel}, \& {Troja}}]{Kuin2019}
{Kuin}, N. P.~M., {Wu}, K., {Oates}, S., {et~al.} 2019, \mnras, 487, 2505,
  \dodoi{10.1093/mnras/stz053}

\bibitem[{{Kulkarni} {et~al.}(2021){Kulkarni}, {Harrison}, {Grefenstette},
  {Earnshaw}, {Andreoni}, {Berg}, {Bloom}, {Cenko}, {Chornock}, {Christiansen},
  {Coughlin}, {Wuollet Criswell}, {Darvish}, {Das}, {De}, {Dessart}, {Dixon},
  {Dorsman}, {El-Badry}, {Evans}, {Ford}, {Fremling}, {Gansicke}, {Gezari},
  {Goetberg}, {Green}, {Graham}, {Heida}, {Ho}, {Jaodand}, {Johns-Krull},
  {Kasliwal}, {Lazzarini}, {Lu}, {Margutti}, {Martin}, {Masters}, {McKernan},
  {Naze}, {Nissanke}, {Parazin}, {Perley}, {Phinney}, {Piro}, {Raaijmakers},
  {Rauw}, {Rodriguez}, {Sana}, {Senchyna}, {Singer}, {Spake}, {Stassun},
  {Stern}, {Teplitz}, {Weisz}, \& {Yao}}]{Kulkarni2021-UVEX}
{Kulkarni}, S.~R., {Harrison}, F.~A., {Grefenstette}, B.~W., {et~al.} 2021,
  arXiv e-prints, arXiv:2111.15608, \dodoi{10.48550/arXiv.2111.15608}

\bibitem[{{Leung} {et~al.}(2020){Leung}, {Blinnikov}, {Nomoto}, {Baklanov},
  {Sorokina}, \& {Tolstov}}]{Leung2020}
{Leung}, S.-C., {Blinnikov}, S., {Nomoto}, K., {et~al.} 2020, \apj, 903, 66,
  \dodoi{10.3847/1538-4357/abba33}

\bibitem[{{Leung} {et~al.}(2021){Leung}, {Fuller}, \& {Nomoto}}]{Leung2021}
{Leung}, S.-C., {Fuller}, J., \& {Nomoto}, K. 2021, \apj, 915, 80,
  \dodoi{10.3847/1538-4357/abfcbe}

\bibitem[{{Li} {et~al.}(2011){Li}, {Chornock}, {Leaman}, {Filippenko},
  {Poznanski}, {Wang}, {Ganeshalingam}, \& {Mannucci}}]{Li2011}
{Li}, W., {Chornock}, R., {Leaman}, J., {et~al.} 2011, \mnras, 412, 1473,
  \dodoi{10.1111/j.1365-2966.2011.18162.x}

\bibitem[{{MacLeod} \& {Loeb}(2020)}]{MacLeod2020}
{MacLeod}, M., \& {Loeb}, A. 2020, \apj, 895, 29,
  \dodoi{10.3847/1538-4357/ab89b6}

\bibitem[{{MacLeod} {et~al.}(2017){MacLeod}, {Macias}, {Ramirez-Ruiz},
  {Grindlay}, {Batta}, \& {Montes}}]{MacLeod2017}
{MacLeod}, M., {Macias}, P., {Ramirez-Ruiz}, E., {et~al.} 2017, \apj, 835, 282,
  \dodoi{10.3847/1538-4357/835/2/282}

\bibitem[{{Margalit} \& {Metzger}(2016)}]{Margalit2016}
{Margalit}, B., \& {Metzger}, B.~D. 2016, \mnras, 461, 1154,
  \dodoi{10.1093/mnras/stw1410}

\bibitem[{{Margalit} \& {Quataert}(2021)}]{Margalit2021}
{Margalit}, B., \& {Quataert}, E. 2021, \apjl, 923, L14,
  \dodoi{10.3847/2041-8213/ac3d97}

\bibitem[{{Margutti} {et~al.}(2013){Margutti}, {Zaninoni}, {Bernardini},
  {Chincarini}, {Pasotti}, {Guidorzi}, {Angelini}, {Burrows}, {Capalbi},
  {Evans}, {Gehrels}, {Kennea}, {Mangano}, {Moretti}, {Nousek}, {Osborne},
  {Page}, {Perri}, {Racusin}, {Romano}, {Sbarufatti}, {Stafford}, \&
  {Stamatikos}}]{Margutti13}
{Margutti}, R., {Zaninoni}, E., {Bernardini}, M.~G., {et~al.} 2013, \mnras,
  428, 729, \dodoi{10.1093/mnras/sts066}

\bibitem[{{Margutti} {et~al.}(2019){Margutti}, {Metzger}, {Chornock}, {Vurm},
  {Roth}, {Grefenstette}, {Savchenko}, {Cartier}, {Steiner}, {Terreran},
  {Margalit}, {Migliori}, {Milisavljevic}, {Alexander}, {Bietenholz},
  {Blanchard}, {Bozzo}, {Brethauer}, {Chilingarian}, {Coppejans}, {Ducci},
  {Ferrigno}, {Fong}, {G{\"o}tz}, {Guidorzi}, {Hajela}, {Hurley}, {Kuulkers},
  {Laurent}, {Mereghetti}, {Nicholl}, {Patnaude}, {Ubertini}, {Banovetz},
  {Bartel}, {Berger}, {Coughlin}, {Eftekhari}, {Frederiks}, {Kozlova},
  {Laskar}, {Svinkin}, {Drout}, {MacFadyen}, \& {Paterson}}]{Margutti2019-cow}
{Margutti}, R., {Metzger}, B.~D., {Chornock}, R., {et~al.} 2019, \apj, 872, 18,
  \dodoi{10.3847/1538-4357/aafa01}

\bibitem[{{Margutti} {et~al.}(2024){Margutti}, {AJ}, {Chornock}, {Wiston},
  {Sfaradi}, {Hammerstein}, {LeBaron}, {Brethauer}, {Sears}, {Migliori}, \&
  {Laskar}}]{Margutti2024-wpp-Nustar}
{Margutti}, R., {AJ}, N., {Chornock}, R., {et~al.} 2024, Transient Name Server
  AstroNote, 278, 1

\bibitem[{{Matsumoto} \& {Metzger}(2022)}]{Matsumoto2022}
{Matsumoto}, T., \& {Metzger}, B.~D. 2022, \apj, 938, 5,
  \dodoi{10.3847/1538-4357/ac6269}

\bibitem[{{Matsumoto} \& {Piran}(2023)}]{Matsumoto23equipartition}
{Matsumoto}, T., \& {Piran}, T. 2023, \mnras, 522, 4565,
  \dodoi{10.1093/mnras/stad1269}

\bibitem[{{Matthews} {et~al.}(2023){Matthews}, {Margutti}, {Metzger},
  {Milisavljevic}, {Migliori}, {Laskar}, {Brethauer}, {Berger}, {Chornock},
  {Drout}, \& {Ramirez-Ruiz}}]{Matthews23}
{Matthews}, D., {Margutti}, R., {Metzger}, B.~D., {et~al.} 2023, Research Notes
  of the American Astronomical Society, 7, 126,
  \dodoi{10.3847/2515-5172/acdde1}

\bibitem[{{Matzner} \& {McKee}(1999)}]{Matzner99}
{Matzner}, C.~D., \& {McKee}, C.~F. 1999, \apj, 510, 379,
  \dodoi{10.1086/306571}

\bibitem[{{Metzger}(2022)}]{Metzger22FBOT}
{Metzger}, B.~D. 2022, \apj, 932, 84, \dodoi{10.3847/1538-4357/ac6d59}

\bibitem[{{Metzger} \& {Piro}(2014)}]{MetzgerPiro14}
{Metzger}, B.~D., \& {Piro}, A.~L. 2014, \mnras, 439, 3916,
  \dodoi{10.1093/mnras/stu247}

\bibitem[{{Metzger} {et~al.}(2014){Metzger}, {Vurm}, {Hasco{\"e}t}, \&
  {Beloborodov}}]{Metzger14}
{Metzger}, B.~D., {Vurm}, I., {Hasco{\"e}t}, R., \& {Beloborodov}, A.~M. 2014,
  \mnras, 437, 703, \dodoi{10.1093/mnras/stt1922}

\bibitem[{{Migliori} {et~al.}(2024){Migliori}, {Margutti}, {Metzger},
  {Chornock}, {Vignali}, {Brethauer}, {Coppejans}, {Maccarone}, {Rivera
  Sandoval}, {Bright}, {Laskar}, {Milisavljevic}, {Berger}, \&
  {Nayana}}]{Migliori24}
{Migliori}, G., {Margutti}, R., {Metzger}, B.~D., {et~al.} 2024, \apjl, 963,
  L24, \dodoi{10.3847/2041-8213/ad2764}

\bibitem[{{Narayan} \& {Yi}(1995)}]{Narayan1995}
{Narayan}, R., \& {Yi}, I. 1995, \apj, 444, 231, \dodoi{10.1086/175599}

\bibitem[{{Nayana} \& {Chandra}(2021)}]{Nayana2021}
{Nayana}, A.~J., \& {Chandra}, P. 2021, \apjl, 912, L9,
  \dodoi{10.3847/2041-8213/abed55}

\bibitem[{{Nayana} {et~al.}(2025){Nayana}, {Margutti}, {Laskar}, {Galvin},
  {Wiston}, {Chornock}, {Sfaradi}, {Metzger}, {Lu}, {Berger}, \&
  {Colle}}]{Nayana2025-TNS}
{Nayana}, A.~J., {Margutti}, R., {Laskar}, T., {et~al.} 2025, Transient Name
  Server AstroNote, 114, 1

\bibitem[{{Ofek} {et~al.}(2025){Ofek}, {Ozer}, {Konno}, {Strasman}, {Chen},
  {Ben-Ami}, {Polishook}, {Krassilchtchikov}, {Garrappa}, {Zimmermann},
  {Segre}, {Horowicz}, {Gal-Yam}, {Shani}, {Fainer}, {Engel}, {Sofer-Rimalt},
  {Ho}, {Shvartzvald}, {Yaron}, {Rybicki}, {Blumenzweig}, {Spitzer}, \&
  {Arad}}]{Ofek2025}
{Ofek}, E.~O., {Ozer}, L., {Konno}, R., {et~al.} 2025, arXiv e-prints,
  arXiv:2508.18359.
\newblock \doarXiv{2508.18359}

\bibitem[{{Offringa} \& {Smirnov}(2017)}]{Offringa2017}
{Offringa}, A.~R., \& {Smirnov}, O. 2017, \mnras, 471, 301,
  \dodoi{10.1093/mnras/stx1547}

\bibitem[{{Ohsuga}(2007)}]{Ohsuga2007}
{Ohsuga}, K. 2007, \pasj, 59, 1033, \dodoi{10.1093/pasj/59.5.1033}

\bibitem[{{Pejcha} {et~al.}(2016{\natexlab{a}}){Pejcha}, {Metzger}, \&
  {Tomida}}]{Pejcha2016a}
{Pejcha}, O., {Metzger}, B.~D., \& {Tomida}, K. 2016{\natexlab{a}}, \mnras,
  461, 2527, \dodoi{10.1093/mnras/stw1481}

\bibitem[{{Pejcha} {et~al.}(2016{\natexlab{b}}){Pejcha}, {Metzger}, \&
  {Tomida}}]{Pejcha2016b}
---. 2016{\natexlab{b}}, \mnras, 455, 4351, \dodoi{10.1093/mnras/stv2592}

\bibitem[{{Pejcha} {et~al.}(2017){Pejcha}, {Metzger}, {Tyles}, \&
  {Tomida}}]{Pejcha2017}
{Pejcha}, O., {Metzger}, B.~D., {Tyles}, J.~G., \& {Tomida}, K. 2017, \apj,
  850, 59, \dodoi{10.3847/1538-4357/aa95b9}

\bibitem[{{Pellegrino} {et~al.}(2022){Pellegrino}, {Howell}, {Vink{\'o}},
  {Gangopadhyay}, {Xiang}, {Arcavi}, {Brown}, {Burke}, {Hiramatsu},
  {Hosseinzadeh}, {Li}, {McCully}, {Misra}, {Newsome}, {Gonzalez}, {Pritchard},
  {Valenti}, {Wang}, \& {Zhang}}]{Pellegrino2022}
{Pellegrino}, C., {Howell}, D.~A., {Vink{\'o}}, J., {et~al.} 2022, \apj, 926,
  125, \dodoi{10.3847/1538-4357/ac3e63}

\bibitem[{{Perley} {et~al.}(2022){Perley}, {Ho}, {Petitpas}, \&
  {Keating}}]{Perley2022-at2022cmc-gcn}
{Perley}, D.~A., {Ho}, A.~Y.~Q., {Petitpas}, G., \& {Keating}, G. 2022, GRB
  Coordinates Network, 31627, 1

\bibitem[{{Perley} {et~al.}(2019){Perley}, {Mazzali}, {Yan}, {Cenko}, {Gezari},
  {Taggart}, {Blagorodnova}, {Fremling}, {Mockler}, {Singh}, {Tominaga},
  {Tanaka}, {Watson}, {Ahumada}, {Anupama}, {Ashall}, {Becerra}, {Bersier},
  {Bhalerao}, {Bloom}, {Butler}, {Copperwheat}, {Coughlin}, {De}, {Drake},
  {Duev}, {Frederick}, {Gonz{\'a}lez}, {Goobar}, {Heida}, {Ho}, {Horst},
  {Hung}, {Itoh}, {Jencson}, {Kasliwal}, {Kawai}, {Khanam}, {Kulkarni},
  {Kumar}, {Kumar}, {Kutyrev}, {Lee}, {Maeda}, {Mahabal}, {Murata}, {Neill},
  {Ngeow}, {Penprase}, {Pian}, {Quimby}, {Ramirez-Ruiz}, {Richer},
  {Rom{\'a}n-Z{\'u}{\~n}iga}, {Sahu}, {Srivastav}, {Socia}, {Sollerman},
  {Tachibana}, {Taddia}, {Tinyanont}, {Troja}, {Ward}, {Wee}, \&
  {Yu}}]{Perley2019}
{Perley}, D.~A., {Mazzali}, P.~A., {Yan}, L., {et~al.} 2019, \mnras, 484, 1031,
  \dodoi{10.1093/mnras/sty3420}

\bibitem[{{Perley} {et~al.}(2024{\natexlab{a}}){Perley}, {Qin}, {Rich},
  {Daddi}, {Collins}, {Gatkine}, {Neill}, {Hinds}, \& {McGurk}}]{Perley2024}
{Perley}, D.~A., {Qin}, Y., {Rich}, R.~M., {et~al.} 2024{\natexlab{a}},
  Transient Name Server AstroNote, 280, 1

\bibitem[{{Perley} {et~al.}(2024{\natexlab{b}}){Perley}, {Qin}, {Rich},
  {Daddi}, {Collins}, {Gatkine}, {Neill}, {Hinds}, \&
  {McGurk}}]{Perley24wppredshift}
---. 2024{\natexlab{b}}, Transient Name Server AstroNote, 280, 1

\bibitem[{{Planck Collaboration} {et~al.}(2020){Planck Collaboration},
  {Aghanim}, {Akrami}, {Ashdown}, {Aumont}, {Baccigalupi}, {Ballardini},
  {Banday}, {Barreiro}, {Bartolo}, {Basak}, {Battye}, {Benabed}, {Bernard},
  {Bersanelli}, {Bielewicz}, {Bock}, {Bond}, {Borrill}, {Bouchet}, {Boulanger},
  {Bucher}, {Burigana}, {Butler}, {Calabrese}, {Cardoso}, {Carron},
  {Challinor}, {Chiang}, {Chluba}, {Colombo}, {Combet}, {Contreras}, {Crill},
  {Cuttaia}, {de Bernardis}, {de Zotti}, {Delabrouille}, {Delouis}, {Di
  Valentino}, {Diego}, {Dor{\'e}}, {Douspis}, {Ducout}, {Dupac}, {Dusini},
  {Efstathiou}, {Elsner}, {En{\ss}lin}, {Eriksen}, {Fantaye}, {Farhang},
  {Fergusson}, {Fernandez-Cobos}, {Finelli}, {Forastieri}, {Frailis},
  {Fraisse}, {Franceschi}, {Frolov}, {Galeotta}, {Galli}, {Ganga},
  {G{\'e}nova-Santos}, {Gerbino}, {Ghosh}, {Gonz{\'a}lez-Nuevo}, {G{\'o}rski},
  {Gratton}, {Gruppuso}, {Gudmundsson}, {Hamann}, {Handley}, {Hansen},
  {Herranz}, {Hildebrandt}, {Hivon}, {Huang}, {Jaffe}, {Jones}, {Karakci},
  {Keih{\"a}nen}, {Keskitalo}, {Kiiveri}, {Kim}, {Kisner}, {Knox},
  {Krachmalnicoff}, {Kunz}, {Kurki-Suonio}, {Lagache}, {Lamarre}, {Lasenby},
  {Lattanzi}, {Lawrence}, {Le Jeune}, {Lemos}, {Lesgourgues}, {Levrier},
  {Lewis}, {Liguori}, {Lilje}, {Lilley}, {Lindholm}, {L{\'o}pez-Caniego},
  {Lubin}, {Ma}, {Mac{\'\i}as-P{\'e}rez}, {Maggio}, {Maino}, {Mandolesi},
  {Mangilli}, {Marcos-Caballero}, {Maris}, {Martin}, {Martinelli},
  {Mart{\'\i}nez-Gonz{\'a}lez}, {Matarrese}, {Mauri}, {McEwen}, {Meinhold},
  {Melchiorri}, {Mennella}, {Migliaccio}, {Millea}, {Mitra},
  {Miville-Desch{\^e}nes}, {Molinari}, {Montier}, {Morgante}, {Moss}, {Natoli},
  {N{\o}rgaard-Nielsen}, {Pagano}, {Paoletti}, {Partridge}, {Patanchon},
  {Peiris}, {Perrotta}, {Pettorino}, {Piacentini}, {Polastri}, {Polenta},
  {Puget}, {Rachen}, {Reinecke}, {Remazeilles}, {Renzi}, {Rocha}, {Rosset},
  {Roudier}, {Rubi{\~n}o-Mart{\'\i}n}, {Ruiz-Granados}, {Salvati}, {Sandri},
  {Savelainen}, {Scott}, {Shellard}, {Sirignano}, {Sirri}, {Spencer},
  {Sunyaev}, {Suur-Uski}, {Tauber}, {Tavagnacco}, {Tenti}, {Toffolatti},
  {Tomasi}, {Trombetti}, {Valenziano}, {Valiviita}, {Van Tent}, {Vibert},
  {Vielva}, {Villa}, {Vittorio}, {Wandelt}, {Wehus}, {White}, {White},
  {Zacchei}, \& {Zonca}}]{Cosmology20}
{Planck Collaboration}, {Aghanim}, N., {Akrami}, Y., {et~al.} 2020, \aap, 641,
  A6, \dodoi{10.1051/0004-6361/201833910}

\bibitem[{{Pursiainen} {et~al.}(2018){Pursiainen}, {Childress}, {Smith},
  {Prajs}, {Sullivan}, {Davis}, {Foley}, {Asorey}, {Calcino}, {Carollo},
  {Curtin}, {D'Andrea}, {Glazebrook}, {Gutierrez}, {Hinton}, {Hoormann},
  {Inserra}, {Kessler}, {King}, {Kuehn}, {Lewis}, {Lidman}, {Macaulay},
  {M{\"o}ller}, {Nichol}, {Sako}, {Sommer}, {Swann}, {Tucker}, {Uddin},
  {Wiseman}, {Zhang}, {Abbott}, {Abdalla}, {Allam}, {Annis}, {Avila}, {Brooks},
  {Buckley-Geer}, {Burke}, {Carnero Rosell}, {Carrasco Kind}, {Carretero},
  {Castander}, {Cunha}, {Davis}, {De Vicente}, {Diehl}, {Doel}, {Eifler},
  {Flaugher}, {Fosalba}, {Frieman}, {Garc{\'\i}a-Bellido}, {Gruen}, {Gruendl},
  {Gutierrez}, {Hartley}, {Hollowood}, {Honscheid}, {James}, {Jeltema},
  {Kuropatkin}, {Li}, {Lima}, {Maia}, {Martini}, {Menanteau}, {Ogando},
  {Plazas}, {Roodman}, {Sanchez}, {Scarpine}, {Schindler}, {Smith},
  {Soares-Santos}, {Sobreira}, {Suchyta}, {Swanson}, {Tarle}, {Tucker},
  {Walker}, \& {DES Collaboration}}]{Pursiainen2018}
{Pursiainen}, M., {Childress}, M., {Smith}, M., {et~al.} 2018, \mnras, 481,
  894, \dodoi{10.1093/mnras/sty2309}

\bibitem[{{Pursiainen} {et~al.}(2025){Pursiainen}, {Killestein},
  {Kuncarayakti}, {Charalampopoulos}, {Warwick}, {Lyman}, {Kotak}, {Leloudas},
  {Coppejans}, {Kravtsov}, {Maeda}, {Nagao}, {Taguchi}, {Ackley}, {Dhillon},
  {Galloway}, {Kumar}, {O'Neill}, {Ramsay}, \& {Steeghs}}]{Pursiainen25}
{Pursiainen}, M., {Killestein}, T.~L., {Kuncarayakti}, H., {et~al.} 2025,
  \mnras, 537, 3298, \dodoi{10.1093/mnras/staf232}

\bibitem[{{Quataert} {et~al.}(2019){Quataert}, {Lecoanet}, \&
  {Coughlin}}]{Quataert2019}
{Quataert}, E., {Lecoanet}, D., \& {Coughlin}, E.~R. 2019, \mnras, 485, L83,
  \dodoi{10.1093/mnrasl/slz031}

\bibitem[{{Renzo} {et~al.}(2020){Renzo}, {Farmer}, {Justham}, {G{\"o}tberg},
  {de Mink}, {Zapartas}, {Marchant}, \& {Smith}}]{Renzo2020}
{Renzo}, M., {Farmer}, R., {Justham}, S., {et~al.} 2020, \aap, 640, A56,
  \dodoi{10.1051/0004-6361/202037710}

\bibitem[{{Rest} {et~al.}(2018){Rest}, {Garnavich}, {Khatami}, {Kasen},
  {Tucker}, {Shaya}, {Olling}, {Mushotzky}, {Zenteno}, {Margheim},
  {Strampelli}, {James}, {Smith}, {F{\"o}rster}, \& {Villar}}]{Rest2018}
{Rest}, A., {Garnavich}, P.~M., {Khatami}, D., {et~al.} 2018, Nature Astronomy,
  2, 307, \dodoi{10.1038/s41550-018-0423-2}

\bibitem[{{Rybicki} \& {Lightman}(1979)}]{Rybicki1979}
{Rybicki}, G.~B., \& {Lightman}, A.~P. 1979, {Radiative processes in
  astrophysics}

\bibitem[{{Sadowski} \& {Narayan}(2015)}]{Sadowski15}
{Sadowski}, A., \& {Narayan}, R. 2015, \mnras, 453, 3213,
  \dodoi{10.1093/mnras/stv1802}

\bibitem[{{Sadowski} \& {Narayan}(2016)}]{Sadowski16}
---. 2016, \mnras, 456, 3929, \dodoi{10.1093/mnras/stv2941}

\bibitem[{{Sari} {et~al.}(1999){Sari}, {Piran}, \& {Halpern}}]{Sari1999}
{Sari}, R., {Piran}, T., \& {Halpern}, J.~P. 1999, \apjl, 519, L17,
  \dodoi{10.1086/312109}

\bibitem[{{Schroeder} {et~al.}(2024){Schroeder}, {Ho}, \&
  {Perley}}]{Schroeder2024}
{Schroeder}, G., {Ho}, A.~Y.~Q., \& {Perley}, D.~A. 2024, Transient Name Server
  AstroNote, 314, 1

\bibitem[{{Sfaradi} {et~al.}(2024){Sfaradi}, {Margutti}, {Farah}, {Wiston},
  {J}, {Bright}, {Chornock}, {LeBaron}, {Hammerstein}, {Brethauer}, {Laskar},
  {Siemion}, {Pollak}, {Sheikh}, {Sears}, \& {Migliori}}]{Sfaradi2024}
{Sfaradi}, I., {Margutti}, R., {Farah}, W., {et~al.} 2024, Transient Name
  Server AstroNote, 290, 1

\bibitem[{{Shvartzvald} {et~al.}(2024){Shvartzvald}, {Waxman}, {Gal-Yam},
  {Ofek}, {Ben-Ami}, {Berge}, {Kowalski}, {B{\"u}hler}, {Worm}, {Rhoads},
  {Arcavi}, {Maoz}, {Polishook}, {Stone}, {Trakhtenbrot}, {Ackermann},
  {Aharonson}, {Birnholtz}, {Chelouche}, {Guetta}, {Hallakoun}, {Horesh},
  {Kushnir}, {Mazeh}, {Nordin}, {Ofir}, {Ohm}, {Parsons}, {Pe'er}, {Perets},
  {Perdelwitz}, {Poznanski}, {Sadeh}, {Sagiv}, {Shahaf}, {Soumagnac}, {Tal-Or},
  {Santen}, {Zackay}, {Guttman}, {Rekhi}, {Townsend}, {Weinstein}, \&
  {Wold}}]{Shvartzvald2024-ULTRASAT}
{Shvartzvald}, Y., {Waxman}, E., {Gal-Yam}, A., {et~al.} 2024, \apj, 964, 74,
  \dodoi{10.3847/1538-4357/ad2704}

\bibitem[{{Sironi} \& {Spitkovsky}(2009)}]{Sironi2009}
{Sironi}, L., \& {Spitkovsky}, A. 2009, \apj, 698, 1523,
  \dodoi{10.1088/0004-637X/698/2/1523}

\bibitem[{{Sironi} \& {Spitkovsky}(2011)}]{Sironi2011}
---. 2011, \apj, 726, 75, \dodoi{10.1088/0004-637X/726/2/75}

\bibitem[{{Smith}(2014)}]{Smith14}
{Smith}, N. 2014, \araa, 52, 487, \dodoi{10.1146/annurev-astro-081913-040025}

\bibitem[{{Spitkovsky}(2008)}]{Spitkovsky2008}
{Spitkovsky}, A. 2008, \apjl, 682, L5, \dodoi{10.1086/590248}

\bibitem[{{Srinivasaragavan} {et~al.}(2024){Srinivasaragavan}, {Ho}, {Perley},
  {Andreoni}, {Rehentulla}, {Qin}, \& {Bellm}}]{Srinivasaragavan2024}
{Srinivasaragavan}, G., {Ho}, A., {Perley}, D., {et~al.} 2024, Transient Name
  Server AstroNote, 276, 1

\bibitem[{{Takahashi} \& {Ohsuga}(2017)}]{Takahashi2017}
{Takahashi}, H.~R., \& {Ohsuga}, K. 2017, \apjl, 845, L9,
  \dodoi{10.3847/2041-8213/aa8222}

\bibitem[{{Tampo} {et~al.}(2020){Tampo}, {Tanaka}, {Maeda}, {Yasuda},
  {Tominaga}, {Jiang}, {Moriya}, {Morokuma}, {Suzuki}, {Takahashi}, {Kokubo},
  \& {Kawana}}]{Tampo2020}
{Tampo}, Y., {Tanaka}, M., {Maeda}, K., {et~al.} 2020, \apj, 894, 27,
  \dodoi{10.3847/1538-4357/ab7ccc}

\bibitem[{{Tan} {et~al.}(2001){Tan}, {Matzner}, \& {McKee}}]{Tan2001}
{Tan}, J.~C., {Matzner}, C.~D., \& {McKee}, C.~F. 2001, \apj, 551, 946,
  \dodoi{10.1086/320245}

\bibitem[{{Tanaka} {et~al.}(2016){Tanaka}, {Tominaga}, {Morokuma}, {Yasuda},
  {Furusawa}, {Baklanov}, {Blinnikov}, {Moriya}, {Doi}, {Jiang}, {Kato},
  {Kikuchi}, {Kuncarayakti}, {Nagao}, {Nomoto}, \& {Taniguchi}}]{Tanaka2016}
{Tanaka}, M., {Tominaga}, N., {Morokuma}, T., {et~al.} 2016, \apj, 819, 5,
  \dodoi{10.3847/0004-637X/819/1/5}

\bibitem[{{Tauris} {et~al.}(2015){Tauris}, {Langer}, \&
  {Podsiadlowski}}]{Tauris2015}
{Tauris}, T.~M., {Langer}, N., \& {Podsiadlowski}, P. 2015, \mnras, 451, 2123,
  \dodoi{10.1093/mnras/stv990}

\bibitem[{{Titarchuk} \& {Seifina}(2021)}]{Titarchuk21}
{Titarchuk}, L., \& {Seifina}, E. 2021, \mnras, 501, 5659,
  \dodoi{10.1093/mnras/staa3961}

\bibitem[{{Tsuna} \& {Lu}(2025)}]{Tsuna25}
{Tsuna}, D., \& {Lu}, W. 2025, \apj, 986, 84, \dodoi{10.3847/1538-4357/add158}

\bibitem[{{Vurm} \& {Metzger}(2021)}]{Vurm21}
{Vurm}, I., \& {Metzger}, B.~D. 2021, \apj, 917, 77,
  \dodoi{10.3847/1538-4357/ac0826}

\bibitem[{{Waxman} \& {Shvarts}(1993)}]{Waxman1993}
{Waxman}, E., \& {Shvarts}, D. 1993, Physics of Fluids A, 5, 1035,
  \dodoi{10.1063/1.858668}

\bibitem[{{Weiler} {et~al.}(2002){Weiler}, {Panagia}, {Montes}, \&
  {Sramek}}]{Weiler2002}
{Weiler}, K.~W., {Panagia}, N., {Montes}, M.~J., \& {Sramek}, R.~A. 2002,
  \araa, 40, 387, \dodoi{10.1146/annurev.astro.40.060401.093744}

\bibitem[{{Weiler} {et~al.}(2007){Weiler}, {Williams}, {Panagia}, {Stockdale},
  {Kelley}, {Sramek}, {Van Dyk}, \& {Marcaide}}]{Weiler2007}
{Weiler}, K.~W., {Williams}, C.~L., {Panagia}, N., {et~al.} 2007, \apj, 671,
  1959, \dodoi{10.1086/523258}

\bibitem[{{Woosley}(2017)}]{Woosley2017}
{Woosley}, S.~E. 2017, \apj, 836, 244, \dodoi{10.3847/1538-4357/836/2/244}

\bibitem[{{Wu} \& {Fuller}(2022)}]{Wu2022}
{Wu}, S.~C., \& {Fuller}, J. 2022, \apjl, 940, L27,
  \dodoi{10.3847/2041-8213/ac9b3d}

\bibitem[{{Yao} {et~al.}(2021){Yao}, {Ho}, {Medvedev}, {Nayana A.}, {Perley},
  {Kulkarni}, {Chandra}, {Sazonov}, {Gilfanov}, {Khorunzhev}, {Khatami}, \&
  {Sunyaev}}]{Yao2021-mrf}
{Yao}, Y., {Ho}, A. Y.~Q., {Medvedev}, P., {et~al.} 2021, arXiv e-prints,
  arXiv:2112.00751.
\newblock \doarXiv{2112.00751}

\bibitem[{{Yoshioka} {et~al.}(2024){Yoshioka}, {Mineshige}, {Ohsuga},
  {Kawashima}, \& {Kitaki}}]{Yoshioka24}
{Yoshioka}, S., {Mineshige}, S., {Ohsuga}, K., {Kawashima}, T., \& {Kitaki}, T.
  2024, \pasj, 76, 1015, \dodoi{10.1093/pasj/psae067}

\bibitem[{{Yuan} {et~al.}(2016){Yuan}, {Wang}, {Lei}, {Gao}, \&
  {Zhang}}]{Yuan2016}
{Yuan}, Q., {Wang}, Q.~D., {Lei}, W.-H., {Gao}, H., \& {Zhang}, B. 2016,
  \mnras, 461, 3375, \dodoi{10.1093/mnras/stw1543}

\end{thebibliography}
\bibliographystyle{aasjournal}

\end{document}